\documentclass[usenatbib]{mnras}

\usepackage{float}
\usepackage{enumitem}

\usepackage[english]{babel}
\usepackage{amsmath}
\usepackage{amssymb}
\usepackage{graphicx}
\usepackage{tikz}
\usepackage{lipsum,adjustbox}
\usetikzlibrary{shapes,arrows}
\usepackage{breqn}
\usepackage{appendix}

\newcommand{\mstar}{$M_{\ast}$}

\newcommand{\aco}{$\alpha_{CO}$}
\newcommand{\cii}{[C {\sc ii}]}
\newcommand{\edit}{ }

\usepackage{etoolbox}
\makeatletter
\newcount\c@additionalboxlevel
\newcount\c@maxboxlevel
\patchcmd\@combinedblfloats{\box\@outputbox}{%
  \stepcounter{additionalboxlevel}%
  \box\@outputbox
}{}{\errmessage{\noexpand\@combinedblfloats could not be patched}}

\AtBeginShipout{%
  \ifnum\value{additionalboxlevel}>\value{maxboxlevel}%
    \typeout{Warning: maxboxlevel might be too small, increase to %
      \the\value{additionalboxlevel}%
    }%
  \fi 
  \@whilenum\value{additionalboxlevel}<\value{maxboxlevel}\do{%
    \typeout{* Additional boxing of page `\thepage'}%
    \setbox\AtBeginShipoutBox=\hbox{\copy\AtBeginShipoutBox}%
    \stepcounter{additionalboxlevel}%
  }%
  \setcounter{additionalboxlevel}{0}%
}
\makeatother
\begin{document}
\title[xCOLD GASS: multivariate CO-to-H$_2$ conversion function]{Deriving a multivariate CO-to-H$_2$ conversion function using the [C\small II\huge]/CO(1-0) ratio and its application to molecular gas scaling relations} 
\author[G. Accurso et al. ]{G. Accurso$^{1}$\thanks{E-mail:
gioacchino.accurso.13@ucl.ac.uk}, A. Saintonge$^{1}$\thanks{E-mail:
a.saintonge@ucl.ac.uk}, B. Catinella$^{2}$, L. Cortese$^{2}$, R. Dav\'e$^{3}$, S. H. Dunsheath$^{4}$,
\newauthor{R. Genzel$^{5}$, J. Gracia-Carpio$^{5}$, T. M. Heckman$^{6}$, Jimmy$^{7}$, C. Kramer$^{8}$, Cheng Li$^{9, 10}$,} 
\newauthor{K. Lutz$^{2}$, D. Schiminovich$^{11}$, K. Schuster$^{12}$, A. Sternberg$^{13}$, E. Sturm$^{5}$, L. J. Tacconi$^{5}$,} 
\newauthor{K. V. Tran$^{7}$ and J. Wang$^{14}$.}  \\
$^{1}$Deptartment of Physics and Astronomy, University College London, Gower Street, London, WC1E 6BT, UK \\
$^{2}$International Center for Radio Astronomy Research, M468, The University of Western Australia, 35 Stirling Highway, Crawley, \\WA 6009, Australia \\
$^{3}$University of the Western Cape, Bellville, Cape Town 7535, South Africa \\
$^{4}$California Institute of Technology, Mail Code 301-17, 1200 E. California Blvd., Pasadena, CA 91125, USA \\
$^{5}$Max-Planck-Institut f\"ur extraterrestrische Physik, 85741, Garching, Germany \\
$^{6}$Center for Astrophysical Sciences, Department of Physics and Astronomy, Johns Hopkins University, Baltimore, MD 21218, USA \\
$^{7}$George P. and Cynthia W.Mitchell Institute for Fundamental Physics and Astronomy, Department of Physics and Astronomy, \\ Texas A$\&$M University, College Station, TX 77843, USA. \\
$^{8}$Instituto de Radioastronomia Milimetrica (IRAM), Av. Divina Pastora 7, Nucleo Central, 18012 Granada, Spain \\
$^{9}$Tsinghua Center for Astrophysics and Physics Department, Tsinghua University, Beijing 100084, China \\
$^{10}$Shanghai Astronomical Observatory, Shanghai 200030, China \\
$^{11}$Department of Astronomy, Columbia University, 550 West 120th Street, New York, NY 10027, USA \\
$^{12}$Institut de Radioastronomie Millimetrique, 300 rue de la Piscine, F-38406, Saint Martin dÕH`eres, France \\
$^{13}$Tel Aviv University, Sackler School of Physics \& Astronomy, Ramat Aviv 69978, Israel \\
$^{14}$CSIRO Astronomy and Space Science, Australia Telescope National Facility, PO Box 76, Epping, NSW 1710, Australia} 

\maketitle

\date{Accepted Date Received Date; in original form Date}


\begin{abstract}
\normalsize We present {\it Herschel} PACS observations of the \cii\ 158$\mu$m emission line in a sample of 24 intermediate mass ($9<\log M_{\ast}/M_{\odot}<10$) and low metallicity ($0.4< Z/Z_{\odot}<1.0$) galaxies from the xCOLD GASS survey. Combining them with IRAM CO(1-0) measurements, we establish scaling relations between integrated and molecular region $L_{\mbox{\cii}}$/$L_{\mbox{CO(1-0)}}$ ratios as a function of integrated galaxy properties. A Bayesian analysis reveals that only two parameters, metallicity and offset from the star formation main sequence, $\Delta$(MS), are needed to quantify variations in the luminosity ratio; metallicity describes the total dust content available to shield CO from UV radiation, while $\Delta$(MS) describes the strength of this radiation field. We connect the $L_{\mbox{\cii}}$/$L_{\mbox{CO(1-0)}}$ ratio to the CO-to-H$_{2}$ conversion factor and find a multivariate conversion function \aco, which can be used up to z$\sim$2.5. {\edit This function depends primarily on metallicity, with a second order dependence on $\Delta$(MS)}. We apply this to the full xCOLD GASS and PHIBSS1 surveys and investigate molecular gas scaling relations. We find a flattening of the relation between gas mass fraction and stellar mass at $\log$\mstar$<10.0$.  While the molecular gas depletion time varies with sSFR, it is mostly independent of mass, indicating that the low L$_{CO}$/SFR ratios long observed in low mass galaxies are entirely due to photodissociation of CO and not to an enhanced star formation efficiency.
\end{abstract}


\section{Introduction}\label{intro}
Observations of the cold interstellar medium (ISM) of normal star-forming galaxies at low and high redshifts have recently helped to establish an ``equilibrium model'' for galaxy evolution. Under this framework, galaxy growth is controlled by the total available gas reservoir through the interplay of gas inflows and outflows, and the efficiency of the star formation process \citep[e.g.][]{1991ApJ...379...52W, 2010ApJ...718.1001B, 2012MNRAS.421...98D,2013ApJ...772..119L}. Trying to understand what triggers and drives star formation, how the chemical enrichment of the ISM proceeds, and what causes the growth of galaxies is intricately linked to understanding their total gas content. Being able to measure the amount and properties of cold gas in a very wide range of galaxies is therefore of great importance to further our understanding of galaxy evolution.

It has been well established that most star formation in the Milky Way and nearby galaxies occurs in dense giant molecular clouds (GMCs), and that most of this molecular gas is in the form of cold H$_{2}$  \citep[e.g.][]{2012ARA&A..50..531K, 1987ApJ...319..730S, 1991ARA&A..29..581Y, 2008AJ....136.2782L}. However, with the molecule lacking a permanent electric dipole moment, cold H$_{2}$ is not directly observable, and it is common practice to instead trace cold molecular gas through the low-lying rotational transitions of the second most abundant molecule, $^{12}$CO \citep{2013seg..book..491S}. The ground rotational transition of CO has a low excitation temperature, 5.53K, and a low critical density, 700 cm$^{-3}$, making it easily excited in cold molecular clouds \citep{1986ApJ...309..326D}. At a wavelength of 2.6mm the ground state falls within the Earth's atmospheric window allowing it to be easily observed from ground based facilities. Given all this, it has become the workhorse tracer to quantify the total molecular gas reservoir in galaxies both near and distant \citep{2013ARA&A..51..207B}. 

Although the CO rotational transitions are optically thick in typical conditions, the information on the total molecular gas mass is contained in the width of the line under the assumption that GMCs are virialized and that the line emission is the superposition of a number of such virialized clouds. The correlation between the velocity-integrated line luminosity L$_{CO}^{'}$ in the $1$ $\rightarrow$ $0$ transition and the total molecular gas mass is given by the empirical relation \citep{1986ApJ...309..326D, 2009MNRAS.394.1857O}:
\begin{dmath}
\centering
M_{H_{2}} = \alpha_{CO}  L_{CO(1-0)}^{'}.
\end{dmath}
Here M$_{H_{2}}$ has units of M$_{\odot}$ and L$_{CO(1-0)}^{'}$ (K km s$^{-1}$ pc$^{2}$), the integrated line luminosity, is related to the observed velocity integrated flux density, I$_{CO} \Delta v$ (Jy km s$^{-1}$), following \citet{1997ApJ...478..144S}:
\begin{dmath}
\centering
L_{CO(1-0)}^{'} = 3.25\times 10^{7} I_{CO} \Delta v  \nu_{obs}^{-2} D_{L}^{2} (1+z)^{-3} 
\label{solomon_equation} 
\end{dmath}
where the frequency is in GHz \citep{1987ApJ...319..730S} and the luminosity distance, D$_{L}$, is in Mpc. Thus $\alpha_{CO}$, the CO-to-H$_{2}$ conversion factor, can be considered a mass to light ratio. Across observations of the Milky Way and nearby star-forming galaxies with approximately solar metallicities, the empirical CO(1-0) conversion factors are consistent with a typically value of 4.36 M$_{\odot}$ (K km s$^{-1})^{-1}$, which includes a 36$\%$ correction for Helium gas \citep{1996A&A...308L..21S, 2010ApJ...710..133A}. 

However CO has been difficult to detect in local low mass galaxies. Sensitivity has limited CO detections to galaxies with $Z/Z_{\odot} \geq $ 0.1 \citep{2009AJ....137.4670L, 2012AJ....143..138S, 2014A&A...563A..31R}. Does this mean that galaxies with even lower metallicities have very little molecular gas, or does CO become a poor tracer of the molecular ISM in low metallicity conditions? Indeed there is strong evidence that not all of the H$_2$ is traced by CO emission; UV radiation from massive stars destroys CO to a cloud depth of a few $A_V$, which can correspond to a significant fraction of the total gas column in low metallicity clouds \citep{1986ApJS...62..109V, 2010ApJ...716.1191W}. While H$_{2}$ is self-shielded from this UV radiation, CO relies on dust shielding and therefore, in low metallicity star forming galaxies which have hard radiation fields and lower dust-to-gas mass ratios, CO is easily photodissociated into C$^{+}$ and O \citep{1995ApJ...454..293P,2006A&A...451..917R}. 

The regions where molecular hydrogen undergoes a dissociation transition into neutral hydrogen are suitably named photodissociation regions (PDRs hereafter) and it is here that CO is also photodissociated. In this case, the CO flux per fixed hydrogen column is less than in high metallicity environments and, in turn, the Galactic conversion factor globally underestimates the true molecular hydrogen content \citep{1996PASJ...48..275A, 2011ApJ...737...12L}.  This molecular gas, not traced by CO emission, has been referred to as the ``dark gas'' \citep{2010ApJ...716.1191W, 2005Sci...307.1292G} and emits brightly in other fine-structure PDR tracers such as [OI] and \cii. In particular, \cii\ is a promising tracer to quantify the total dark molecular gas reservoir. It was first used in a low metallicity dwarf galaxy by \citet{1997ApJ...483..200M} and is the focus of this work. 

The \cii\ 158$\mu$m emission line is one of the strongest coolants of the interstellar medium and can contribute up to a few percent of the total FIR emission from a galaxy \citep{1985ApJ...291..722T}. Ionised carbon has a lower ionisation potential than hydrogen (11.3 eV instead of 13.6 eV), and the \cii\ line lies 92K above the ground state with a critical density for collisions with neutral hydrogen of $3 \times 10^3$ cm$^{-3}$ \citep{1999ApJ...527..795K}. This means \cii\ is produced not only in PDRs, but also in the ionised and atomic phases of the ISM.  Measurements of the \cii\ emission originating from the PDRs combined with measurements of CO (which can only arise in PDRs), can be used {\bf a)} to trace the total molecular reservoir on galaxy wide scales and {\bf b)} to quantify the nature of variations of the conversion function. We here use the nomenclature of a conversion `function' because this quantity does vary as a function of ISM properties.

There have been a significant effort, both observationally and theoretically, made to establish a robust prescription for the conversion function \citep[e.g.][]{1995ApJ...448L..97W, 1997A&A...328..471I, 2002A&A...384...33B, 2003A&A...404..495I, 2011ApJ...737...12L, 2012ApJ...746...69G, 2012AJ....143..138S, 2013ARA&A..51..207B, 2013ApJ...777....5S}. However a consensus has not yet been reached,  and some empirical methods rely on the assumption of a universal gas depletion timescale. To disentangle possible systematic variations in depletion time from the issues of CO in low metallicity environments, a possibility is to use multi-wavelength observations to investigate the properties of the ISM \citep[e.g.][]{2012A&A...548A..22M}.

By measuring carbon both in its molecular and ionised form in PDRs in a statistically robust sample of galaxies, with a full suite of multi-wavelength observations, it is possible to investigate any secondary dependencies on the conversion function. Such a dataset would allow us to fully parametrise and quantify variations in the conversion function with global galaxy properties. In this paper we present CO(1-0) and \cii\ observations of 24 intermediate mass ($9<\log M_{\ast}/M_{\odot}<10$) and low metallicity ($0.4<Z/Z_{\odot}<1.0$) galaxies from the xCOLD GASS survey. We combine this with spectroscopic CO(1-0) and \cii\ data from the Dwarf Galaxy Survey (DGS hereafter) \citep{2014A&A...564A.121C, 2013PASP..125..600M, 2014A&A...563A..31R} to probe galaxies that are even more metal-poor.

In Section \ref{survey_description} we present an overview of the survey and details regarding the data reduction of the IRAM CO(1-0), Herschel \cii\ observations and auxiliary data from WISE, GALEX and SDSS. In Section \ref{obs_results_not_corrected} we provide scaling relations between integrated L$_{\mbox{[C\scriptsize II]}}$/L$_{\mbox{CO(1-0)}}$ and global galaxy parameters. In Section \ref{contamination} we explore the contribution of the \cii\ emission from non-molecular regions and proceed to use a Bayesian Inference method and radiative transfer modelling to retrieve a prescription for the conversion function in Sections \ref{statsmethod} and \ref{going_to_alpha}. Finally we discuss the implications of our new prescription on molecular gas scaling relations in Section \ref{discussion}.

Throughout this paper we use a standard flat $\Lambda \mbox{CDM}$ cosmology with $H_{0} = 70$ km s$^{-1}$ Mpc$^{-1}$.

\section{Survey Description and Sample Selection}\label{survey_description}

The extension to COLD GASS is a randomly selected sample of  galaxies from the regions of overlap between the SDSS \citep{2002AJ....123..485S}, GALEX \citep{2005ApJ...619L...1M}, WISE \citep{2010AJ....140.1868W} and ALFALFA HI \citep{2005AJ....130.2598G} surveys. It is an unbiased sub-sample of all the galaxies in the redshift range 0.01$<$ z $<$0.02 and is stellar mass selected (9$<$ log M$_{*}$/M$_{\odot}$ $<$ 10). There is no other selection criteria based on colour, star formation rate, etc. The SDSS data provide us with optical imaging and spectroscopy over the central 3'' of our galaxies and with the GALEX and WISE data we have FUV, NUV, 3.4$\mu$m, 4.5$\mu$m, 12$\mu$m and 22$\mu$m photometry. Moreover, we have HI fluxes from observations at the Arecibo observatory as part of the GASS survey \citep{2010MNRAS.403..683C, 2013MNRAS.436...34C}. The full xCOLD GASS sample contains 133 galaxies, which will be presented in full alongside all the CO(1-0) IRAM observations in Saintonge et al. 2017. 

Our target selection for Herschel PACS observations involved removing from the total sample of 133 xCOLD GASS objects the passive elliptical galaxies, where \citet{2011MNRAS.415...32S} showed that molecular gas contributes insignificantly to the mass budget (M$_{\mbox{H_{2}}}$/M$_{*}$ $<$ 0.2\%), and one galaxy which already had adequate PACS observations. After this a mass-selected sample of 103 galaxies remain which allows for roughly 20 galaxies in 5 stellar mass bins between 9$<$ log M$_{*}$/M$_{\odot}$ $<$ 10. However the Herschel telescope exhausted its supply of liquid helium coolant midway through our observations and so, out of those 103 galaxies which were originally proposed, we here present the 24 galaxies for which the \cii\ 158$\mu$m line was observed with PACS. 

In order to provide a statistically robust prescription of the conversion function it is imperative to probe as large a parameter space as possible, hence we combine our data with literature data if possible. Unfortunately multi-wavelength data sets with accurate galaxy parameter measurements and observations of both CO and \cii\ are rare. \citet{2014A&A...564A.121C} do present data for a compilation of galaxies from the DGS which have good CO(1-0) and \cii\ observations, which we therefore add to the xCOLD GASS objects. We are interested in objects with total galaxy-wide integrated detections in CO(1-0) and [C\scriptsize II\normalsize]; we therefore have to eliminate all objects which have optical diameters greater than 47'' (the PACS IFU map size), and objects which have resolved interferometric CO observations as aperture corrections are impossible due to the unresolved nature of the Herschel PACS data. 

Overall we are able to add seven extremely low metallicity galaxies, $0.2Z_{\odot}< Z < 0.5Z_{\odot}$, from the DGS survey; these are Haro 11, Mrk 930, Haro 3, Mrk 1089, UM 448, Haro 2 and II Zw 40. Derived galaxy parameters are found within \citet{2013PASP..125..600M} while observed [C\scriptsize II\normalsize]/CO(1-0) ratios, with aperture corrections, and star formation rates are calculated here. We fold these galaxies into our sample and derive all necessary measurements consistently as detailed below.

\subsection{Optical, UV \& IR data}\label{opticaluvir}
For the xCOLD GASS galaxies parameters such as redshifts, sizes and magnitudes are retrieved from the SDSS DR7 database \citep{2009ApJS..182..543A}. We retrieve stellar masses and emission line fluxes from the MPA-JHU catalogue\footnote{The data catalogues are available from
{\color{blue} http://www.mpa-garching.mpg.de/SDSS/}} where calculations were performed using the methods presented in \citet{2004ApJ...613..898T} and \citet{2003MNRAS.341...33K}. With these, we then calculate the gas-phase metallicity of the galaxies using the prescription from \citet{2004MNRAS.348L..59P} (PP04 hereafter):
\begin{dmath}
12 + \log(\mbox{O}/\mbox{H}) = 8.73 - 0.32 \log \left ( \frac{\mbox{[OIII]}}{\mbox{H}_{\beta}} / \frac{\mbox{[NII]}}{\mbox{H}_{\alpha}} \right)
\end{dmath}
which was calibrated down to the low metallicity range probed here. On this scale a value of $12 + \log(O/H) = 8.69$ corresponds to solar metallicity following the oxygen abundance of \citet{2009ARA&A..47..481A}. The stellar mass surface density is defined as:
\begin{dmath}
\centering
\mu_{*} = \frac{\mbox{M$_{*}$}}{2\pi \mbox{R}^{2}_{50,z}} 
\end{dmath}
where R$_{50,z}$ is the z-band 50\% flux intensity Petrosian radius, in kiloparsecs, taken from the SDSS database.

For the DGS galaxies stellar masses and redshifts are taken from the catalogue presented in \citet{2013PASP..125..600M}, with an explanation on how these quantities were derived, while metallicities are taken from the literature and converted to PP04 units using the methods from \citet{2008ApJ...681.1183K}.

To account for unobscured star formation we use GALEX FUV and NUV images retrieved from the public GALEX All-sky and Medium Imaging surveys \citep[AIS and MIS respectively,][]{2005ApJ...619L...1M}, and for the obscured star formation we make use of WISE imaging. For both GALEX and WISE maps we perform aperture photometry using a similar data reduction technique to \citet{2010MNRAS.401..433W}. Total SFRs from the combination of UV and IR photometry are then calculated as presented in \citet{2016arXiv160705289S} and we do this for both the xCOLD GASS and DGS samples.

From this it is possible to calculate, for both xCOLD GASS and DGS galaxies, the effective UV attenuation, A$_{IRX}$, where the log quantity is defined in \citet{2013ApJ...778....2S} as:
\begin{dmath}\label{exctinction_correction}
\centering
A_{\mbox{IRX}} = \left(\frac{\mbox{SFR}_{\mbox{IR}}}{\mbox{SFR}_{\mbox{UV}}} + 1\right)^{2.5}
\end{dmath}. 
This is another important quantity which may correlate with L$_{\mbox{[C\scriptsize II]}}$/L$_{\mbox{CO(1-0)}}$ - we express it in this form, as opposed to the log quantity, for mathematical convenience when performing the statistical analysis described later in Section \ref{statsmethod}. We also want to allow for the possibility of a redshift dependence in our conversion function and so we measure the distance off the main sequence for each galaxy. Using the analytical definition of the main sequence by \citet{2012ApJ...754L..29W}, where:
\begin{dmath}
\centering
\log (\mbox{sSFR}_{ms}(z, M_{*})) = -1.12 + 1.14z -0.19z^{2} - (0.3 + 0.13z) \times (\log M_{*} - 10.5)  [\mbox{Gyr}^{-1}]
\end{dmath} 
with $z$ and M$_{\ast}$ denoting redshift and stellar mass, we can then define the distance off the main sequence as:
\begin{dmath}\label{deltamsdefinition}
\centering
\Delta(\mbox{MS}) = \frac{\mbox{sSFR}_{\mbox{measured}}}{\mbox{sSFR}_{ms}(z, M_{*})}
\end{dmath}
which is applicable up to z$\sim$2.5, as stated in \citet{2012ApJ...754L..29W}.

\subsection{DSS observations and data reduction}
To measure {\it r}-band magnitudes in a homogenous way between the xCOLD GASS and DGS samples we perform aperture photometry here; the DGS galaxies do not have SDSS photometric measurements unlike the former and so we utilise the ESO DSS (Digital Sky Survey). {\it R}-band images were downloaded from the ESO DSS Online Archive\footnote{\url{http://archive.eso.org/dss/dss}}. Photometry was then carried out using the same apertures as for the GALEX and WISE images, allowing for the calculation of consistent, aperture-matched NUV$-r$ colours. 

\begin{figure*}
\centering 
\includegraphics[scale=0.8]{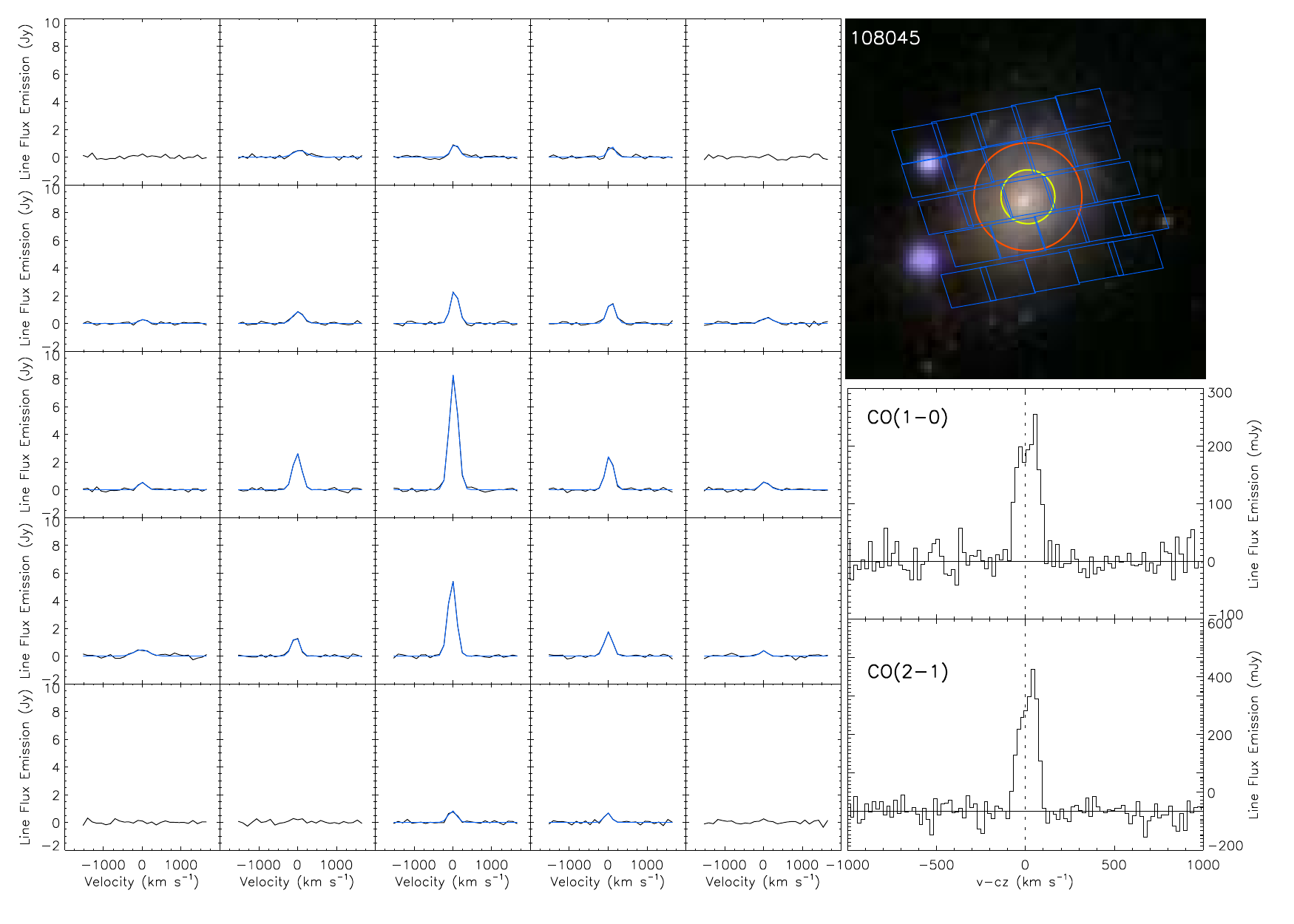}
\caption{Example of the IRAM and {\it Herschel} data products for an example galaxy. {\it Left:}  Spectra centered on the \cii\ 158$\mu$m emission line for each of the 16 PACS spaniels, with the best Gaussian fit to the lines detected with S/N$>3$ also shown. {\it Top right:} SDSS image showing the position of the PACS spaniels (blue squares) and of the IRAM beam which was used to measure the CO(1-0) and CO(2-1) lines, in red and yellow, respectively. {\it Bottom right:} CO(1-0) and CO(2-1) spectra from the IRAM 30m.}
\label{example_galaxy_herschel_iram_data} 
\end{figure*}

\subsection{Herschel observations and data reduction}\label{herschel_obs}
Twenty-four of the galaxies within the xCOLD GASS survey were observed with the PACS spectrometer \citep{2010A&A...518L...2P} onboard Herschel \citep{2010A&A...518L...1P} as part of the programme OT2\textunderscore asainton\textunderscore 1, P.I. A. Saintonge. The seven DGS galaxies were observed as part of a guaranteed time key programme.

The PACS array is composed of 5x5 square spatial pixels each of side 9.4'', covering a total field of view of 47''. The observations were carried out in line-spectroscopy mode. The data were reduced using the Herschel Interactive Processing Environment (HIPE) \citep{2010ASPC..434..139O}. We used standard scripts of the PACS spectrometer pipeline to reduce the data from the raw product to its level 2 processed form. From this, line fitting was done using IDL scripts, written by the first author, employing routines within the IDL Astronomy User's Library (\url{http://idlastro.gsfc.nasa.gov/}).  The lines are well fitted with a single gaussian using the IDL fitting routine GAUSSFIT. The signal from each spaxel of the PACS array is fitted with a second order polynomial with the addition of a Gaussian for the baseline and emission line respectively. From this we calculate the S/N in each spaxel. To calculate the total flux for each individual galaxy, we add up the integrals of the fitted Gaussians from all the spaxels which have a S/N $>$ 3. For total uncertainties we assume a 30\% error of the total \cii\ line flux, due to calibration errors \citep{2010A&A...518L...2P}. With Herschel all 24 xCOLD GASS galaxies show a clear detection in \cii\ with a S/N $>$ 3, while all seven DGS galaxies also show clear detections.

We want to directly compare the \cii\ line luminosities with the CO(1-0) luminosity and to do this we convolve the PACS spaxel array with the IRAM CO(1-0) beam, detailed in Section \ref{iram_obs}. We approximate the 2-D IRAM beam point spread function as a Gaussian, with a FWHM of 21.4'', where the peak is normalised to one\footnote{The IRAM beam width and beam efficiencies used in these calculation can be found at \url{http://www.iram.es/IRAMES/mainWiki/Iram30mEfficiencies}.}. We ensure the peak of this Gaussian is set to one to give maximum weighting to the central spaxel. We do this for all the xCOLD GASS objects. For the DGS objects we do different aperture corrections depending on the size of the beam PSF used for each individual CO(1-0) measurement \citet{2014A&A...563A..31R}.

\subsection{IRAM observations and data reduction}\label{iram_obs}

Observations were carried out at the IRAM 30m Telescope, as part of the xCOLD GASS survey, using the Eight Mixer Receiver (EMIR) \citep{2012A&A...538A..89C} to observe the CO(1$-$0) line, which has a rest frequency of 115.27 GHz. EMIR allows for observations on two sidebands with 8GHz bandwidth per sideband per polarisation.  The second band was tuned to  cover the redshifted CO(2-1)\footnote{The analysis of this paper will focus on the CO(1-0) data only. The CO(2-1) data will be presented in a future publication.}. The observing strategy was identical to that described in \citet{2011MNRAS.415...61S} for the initial COLD GASS survey. 

The xCOLD GASS data are reduced using the Continuum and Line Analysis Single-dish Software (CLASS\footnote{\url{http://www.iram.fr/IRAMFR/GILDAS/doc/html/class-html/class.html}}) which is part of the GILDAS software. All scans are visually inspected and those with anomalous features, such as distorted baselines or increased noise due to poor atmospheric conditions or high water vapour levels, are discarded. The individual scans for a single galaxy are baseline-subtracted, using a first order fit, and then combined. This average spectrum is then binned to achieve a resolution of $\sim$20kms$^{-1}$ and the rms is obtained and recorded. The flux of the CO(1$-$0) line is measured by adding the signal within an appropriately defined frequency window; in the case of a detection this window is set by hand to match the observed line profile. For the null detection scenario the window is set to a width of 200kms$^{-1}$ or to the full width of the HI line. Uncertainties on the CO(1-) line flux is calculated as:
\begin{dmath}\label{obs_uncertainities}
\centering
\epsilon_{obs} = \frac{\sigma_{rms}W50_{CO}}{ \sqrt{W50_{CO}\Delta w^{-1}_{ch}} }
\end{dmath}
where $\Delta$ w$_{ch}$, the width of each spectral channel, is equal to $\sim$21km s$^{-1}$ with mild variations due to differing redshifts. The rms noise per spectral channel is denoted by $\sigma$$_{rms}$ and W50$_{CO}$ is the FWHM of the CO(1$-$0) line. 

As mentioned in Section \ref{herschel_obs}, we convolved the PACS spaxel array with the IRAM CO(1-0) beam to measure the $L_{\mbox{[C\scriptsize II]}}$/$L_{\mbox{CO(1-0)}}$ \normalsize ratio over the footprint of the IRAM beam (or for the different beam FWHMs for the CO(1-0) observations in the DGS sample). As the angular size of the xCOLD GASS galaxies is larger than the IRAM beam, small aperture corrections need to be applied to obtain total CO fluxes. We calculated these aperture corrections as in \citet{2012ApJ...758...73S}, and apply them to obtain total molecular gas masses when investigating the molecular gas scaling relations, in Section \ref{discussion}.

We plot the IRAM and Herschel spectroscopic data for an example galaxy in Figure \ref{example_galaxy_herschel_iram_data}. Similar plots for all of our 24 galaxies can be found in Appendix \ref{AppA}. For the DGS galaxies, we use the CO(1-0) fluxes compiled by \citet{2014A&A...563A..31R}. The measured \cii\ and CO(1-0) line luminosities, as well as key physical parameters are given for the xCOLD GASS objects in Table \ref{sdss_data_table_cg}, and for the DGS galaxies in Table \ref{sdss_data_table_dgs}.

\begin{table*}
  \begin{tabular}{@{}ccccccccc@{}}
  \hline
   xCOLD GASS ID      &  Log M$_{*}$  & Redshift & SFR &  Metallicity  &  Log [C\scriptsize II\normalsize] & Log CO(1-0) & Log sSFR & f$_{\mbox{\scriptsize{[C\tiny II\scriptsize] mol}}}$  \\   
   & [\mbox{M}$_{\odot}$] & & [\mbox{M}$_{\odot}$ \mbox{yr}$^{-1}$] & [O/H] & [\mbox{L}$_{\odot}$] & [\mbox{L}$_{\odot}$] &  [\mbox{yr}$^{-1}$] & \%  \\
 \hline
107026 & 9.37 & 0.0167 & 0.34 & 8.55 & 6.65 & 2.92 & -9.84 & 67.21 \\
108064 & 9.56 & 0.015 & 0.75 & 8.64 & 7.05 & 3.45 & -9.68 & 64.4 \\
108080 & 9.31 & 0.0153 & 0.05 & 8.7 & 5.81 & 2.99 & -10.61 & 76.8 \\
108093 & 9.94 & 0.0168 & 0.68 & 8.78 & 7.48 & 3.98 & -10.11 & 71.3 \\
108113 & 9.88 & 0.0199 & 5.12 & 8.79 & 6.99 & 3.75 & -9.17 & 53.62 \\
108129 & 9.47 & 0.0175 & 0.19 & 8.45 & 6.39 & 3.11 & -10.19 & 72.37 \\
108142 & 9.61 & 0.019 & 0.71 & 8.7 & 7.27 & 3.36 & -9.76 & 65.84 \\
108147 & 9.64 & 0.0197 & 0.62 & 8.67 & 7.14 & 3.43 & -9.85 & 67.37 \\
101037 & 9.31 & 0.0164 & 0.27 & 8.61 & 6.93 & 3.22 & -9.88 & 67.86 \\
108050 & 9.75 & 0.0162 & 0.76 & 8.82 & 7.32 & 3.76 & -9.87 & 67.7 \\
108021 & 9.64 & 0.0173 & 0.37 & 8.82 & 7.2 & 3.86 & -10.07 & 70.75 \\
109066 & 9.99 & 0.0149 & 0.14 & 8.71 & 6.79 & 3.38 & -10.84 & 78.42 \\
109101 & 9.01 & 0.0104 & 0.15 & 8.52 & 6.31 & 2.31 & -9.83 & 67.04 \\
109038 & 9.21 & 0.0116 & 0.66 & 8.4 & 7.24 & 3.11 & -9.39 & 58.62 \\
109139 & 9.59 & 0.0188 & 0.93 & 8.78 & 7.25 & 3.6 & -9.62 & 63.28 \\
109092 & 9.7 & 0.018 & 0.58 & 8.68 & 7.12 & 3.58 & -9.94 & 68.82 \\
109072 & 9.68 & 0.0189 & 0.73 & 8.79 & 7.22 & 3.6 & -9.82 & 66.87 \\
109010 & 9.75 & 0.0102 & 0.69 & 8.66 & 6.87 & 3.54 & -9.91 & 68.34 \\
109102 & 9.38 & 0.0117 & 0.44 & 8.33 & 6.82 & 2.93 & -9.74 & 65.49 \\
110038 & 10.0 & 0.0166 & 0.51 & 8.71 & 7.03 & 3.72 & -10.29 & 73.6 \\
109109 & 9.08 & 0.0196 & 0.27 & 8.61 & 6.63 & --- & -9.65 & 63.85 \\
108054 & 9.93 & 0.0125 & 0.65 & 8.66 & 6.85 & 3.65 & -10.12 & 71.44 \\
108045 & 10.08 & 0.015 & 1.76 & 8.69 & 7.62 & 4.22 & -9.83 & 67.04 \\
109028 & 10.07 & 0.0178 & 1.36 & 8.73 & 7.56 & 4.17 & -9.94 & 68.82 \\
 \hline
\end{tabular}
\caption{Derived observational quantities for the xCOLD-GASS galaxies.}
\label{sdss_data_table_cg} 
\end{table*}

\begin{table*}
  \begin{tabular}{@{}ccccccccc@{}}
  \hline
   DGS Name      &  Log M$_{*}$  & Redshift & SFR &  Metallicity  &  Log [C\scriptsize II\normalsize] & Log CO(1-0) & Log sSFR & f$_{\mbox{\scriptsize{[C\tiny II\scriptsize] mol}}}$  \\   
   & [\mbox{M}$_{\odot}$] & & [\mbox{M}$_{\odot}$ \mbox{yr}$^{-1}$] & [O/H] & [\mbox{L}$_{\odot}$] & [\mbox{L}$_{\odot}$] &  [\mbox{yr}$^{-1}$] & \%  \\
 \hline
Haro 11 & 10.24 & 0.0206 & 37.53 & 8.32 & 8.08 & 3.7 & -8.67 & 40.3 \\
Mrk 930 & 9.44 & 0.0183 & 2.78 & 8.08 & 7.44 & 2.81 & -9.0 & 49.39 \\
Haro 3 & 9.49 & 0.0031 & 0.55 & 8.12 & 6.49 & 2.42 & -9.75 & 65.66 \\
Mrk 1089 & 10.28 & 0.0134 & 5.72 & 8.3 & 7.59 & 3.8 & -9.52 & 61.32 \\
UM 448 & 10.62 & 0.0186 & 9.53 & 8.1 & 8.14 & 3.93 & -9.64 & 63.66 \\
Haro 2 & 9.56 & 0.0048 & 1.24 & 8.39 & 6.89 & 3.29 & -9.47 & 60.3 \\
II Zw 40 & 8.6 & 0.0026 & 0.57 & 7.92 & 6.41 & 1.92 & -8.84 & 45.13 \\
 \hline
\end{tabular}
\caption{Derived observational quantities for the seven DGS galaxies.}
\label{sdss_data_table_dgs} 
\end{table*}

\section{Observational Results: \cii/CO Scaling Relations}\label{obs_results_not_corrected}
\begin{figure*}
  \centering 
  \includegraphics[page=1, scale=1.0]{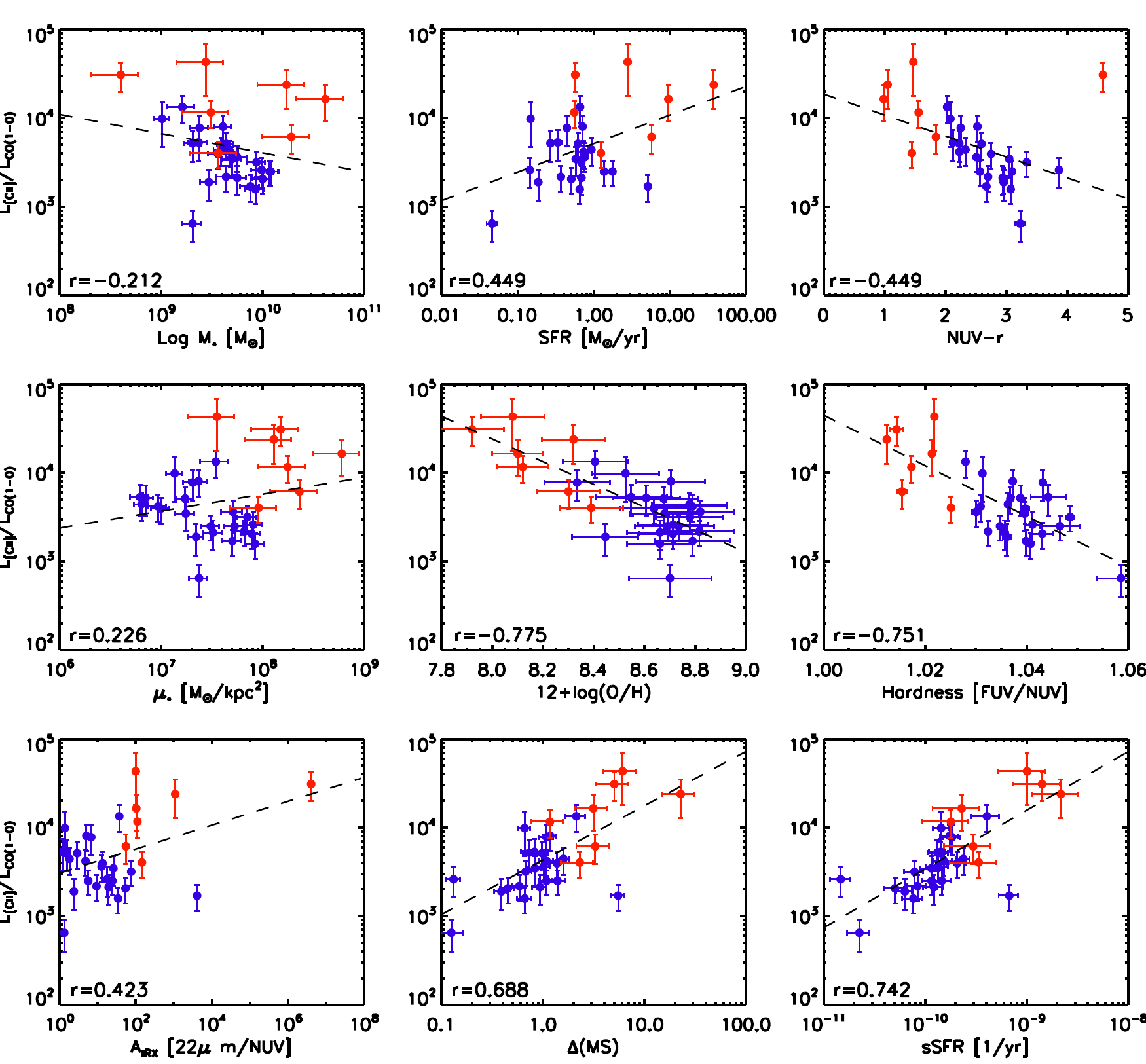}
  \caption{Scaling relations between the $L_{\mbox{\cii}}$/$L_{\mbox{CO(1-0)}}$ ratio and global properties for the 23 xCOLD GASS objects (blue symbols) and the seven galaxies from the DGS (red symbols). The Pearson correlation coefficient and the best-fit linear relation are shown in each panel.}
  \label{scaling_relations} 
\end{figure*}

In Figure \ref{scaling_relations}, we present the L$_{\mbox{[C\scriptsize II\normalsize]}}$/L$_{\mbox{CO(1-0)}}$ scaling relations for the 23 galaxies from the xCOLD GASS sample and the seven galaxies from the Dwarf Galaxy Survey, against a range of parameters describing the global physical properties of the galaxies. Although there is evident scatter, a clear dependence of L$_{\mbox{[C\scriptsize II\normalsize]}}$/L$_{\mbox{CO(1-0)}}$ on many of the parameters is observed. We use the Pearson correlation coefficient, $r$, to quantify the strength of the dependence; a more refined statistical approach is presented later in Section \ref{statsmethod}. 

The L$_{\mbox{[C\scriptsize II\normalsize]}}$/L$_{\mbox{CO(1-0)}}$ depends most strongly on parameters which {\bf a)} describe the strength of the UV radiation field and {\bf b)} describe the ability of the CO molecule to shield itself, via dust, from the UV radiation impinging on the surface of the the molecular regions deep inside the PDR. A clear dependence of L$_{\mbox{[C\scriptsize II\normalsize]}}$/L$_{\mbox{CO(1-0)}}$ on the colour of the systems, parametrised by NUV-r photometry, is observed, as well as the specific star formation rate, distance from the main sequence, UV field hardness, and the gas-phase metallicity. The two former quantities are directly linked because NUV-{\it r} is a good proxy for sSFR since it relates a quantity tracing ongoing star formation activity (NUV magnitudes) and a quantity sensitive to the older stellar population ({\it r}-band magnitudes). Their differences arise as the sSFR takes into account internal dust attenuation while NUV-{\it r} is not dust corrected. Furthermore the gas-phase metallicity can be seen as a proxy for the total dust to gas mass fraction (via a metallicity dependent dust to gas ratio). It is this dust which shields the CO molecule from the UV radiation, with decreasing metal content the systems become more dust poor allowing the UV radiation to penetrate deeper into the molecular clouds.

Overall the strongest dependencies in Figure \ref{scaling_relations} are on quantities that are sensitive to the amount of UV radiation penetrating into the molecular clouds, photodissociating CO into its ionised form. Of the four parameters mentioned above the dependence on metallicity is strongest and will be further justified with the full multi-parameter Bayesian treatment in Section \ref{statsmethod}. 

Conversely the L$_{\mbox{[C\scriptsize II\normalsize]}}$/L$_{\mbox{CO(1-0)}}$ ratio does not depend strongly on parameters which describe the masses and structural properties of the galaxies. A low correlation is observed with the stellar mass of the systems, and with the morphology as measured by the stellar mass surface density, $\mu_{*}$. This implies that CO photodissociation is happening on the small scales of molecular clouds as the large scale global properties of galaxies have a low correlation; global properties are likely less important than local ones.

Finally, rather interestingly, the L$_{\mbox{[C\scriptsize II\normalsize]}}$/L$_{\mbox{CO(1-0)}}$ ratio does not depend strongly on the extinction, calculated from Equation \ref{exctinction_correction} which is a measure of dust emission versus emission from young stars. Importantly it is a tracer of gas density and metallicity; the clear dependance on metallicity, explained above, suggests that the gas density may not correlate strongly with L$_{\mbox{[C\scriptsize II\normalsize]}}$/L$_{\mbox{CO(1-0)}}$. We speculate that the reason for this is that on small scales these low mass, low metallicity galaxies are flocculent with very clumpy structures, hence an averaged density across the galaxy washes out average variations of the L$_{\mbox{[C\scriptsize II\normalsize]}}$/L$_{\mbox{CO(1-0)}}$ ratio on small scales. An alternative explanation for the low correlation with extinction is because, in particular at low stellar mass, NUV and 22$\mu$m emission do not originate from the same H\scriptsize II\normalsize \ regions \citep{2010A&A...518L..55G}, and so we are not tracing the extinction of the same star-forming clouds. It is difficult to disentangle these two effects, assuming both play a role, or to rule either one out.

Our results qualitatively imply that metallicity, colour, sSFR, $\Delta$(MS) and/or hardness of the UV radiation field are responsible for variations in the L$_{\mbox{[C\scriptsize II\normalsize]}}$/L$_{\mbox{CO(1-0)}}$ ratio as the latter three parameters are responsible for the total amount of UV radiation impinging onto the surface layer of molecular clouds. As these three parameters are correlated we may not need all three to statistically parametrise a prescription for L$_{\mbox{[C\scriptsize II\normalsize]}}$/L$_{\mbox{CO(1-0)}}$ and hence the conversion function; a rigorous statistical approach, employing Bayesian methods, is shown below in Section \ref{statsmethod}.  

\begin{figure*}
  \centering 
  \includegraphics[page=1, scale=1.0]{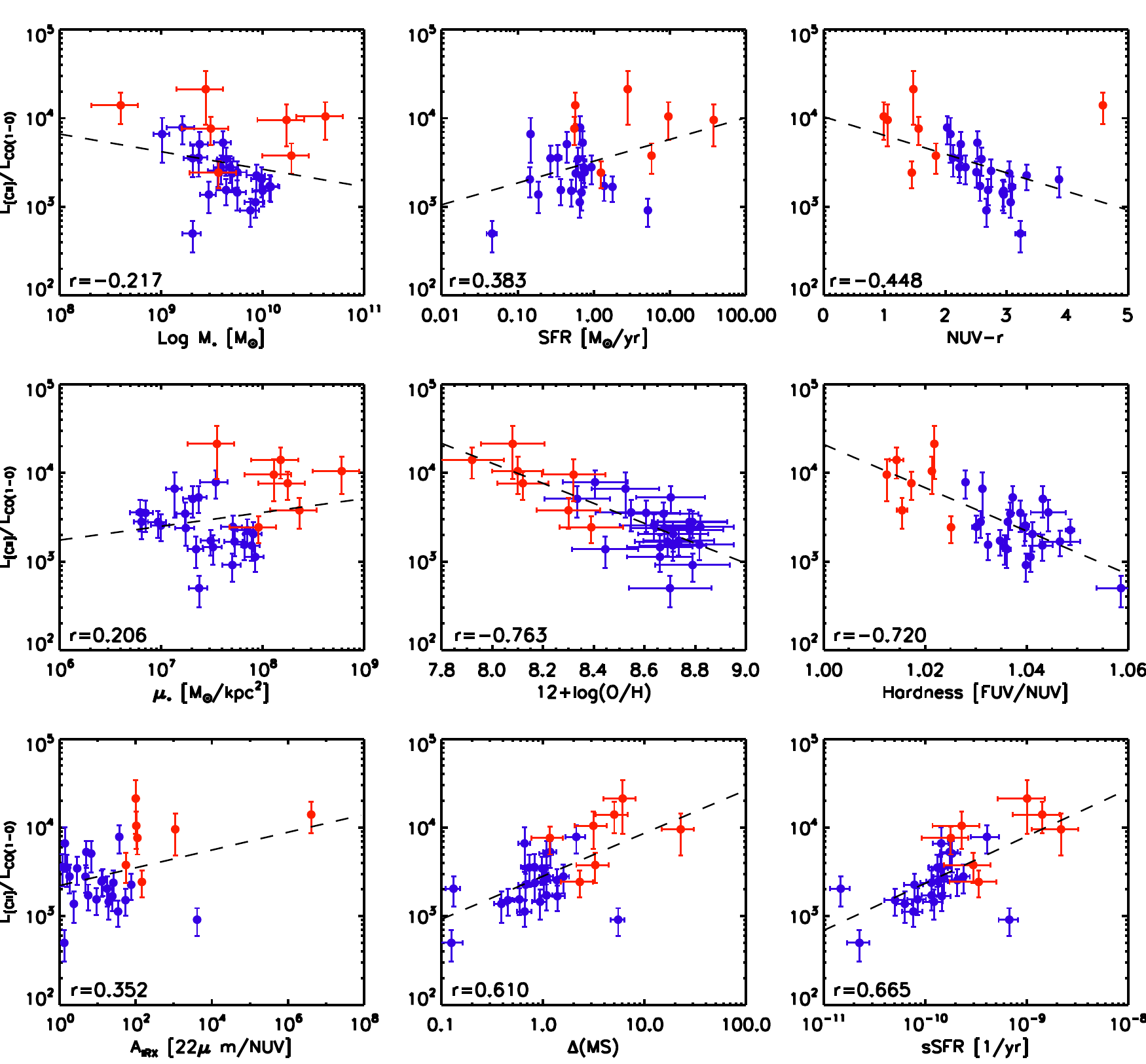}
  \caption{Same scaling relations as presented in Fig. \ref{scaling_relations}, but with the \cii\ luminosities corrected for emission not originating from molecular regions. }
  \label{scaling_relations_with_correction} 
\end{figure*}

\subsection{Contamination of the \cii\ emission by non-PDR sources}\label{contamination}

The scaling relations in Figure \ref{scaling_relations} show integrated measurements across all phases of the ISM, however to accurately measure variations of the conversion function we must restrict our analysis to variations in the L$_{\mbox{[C\scriptsize II\normalsize]}}$/L$_{\mbox{CO(1-0)}}$ ratio from molecular regions only. The CO(1-0) emission line only originates from molecular regions while [C\scriptsize II\normalsize] emission originates from all phases of the ISM; therefore we must clean for the contaminant \cii\ emission arising from non-molecular regions.

Galaxy wide observations of the \cii\ fraction emerging from molecular regions are uncommon and unfortunately an empirical prescription for this quantity does not exist, even though it is becoming apparent that the fraction of \cii\ emission across galaxies can vary. \citet{2013A&A...554A.103P} have studied this quantity across the Milky Way, as part of the GOT C$^{+}$ survey, and found the fraction to be 75\%. Further observations have been done which accurately measure the fraction of \cii\ emerging from ionised gas regions, using the method of \citet{2006ApJ...652L.125O}. This utilises the [C\scriptsize II\normalsize]/[N\scriptsize II\normalsize]205$\mu$m and [N\scriptsize II\normalsize]122$\mu$m/[N\scriptsize II\normalsize]205$\mu$m ratios. A study of NGC 891 by \citet{2015A&A...575A..17H} used this method and found that between 15$\%$-65$\%$ of [C\scriptsize II\normalsize] emission can arise from ionised regions. A further study of the star forming region BCLMP 302 in M33 by \citet{2011A&A...532A.152M} found that 20$\%$-30$\%$ of [C\scriptsize II\normalsize] emission arose from ionised regions while \citet{2013A&A...553A.114K} find that the contribution from the HI gas rises to about 40\% in the outskirts of M33. Therefore, to study variations in the conversion function, it is crucial to remove non-molecular [C\scriptsize II\normalsize] emission from our analysis.

Recent work by \citet{accurso17} provides a prescription to predict the fraction of [C\scriptsize II\normalsize] emission originating from molecular regions as a function of galaxy properties. A new multi-phase 3D radiative transfer interface was used through the coupling of {\sc starburst99} \citep{1999ApJS..123....3L, 2010ApJS..189..309L}, a Stellar Spectrophotometric code, with {\sc mocassin} \citep{2003MNRAS.340.1136E, 2005MNRAS.362.1038E}, a photoionisation code, and {\sc 3d-pdr} \citep{2012MNRAS.427.2100B}, an astrochemistry code. A detailed explanation of the work can be found in \citet{accurso17}, but a summary follows. 

{\edit Individual spherically symmetric star forming regions, with a central population of stars surrounded by gas, were simulated as follows. First, the stellar radiation density field, coming from the stellar population centrally located with the star forming regions, is created using {\sc sb99}. From this output stellar spectrum, the luminosity, temperature and number of ionising photons of the source are calculated; these quantities are then used as in- put parameters for the 3D photoionisation code {\sc mocassin.}, which is used to simulate the ionised gas surrounding the stellar population. The important output of the {\sc mocassin} code is the SED of the ionised gas, dust and stars emerging from the ionised region. This can be used to calculate the strength of the radiation field, G$_{0}$, which impinges onto the PDR surface, by integrating the {\sc mocassin} SED in the far-UV range between 912 and 2400\AA. This value of G$_{0}$ is used as an input into {\sc 3d-pdr} which simulates the ionised and molecular gas. Constant hydrogen number density profiles were assumed in each ISM region and the dust to gas ratio scaled with the input metallicity. 

To jump from these simulated star forming regions to the ISM of entire galaxies it was assumed that the physical conditions found in each cloud, for a given set of input parameters, can represent the average physical conditions found on galaxy-wide scales for galaxies with similar physical properties. Under this assumption a whole galaxy can be considered to be built up from a number of identical star forming regions, hence it provides a larger whole-galaxy model of the ISM. A summary of the input parameters was given in Table 2 of the aforementioned paper and they do consider the metallicity and sSFR regimes probed here.}

Through a Bayesian Inference method, several analytical functions were provided which can be used for this work. Specifically, as we do not have H\footnotesize II \normalsize region electron densities for the xCOLD GASS and DGS samples, we will use the prescription involving sSFR only, namely:
\begin{dmath}
\centering
f_{\mbox{\scriptsize{[C\tiny II\scriptsize], mol}}} = -6.224 -1.235\psi - 0.0543\psi^2
\label{ssfr_quad_bayes}
\end{dmath}.
where $f_{\mbox{\scriptsize{[C\tiny II\scriptsize], mol}}}$ is the fraction of [C\scriptsize II\normalsize] emission originating from molecular regions {\edit where at least 1\% of gas is in the form of H$_{2}$,} and $\psi =$ log(sSFR). 
The error on this quantity is  $\sigma_{f_{\mbox{\scriptsize{[C\tiny II\scriptsize], mol}}}} =$ 0.063, ({\edit the rms uncertainity on the fitting}). The physical motivation for equation \ref{ssfr_quad_bayes} can be found in \citet{accurso17}. We apply this to the 23 xCOLD GASS and seven DGS galaxies used in this work to estimate the relative fraction of [C\scriptsize II\normalsize] emission arising from molecular regions only. These values are in the range of 79-48\% as shown in Tables \ref{sdss_data_table_cg} and \ref{sdss_data_table_dgs}. {\edit In these tables we also provide the integrated CO(1-0) and uncorrected [C\scriptsize II\normalsize] luminosities for readers who are trying to model galaxies and want to match the uncorrected results.}

{\edit The simulations of \citet{accurso17} do not consider the diffuse neutral medium owing to its extremely low density, however they do have an atomic gas component which represents the CNM. The \cii\ emission originating from this atomic component will not be included in the $f_{\mbox{\scriptsize{[C\tiny II\scriptsize], mol}}}$ term, calculated from the simulations, and therefore \citet{accurso17} does consider a CNM like component when transitioning from ionised to molecular states; however this is not of interest to this paper as we only need the fraction of \cii\ emission originating from molecular regions.}

\subsection{Corrected [C\scriptsize II\normalsize]/CO Scaling Relations}
In Figure \ref{scaling_relations_with_correction} we present the L$_{\mbox{\cii}}$/L$_{\mbox{CO(1-0)}}$ scaling relations for molecular regions by using the estimated fractions of ionised carbon emission arising from this phase of the ISM. As can be seen the correction for contaminant ionised carbon emission does not effect the qualitative trends shown in Section \ref{obs_results_not_corrected}. There are mild changes in the statistical trends seen as, in most cases, the measure of correlation decreases as expected as the L$_{\mbox{\cii}}$/L$_{\mbox{CO(1-0)}}$ numerical values have now decreased. Our corrected scaling relation still imply that metallicity, colour, sSFR, $\Delta$(MS) and/or hardness of the UV radiation field are responsible for variations in the conversion function. {\edit Note that the Pearson correlation coefficients and best-fititng linear relations in Fig. \ref{scaling_relations_with_correction} are only intended to help with the interpretation; only the data themselves enter in the Bayesian analysis described below}.

\section{Bayesian Inference}\label{statsmethod}
Following on from the observational scaling relations, shown in Figure \ref{scaling_relations_with_correction}, we now want {\bf a)} to parametrise an analytic expression for the \cii/CO luminosity ratio from molecular regions as a function of galaxy parameters and {\bf b)} to determine the minimum number of parameters needed to provide a statistically robust fit to our data. With eight galaxy parameters\footnote{We do not include sSFR in our fitting procedure as this is very similar to $\Delta$(MS) which itself can be used to include a redshift dependence applicable up to z$\sim$2.5, the highest redshift probed in the $\Delta$(MS) prescription of \citet{2012ApJ...754L..29W}.} available we fit models with different number of free variables using all the possible combinations of parameters e.g we have ${^8C_k}$ number of models when we are fitting for k number of parameters. We perform a Bayesian interference method to find the best fit relations and to find the minimum number of variables needed to fit the data. Bayesian interference fitting methods have been successfully employed in several, wide-ranging, astrophysical scenarios from the derivation of the extinction law in the Perseus molecular cloud \citep{2013MNRAS.428.1606F} and Type Ia supernova light curve analysis \citep{2011ApJ...731..120M} to the extragalactic Kennicutt-Schmidt relation \citep{2013MNRAS.430..288S} and the formation and evolution of Interstellar Ice \citep{2014ApJ...794...45M}. For a more in depth explanation of the Bayesian regression fitting method we refer the reader to \citet{2007ApJ...665.1489K} and restrict ourselves here to the basic concepts. 

The first step is to assume that the measurement uncertainty associated with each [C\scriptsize II\normalsize]/CO observation, for each galaxy, ($y_{i}$ hereafter) is normally distributed. Therefore $y_{i}$ is a random variable distributed like:
\begin{dmath}
\centering
y_{i} = \mathcal{N}(y_{true, i}, \sigma_{y_{i}}^{2})
\label{normaldis}
\end{dmath}.
where $\sigma_{y_{i}}$ is the measurement uncertainty associated with the observable y on the i$^{th}$ galaxy. As can be seen in Figure \ref{scaling_relations_with_correction} all the observed scaling relations show evidence of either no correlation or a linear correlation between the log variables. We therefore use power law models in linear space as none of the above plots, in log space, show higher polynomial behaviour such that:
\begin{dmath}
\centering
y_{i} = 10^{\alpha}(x^{j}_{i})^{\beta_{j}} 
\label{linearmodel}
\end{dmath}.
Where $\alpha$ and $\beta_{j}$ are free variables to be found. We can say that the probability of observing our data, given the true value of the observables and the measurement uncertainties is:
\begin{dmath}
\centering
P(y_{i} | y_{true, i}, \sigma_{y_{i}}) = \frac{1}{\sqrt{2\pi \sigma_{y_{i}}^2}} \exp \left(-\frac{(y_{i} - 10^{\alpha}(x^{j}_{i})^{\beta_{j}} )^2}{2\sigma_{y_{i}}^2 }\right) 
\end{dmath}.

The next assumption to make is that all our [C\scriptsize II\normalsize]/CO observables are independent, e.g each $y_{i}$ and $x^{j}_{i}$ are independent, which is perfectly reasonable as the result from one galaxy will not affect the result from another. With this, and using the definition of independent probabilities, we can simply multiply all the individual probabilities together. Therefore this product of probabilities is our Likelihood, denoted L. By taking the log-likelihood, $\mathcal{L}$, the product returns back to a sum so:
\begin{dmath}
\centering
\label{likelihoodequation}
\mathcal{L} = -\frac{N}{2} \mbox{ln}(2\pi) - \frac{1}{2}\sum\limits_{i=1}^N \mbox{ln}(\sigma_{y_{i}}^2) - \frac{1}{2}\sum\limits_{i=1}^N \frac{(y_{i} - 10^{\alpha}(x^{j}_{i})^{\beta_{j}})^2}{ \sigma_{y_{i}}^2},
\end{dmath}
where N is the sample size. Maximising this log-likelihood for all of our models, at fixed number of free parameters, will provide us with the best fitting models for a given number of free parameters.

\subsection{Model comparison and sampling methods}
We aim to maximise the likelihood for our fits where the number of degrees of freedom varies. To compare likelihoods from models with different numbers of free parameters we use two different methodologies. Firstly we employ the Akaike Information Criterion (AIC) \citep{RePEc:eee:econom:v:16:y:1981:i:1:p:3-14}: 
\begin{dmath}
\centering
AIC = -2 \mathcal{L} + 2p + \frac{2p(p+1)}{N-p-1}
\end{dmath}.
where p is the number of free parameters and N is the sample size and the preferred model is that which minimises AIC. We also use a second method by employing the Bayesian Information Criterion (BIC) \citep{schwarz1978}:
\begin{dmath}
\centering
BIC = -2 \mathcal{L} + p \mbox{Log} (N) 
\end{dmath}
where p and N are defined as above. We compare the results of both these two methodologies to check for agreement and to ensure our results are not biased depending on what information criterion is employed.

A direct derivation/solution for the parameters which maximise the likelihood function, Equation \ref{likelihoodequation}, is computationally expensive and so, to efficiently and effectively sample the full parameter space, we use the well tested Python implementation of the affine-invariant ensemble sampler for Markov Chan Monte Carlo (MCMC) {\sc emcee} \footnote{An example of the code can be found at http://dan.iel.fm/emcee/current/} \citep{emcee}.

\subsection{Statistical results}
\begin{figure}
\includegraphics[scale=0.56]{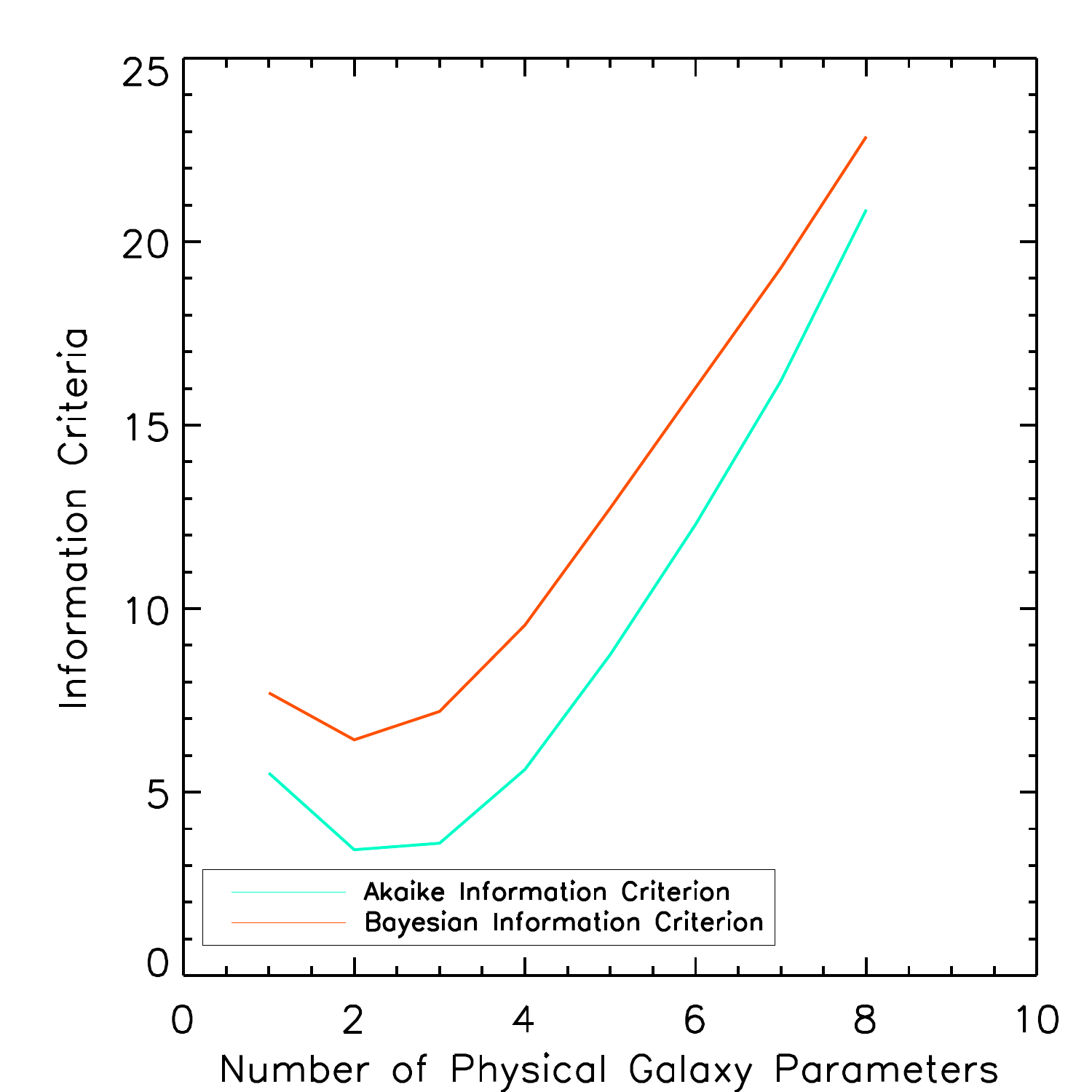}
\caption{Akaike Information Criterion (AIC; turquoise) and the Bayesian Information Criteria (BIC; orange) for the best fitting models when varying the number of galaxy parameters. Both tests indicate that two galaxy parameters are optimal to describe the variations in  L$_{\mbox{\cii}}$/L$_{\mbox{CO(1-0)}}$.}
\label{bic_aic_comparisons} 
\end{figure}

In total we run $\sum\limits_{i=1}^8$ ${^8C_{i}}$ models with the results presented below. By maximising the likelihood for each number of free parameters (from one to eight), and then comparing models with different sample sizes, and using the BIC and AIC, we find that only two parameters are necessary to fully explain the trends seen in Figure \ref{scaling_relations_with_correction}. We show in Fig \ref{bic_aic_comparisons} variations of the BIC and AIC with different number of free parameters. Both the AIC and BIC method agree that only two free parameters are necessary to provide a statistically robust fit to the data and so our choice of information criterion is extraneous.

\begin{figure*}
\includegraphics[scale=0.45]{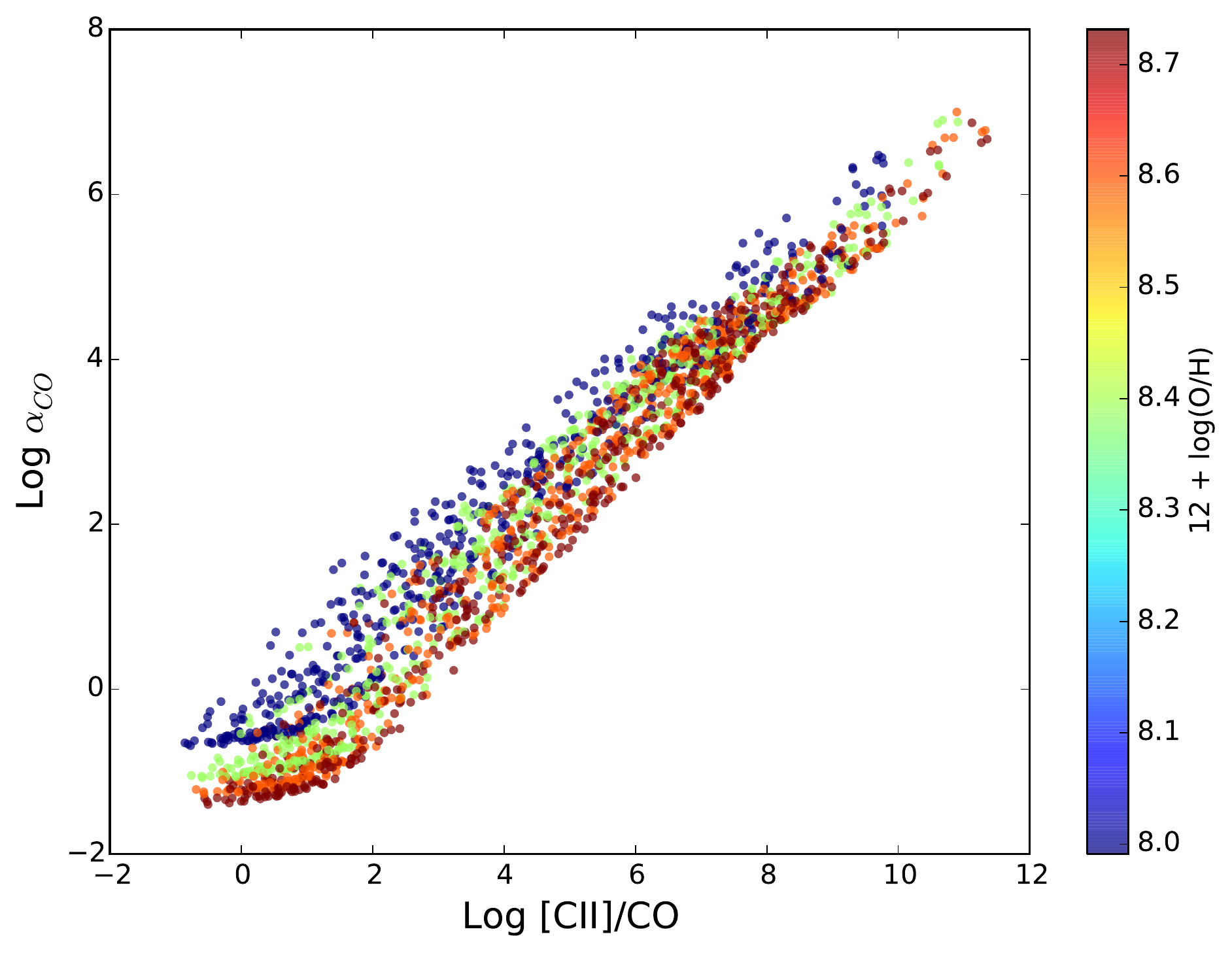}
\includegraphics[scale=0.45]{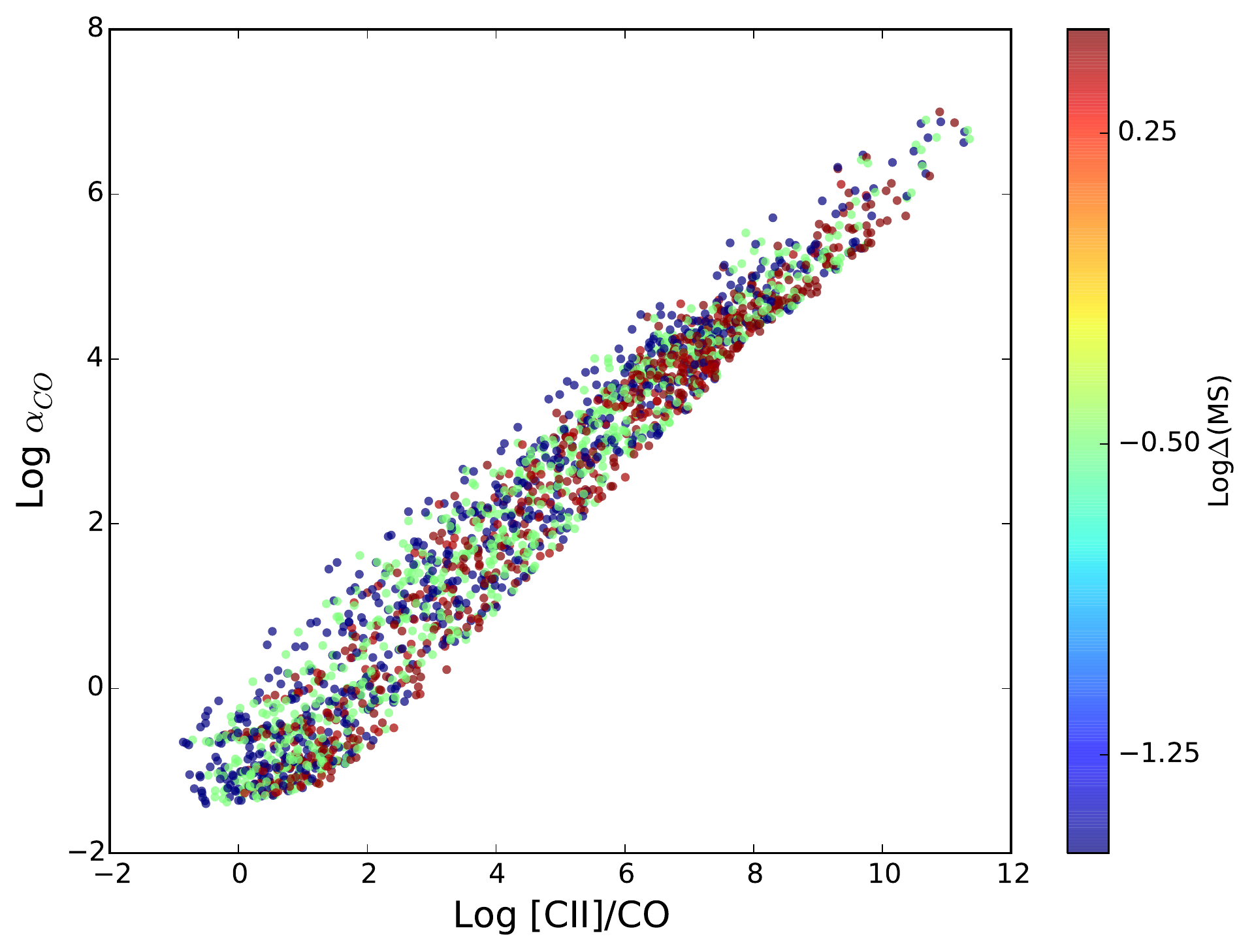}
\caption{Variations of the molecular region L$_{\mbox{\cii}}$/L$_{\mbox{CO(1-0)}}$ against $\alpha_{CO}$ for the 8640 simulated clouds in \citet{accurso17}. Clouds are colour-coded by metallicity (left) and $\Delta$(MS) (right).} 
\label{alpha_cii_co_paper_plots} 
\end{figure*}

For the two parameter case both the AIC and BIC retrieve metallicity and $\Delta$(MS) as the two galaxy parameters needed to provide a good fit to the data (the two parameter model which has the highest likelihood value in both cases). The best fitting analytical prescription is:
\begin{dmath}\label{CIICO_prescription}
\log \left( \frac{\mbox{[C\scriptsize II\normalsize]}}{\mbox{CO}}\right) (\pm 0.223\ \mbox{dex}) = 11.204 + 0.230 \log \Delta(\mbox{MS}) - 0.905[\mbox{12} +\log(\mbox{O/H})] 
\end{dmath}
which has a regression correlation coefficient of 0.801 meaning that these two parameters alone account for 80.1\% of the correlation, with an error on the predicted $\log \left( \frac{\mbox{[C\scriptsize II\normalsize]}}{\mbox{CO}}\right)$ value of 0.223. As the number of free parameters increases (from one to eight) the log posterior and log likelihoods also decrease; hence doing a likelihood ratio analysis would have been insufficient to recover our best fitting two parameter relation as this would have returned the eight parameter case as the best relation as these have the smallest posteriors. We stress that our aim was to find a relation which can explain the correlation with as few observables as possible. Only by employing the AIC and BIC, as above, were we able to find our best fitting relation with only two parameters. 

{\edit We also fit solely to the xCOLD GASS objects to demonstrate that our result is secure even when including the DGS galaxies. For the xCOLD GASS objects we find:
\begin{dmath}\label{xCOLD_GASS_CIICO_prescription}
\log \left( \frac{\mbox{\cii}}{\mbox{CO}}\right) = 12.21 + 0.266 \log \Delta(\mbox{MS}) - 1.0189[\mbox{12} +\log(\mbox{O/H})],  
\end{dmath}
which differs from \ref{CIICO_prescription} by no more than 25\%, well within the observational scatter.}

\section{Radiative Transfer modelling - connecting [C\scriptsize II\normalsize]/CO and $\alpha_{CO}$}\label{going_to_alpha}

The objective in obtaining the above scaling relations is to be able to derive a parametrisation for the conversion function. There are two main parameters responsible for variations in the L$_{\mbox{[C\scriptsize II\normalsize]}}$/L$_{\mbox{CO(1-0)}}$ fraction, namely metallicity and $\Delta$(MS), and so it is necessary to understand how to go from a L$_{\mbox{[C\scriptsize II\normalsize]}}$/L$_{\mbox{CO(1-0)}}$ parametrisation to one for $\alpha$$_{\mbox{CO}}$.

To better understand the astrochemical reactions involved in the photodissociation of CO we inspect the chemical reaction database employed in the PDR code {\sc 3d-pdr} \citep{2012MNRAS.427.2100B}. The main reactions involved in photodissociation of CO arise from the interaction of molecular species with cosmic rays and UV photons. The dominant reactions\footnote{By this we mean the reactions with the highest reaction rates which provide the main routes for creation and destruction of molecular species} involving UV photons and CO are:
\begin{dmath}
\centering
\mbox{CO} + \gamma \rightarrow \mbox{O} + \mbox{C} 
\end{dmath}
\begin{dmath}
\centering
\mbox{C} + \gamma \rightarrow \mbox{C}^{+} + \mbox{e}^{-} 
\end{dmath}
indicating that, whenever CO interacts with UV photons, neutral carbon is formed first as an intermediate species, but then quickly gets ionised as the reaction rates are of the same order\footnote{The reaction rates for these two are 2.0x10$^{-10}$ s$^{-1}$ and 3.0x10$^{-10}$ s$^{-1}$ respectively.}; hence only ionised carbon forms when CO reacts with UV photons. Moreover the reaction of CO with cosmic rays (denoted CRP) happens either via the formation of a free radical, namely He$^{+}$, as follows,
\begin{dmath}
\centering
\mbox{He} + \mbox{CRP} \rightarrow   \mbox{He}^{+}  + \mbox{e}^{-}
\end{dmath}
\begin{dmath}
\centering
\mbox{He}^{+} + \mbox{CO} \rightarrow  \mbox{O} + \mbox{C}^{+} + \mbox{He}
\end{dmath}
\begin{dmath}
\centering
\mbox{He}^{+} + \mbox{CO} \rightarrow \mbox{O}^{+} + \mbox{C} + \mbox{He}
\end{dmath}
or directly as,
\begin{dmath}
\centering
\mbox{CO} + \mbox{CRP} \rightarrow   \mbox{O} + \mbox{C}.
\end{dmath}
These show how, via cosmic rays, the photodissociation of CO can lead to the formation of neutral and ionised carbon species. Recent work by \citet{2015ApJ...803...37B} has shown that ionised carbon is the main product formed through the photodissociation of CO, but knowledge of how much neutral carbon forms is crucial to accurately constrain a prescription of $\alpha_{\mbox{CO}}$ as we only have observed CO and [C\scriptsize II\normalsize], lacking C\scriptsize I\normalsize \ observations.

To describe the formation of neutral carbon we rely on the multi-phase ISM numerical simulations performed in \citet{accurso17}. We take their simulated low-redshift cloud results and here plot in Figure \ref{alpha_cii_co_paper_plots} the variations of the molecular region L$_{\mbox{[C\scriptsize II\normalsize]}}$/L$_{\mbox{CO(1-0)}}$ against $\alpha_{CO}$ for the 2160 simulated clouds with colours indicating varying metallicities and $\Delta$(MS) of the clouds. {\edit \citet{accurso17} do not discuss $\alpha_{CO}$, hence we calculate this here. For each simulated cloud we calculate the total molecular hydrogen mass and total CO(1-0) luminosity and hence can accurately obtain an $\alpha_{CO}$ value.} We find that these two parameters, along with L$_{\mbox{[C\scriptsize II\normalsize]}}$/L$_{\mbox{CO(1-0)}}$, account for 98.4\% of the total correlation with $\alpha_{CO}$. We therefore perform a three dimensional linear fit to the above data, using the Python SciPy routine {\sc curve-fit}, and find
\begin{dmath}
\centering
\log \alpha_{\mbox{CO}} = 0.742 \log \frac{L_{\mbox{[C\scriptsize II\normalsize]}}}{L_{\mbox{CO(1-0)}}} - 0.944[12 + \log(\mbox{O}/\mbox{H})] - 0.109 \log \Delta(\mbox{MS}) + 6.439
\end{dmath}.
The errors on the estimated parameters are less than 0.01 dex and hence are negligible, owing to the high level of correlation retrieved from the fitting, and therefore can be ignored. {\edit We emphasise that this fit was done to the simulated data, hence the high degree of certainty in the $\alpha_{CO}$ calculation.}This has been normalised to the known Milky Way values for $\alpha_{\mbox{\scriptsize{CO}}}^{solar}$ = 4.36 M$_{\odot}$ (K km s$^{-1})^{-1}$ (helium corrected) and (L$_{\mbox{[C\scriptsize II\normalsize]}}$/L$_{\mbox{CO(1-0)}}$)$^{solar}$ = 1400 from \citet{1991ApJ...373..423S}. We then combine this with Equation \ref{CIICO_prescription} to finally arrive at a prescription for the CO conversion function:
\begin{dmath}\label{conversion_factor_equation}
\centering
\log \alpha_{\mbox{CO}} (\pm 0.165\ \mbox{dex}) = 14.752 - 1.623[12 + \log(\mbox{O}/\mbox{H})] + 0.062 \log \Delta(\mbox{MS}),
\end{dmath}
where \mbox{12} +\mbox{log(O/H)} is the metallicity in PP04 units, with an error on the predicted $\log \alpha_{\mbox{CO}}$ estimate of $\pm 0.165$. Metallicity predominantly drives variations in $\log \alpha_{\mbox{CO}}$ with $\Delta$(MS) playing a secondary, though statistically important, role.

The result of this study is therefore a new CO-to-H$_{2}$ conversion function which involves metallicity and $\Delta$(MS), and which is applicable applicable on galaxy-wide scales for star forming galaxies up to z$\sim$2.5 (the highest redshift constrained by Eq. \ref{deltamsdefinition}). The dependence on metallicity is consistent with other conversion functions as will be shown further in Section \ref{comparison}. {\edit The secondary dependence of \aco\ on $\Delta$(MS), or more generally speaking on a parameter relating to the strength of the UV radiation field, does not come as a surprise as it has been previously observed or predicted by simulations and theoretical models \citep[e.g.][]{1997A&A...328..471I,2010ApJ...716.1191W,sandstrom13,2015MNRAS.452.2057C}. The interesting result here is that the dependence of \aco\ and both metallicity and $\Delta$(MS) was not forced, it came out naturally from the data through the Bayesian analysis technique. }

\subsection{Caveat - Suitability across the main sequence}

{\edit There are two different well-studied regimes where \aco\ is know to show deviations from the Galactic value. The first is the situation of low metallicities discussed above, when CO suffers from photodissociation from the UV radiation field. The second regime is the one where the gas is affected by dynamical effects (mostly galaxy mergers),  making the ISM mostly molecular and changing gas temperatures and line widths. This has been well studied in the case of local ULIRGS, with \citet{1997ApJ...478..144S} reporting that an \aco\ value of 1.0 M$_{\odot}$ (K km s$^{-1})^{-1}$ is more appropriate than the Galactic value.  There will be galaxies which ``suffer" from both effects, and therefore the two should ideally be studied simultaneously. While this has been done in a small number of studies \citep[see][and references thereif]{2013ARA&A..51..207B}, it is unfortunately not possibly here as we do not have a sample of galaxies which simultaneously have low metallicities and dense ISMs, and the appropriate CO and \cii\ data products.}

Therefore, the conversion factor given as Eq.  \ref{conversion_factor_equation} should only be applied to galaxies within the parameter space constrained here, i.e for those with $7.9<12 + \log(\mbox{O/H})<8.8$ and $-0.8<\log \Delta(\mbox{MS})<1.3$. For galaxies with $12 + \log(\mbox{O/H})>8.8$ we recommend using the value obtained when using $12 + \log(\mbox{O/H})=8.8$, and for galaxies with $\log \Delta(\mbox{MS})<-0.8$ we recommend using the value obtained by setting $\log \Delta(\mbox{MS})=-0.8$. Because of a lack of data for galaxies below the main sequence, this is the most conservative thing to do, as opposed to allowing the trends to continue into regimes where it was not constrained. The conversion function will level out to a constant value, roughly equal to the galactic conversion factor value, consistently with other studies of early-type passive galaxies \citep{2014MNRAS.444.3427D}. 

Similarly, Eq.  \ref{conversion_factor_equation} should not be applied to galaxies well above the main sequence, where dynamical effects may be compressing the ISM, leading to higher gas densities and dust temperatures. For galaxies which show evidence of being in this regime (e.g. \citet{2012ApJ...758...73S} use the selection criteria of L$_{\mbox{FIR}}$/L$_{\odot}>$ 11.0 and S$_{60}$$\mu$m/S$_{100}$$\mu$m $>$ 0.5), we recommend the use of a lower value of $\alpha_{CO}=1$~M$_{\odot}$ (K km s$^{-1})^{-1}$. A major consequence of this is a discontinuity of $\alpha_{CO}$; when moving up through the main sequence $\alpha_{CO}$ increases and suddenly decreases as $\log \Delta(\mbox{MS})>1.3$. This has also been recently suggested by \citet{2014ApJ...793...19S} who find that $\alpha_{CO}$ increases with increasing sSFR and then suddenly decreases when entering the starburst regime. We speculate this is because ULIRGs and starburst galaxies have different physical environments to main sequence galaxies; while their sSFRs are much higher their densities are also significantly higher, hence CO is more easily shielded in these denser environments, leading to a rapid decline in $\alpha_{CO}$.

\subsection{Caveat - SDSS Metallicities}
Metallicity gradients of different amplitudes have been observed in galaxies similar to those studied here \citep{2016A&A...588A..91M, 2016MNRAS.456.2982T, 2016arXiv160301139W, 2015MNRAS.448.2030H} meaning that the SDSS fiber spectroscopy for our galaxies, which only probes the central 3'', may not provide reliable metallicity measurements to use in the analysis of our [C\scriptsize II\normalsize]/CO data. Star formation histories for these low mass objects are known to be bursty while ongoing star formation activity is known to be inhomogeneous with activity unevenly scattered across the galaxy \citep{2015ApJ...808L..49G, 2015MNRAS.451..839D}. Hence, to accurately measure metallicity gradients we would need IFU data to get accurate spatial sampling out to the extended regions of our galaxies without {\it a priori} information of the location of the HII regions. IFU spectroscopy would be essential to determine the accurate light-weighted, integrated metallicity measurements over the area of the galaxies probed by the IRAM and Herschel observations. Obtaining IFU spectroscopy of our targets may reduce the scatter found in the metallicity dependent scaling relations and may further refine our conversion function.

\begin{figure}
\includegraphics[scale=0.7]{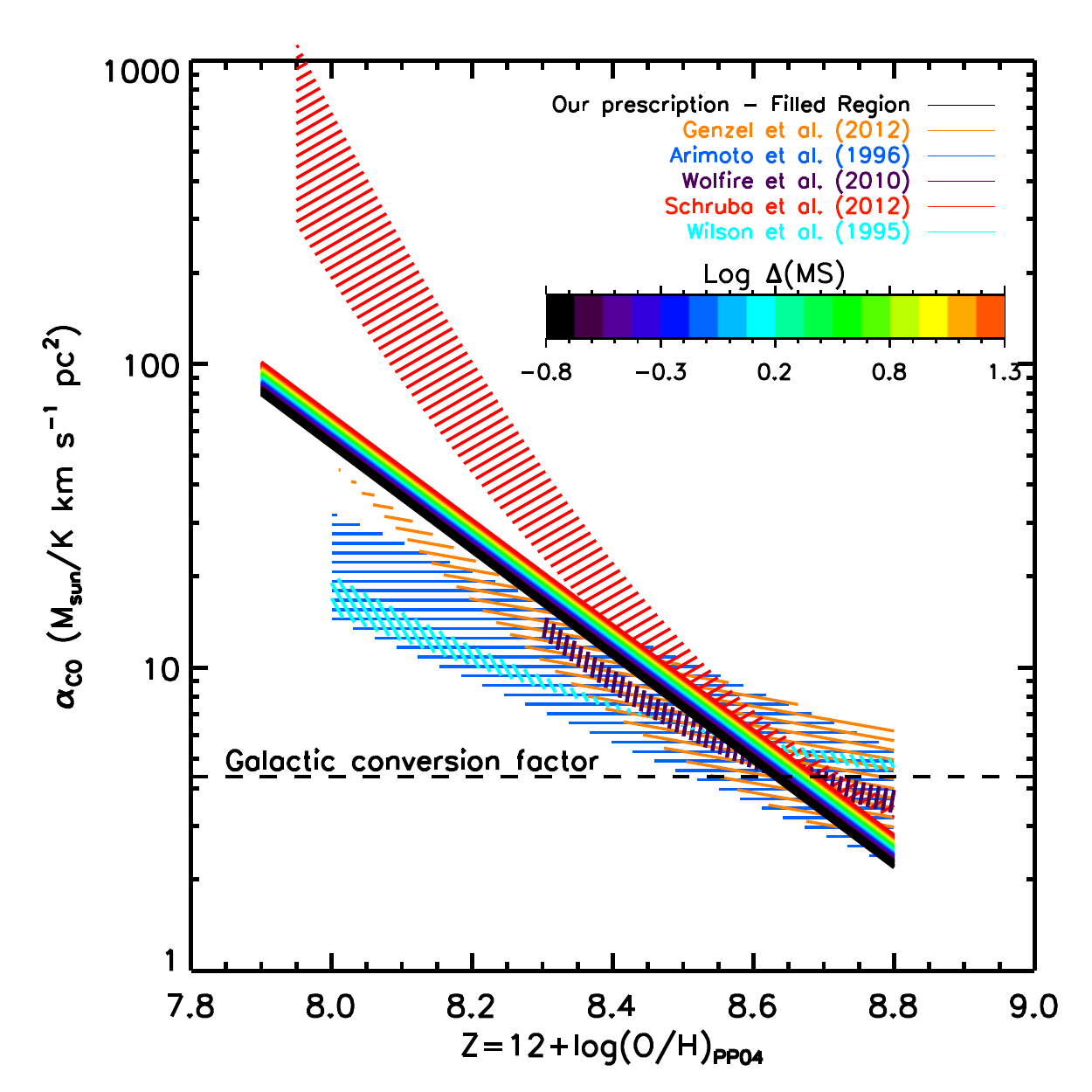}
\caption{The new \aco\ conversion function as given in Eq. \ref{conversion_factor_equation}, which depends on both metallicity and $\Delta$(MS). As a comparison we show several commonly-used metallicity-dependent \aco\ prescriptions. } 
\label{conversion_factor_cg} 
\end{figure}

\subsection{Comparison with previous studies}\label{comparison}
Studies concerning the CO-to-H2 conversion function have been ongoing for decades with a plethora of varying prescriptions; some of the earliest work was carried out by \citet{1987ApJ...319..730S}, \citet{1985ApJ...291..722T} and \citet{1988ApJ...334..771V} and so, to interpret our prescription, we compare it with a selection of other calibration derived using different methodologies. We plot our two parameter conversion function in Figure \ref{conversion_factor_cg}, where colour in the multi-coloured filled surface denotes varying $\Delta(\mbox{MS})$. As can be seen, metallicity predominantly drives variations in our conversion function with $\Delta(\mbox{MS})$ playing a minor, though statistically important, role. Some well-established \aco\ prescriptions are also shown in Figure \ref{conversion_factor_cg}. 

We first compare our prescription to the single variable metallicity dependent conversion functions of \citet{2012AJ....143..138S} and \citet{2012ApJ...746...69G}. These were derived by assuming a constant molecular gas depletion time and then using an inverse star formation law. Our prescription is approximately consistent with both around the high metallicity range, but then predicts different conversion factors for 12 + log(O/H)$\lesssim$8.25. The divergence between the two aforementioned prescriptions is because {\edit in addition to using different galaxy samples as calibrators, they assumed different values for the  molecular gas depletion time (1.8 and 1.0 Gyrs respectively). While diverging at the low metallicity end, these metallicity-dependent prescriptions would be adequate in the metallicity range covered by xCOLD GASS sample.}

The novelty of our approach means that we are able to attribute the scatter found in these previous metallicity-only dependent relations to the position off the main sequence. Those systems with stronger radiation fields (due to higher $\Delta(\mbox{MS})$ values), at fixed metallicity, will have higher conversion factors. Also galaxies which are metal poor, at a fixed position on the main sequence, will have higher conversion factors than their metal rich counterparts. 

We next compare our prescription to another metallicity dependent conversion function based on the PDR modelling of \citet{2010ApJ...716.1191W}. The lowest metallicity employed in their grid of models was 12 + log(O/H) $=$ 8.38. The trend in both conversion functions are in broad agreement for metallicities above 12 + log(O/H) $=$ 8.38 to solar, however they are offset from one another with our prescription predicting higher conversion factors. Their grid of models allowed for a varying ionisation field (for which the sSFR/$\Delta$(MS) contributes towards) and varying metallicities, but the parametrisation only involved metallicity.

We now move onto a comparison with \citet{1995ApJ...448L..97W} and \citet{1996PASJ...48..275A} who both measured virialised masses of molecular clouds to infer the dependence of their conversion function with metallicity. The resulting prescription is also in broad agreement with the trends seen in the \citet{2011ApJ...737...12L} sample, who used a dust-to-gas ratio method. As can be seen in Figure \ref{conversion_factor_cg}, the trend in both conversion functions are in broad agreement for metallicities above 12 + log(O/H) $=$ 8.25 to solar. However, they are offset from one another with our prescription predicting conversion factors two to three times higher. We speculate that this is due to the integrated nature of our observations versus the highly resolved, cloud scale-resolution, of \citet{1995ApJ...448L..97W} and \citet{1996PASJ...48..275A}. On galaxy-wide scales, it is possible to see more diffuse H$_{2}$ molecular gas which is traced by [C\scriptsize II\normalsize] and C\scriptsize I\normalsize, as opposed to CO, hence the need for higher conversion factors to account for this. 

Overall our prescription agrees well with others in the literature that have used integrated observations and assume a constant depletion time. Our prescription does not agree as well with those that have used a) a dust to gas ratio method, b) numerical modelling c) virialised gas mass estimates with observational resolution down to cloud scales. This is because our prescription accounts for the diffuse H$_{2}$ gas, not found in individual clouds and GMCs, which is better traced by ionised and neutral carbon versus its molecular counterpart.

\section{Bridging low and high redshift molecular gas studies}\label{discussion}
We apply our new prescription to the full xCOLD GASS sample (including the new low mass galaxy sample and the higher mass objects presented in \citet{2011MNRAS.415...32S}) and the PHIBSS1 sample \citep{2013ApJ...768...74T}. Star formation rates, stellar masses, CO luminosities and redshifts for the PHIBSS1 sample are taken from \citet{2013ApJ...768...74T}. Metallicities are estimated using the Fundamental Mass-Metallicity Relation (FMR) from \citet{2010MNRAS.408.2115M}. The aim of this is to investigate the consequences of our conversion function on gas scaling relations and how these change when going from a standard constant conversion factor to our prescription presented in Equation \ref{conversion_factor_equation}. We do not include the DGS objects in this analysis as we are only interested in the statistical trends of the molecular gas scaling laws; we need a complete sample of galaxies at different redshifts and so use the xCOLD GASS and PHIBSS1 samples. We leave a more thorough and statistically robust treatment of the scaling relations for a future publication.
\begin{figure}
\includegraphics[scale=0.45]{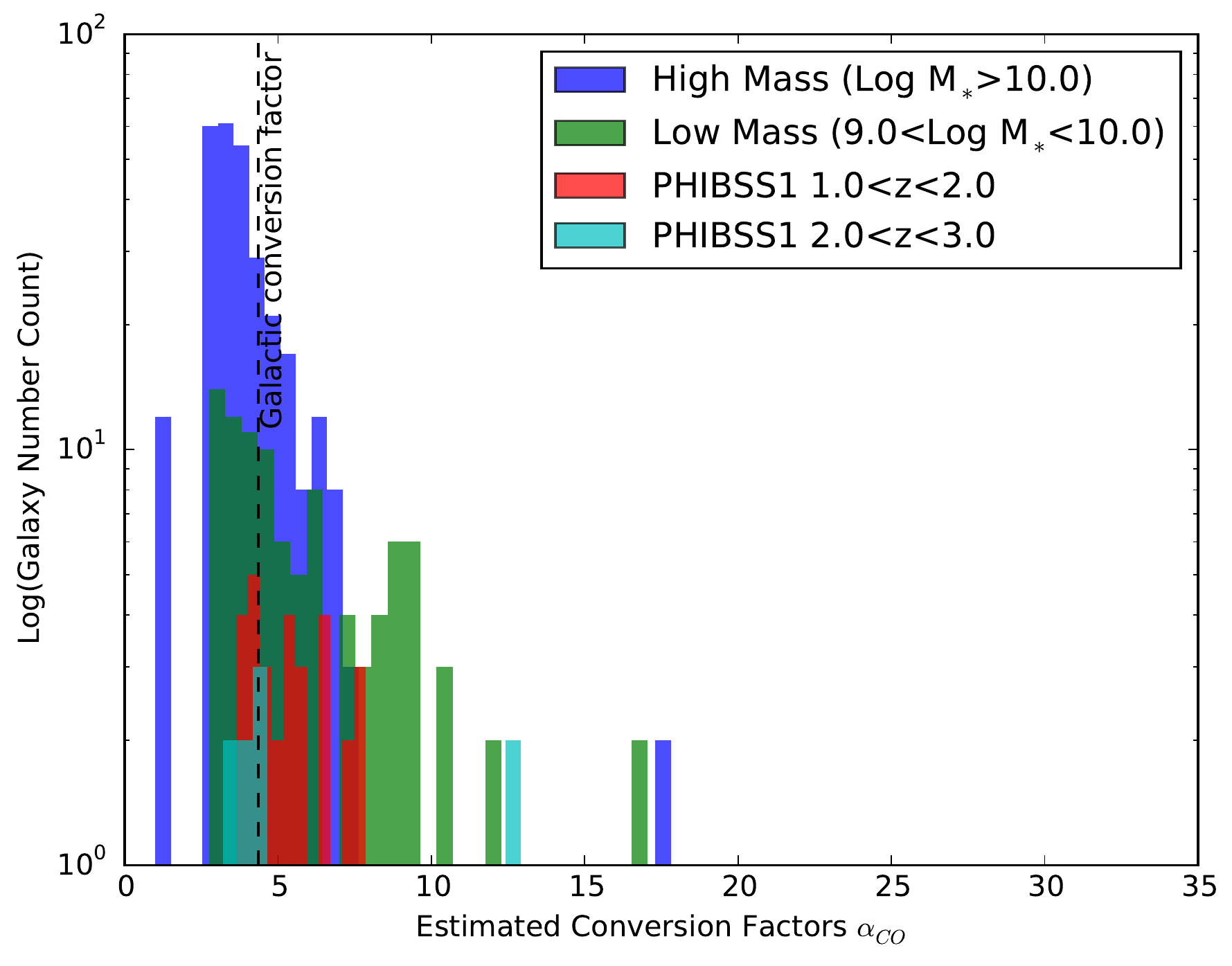}
\caption{Distribution of conversion factors for the xCOLD GASS and PHIBSS1 galaxies. For the xCOLD GASS high mass sample, the distribution is as expected centered on the Galactic value. The xCOLD GASS low mass sample, due to their lower metallicities need higher conversion factors, similar to the PHIBSS1 galaxies that all have metallicities less than solar.}
\label{conversion_factor_distribution} 
\end{figure}

We first plot in Figure \ref{conversion_factor_distribution} a histogram showing the distribution of conversion factors for the high and low mass xCOLD GASS galaxies and the PHIBSS1 samples. For the original COLD GASS galaxies the distribution peaks slightly below the galactic value ($\sim$ 4.36 M$_{\odot}$ (K km s$^{-1})^{-1}$) with other values clustered closely around this, hence we expect the low redshift molecular gas scaling relations to remain relatively unchanged for M$_{*}$$>$10$^{10}$. However for the lower mass galaxies the conversion factors extend to high values. Furthermore, the predicted conversion factors for the high redshift galaxies are at most a factor of two larger or smaller than the galactic conversion factor value so we do not expect a major change in the gas scaling laws at high redshifts.

\begin{figure*}
\includegraphics[scale=0.85]{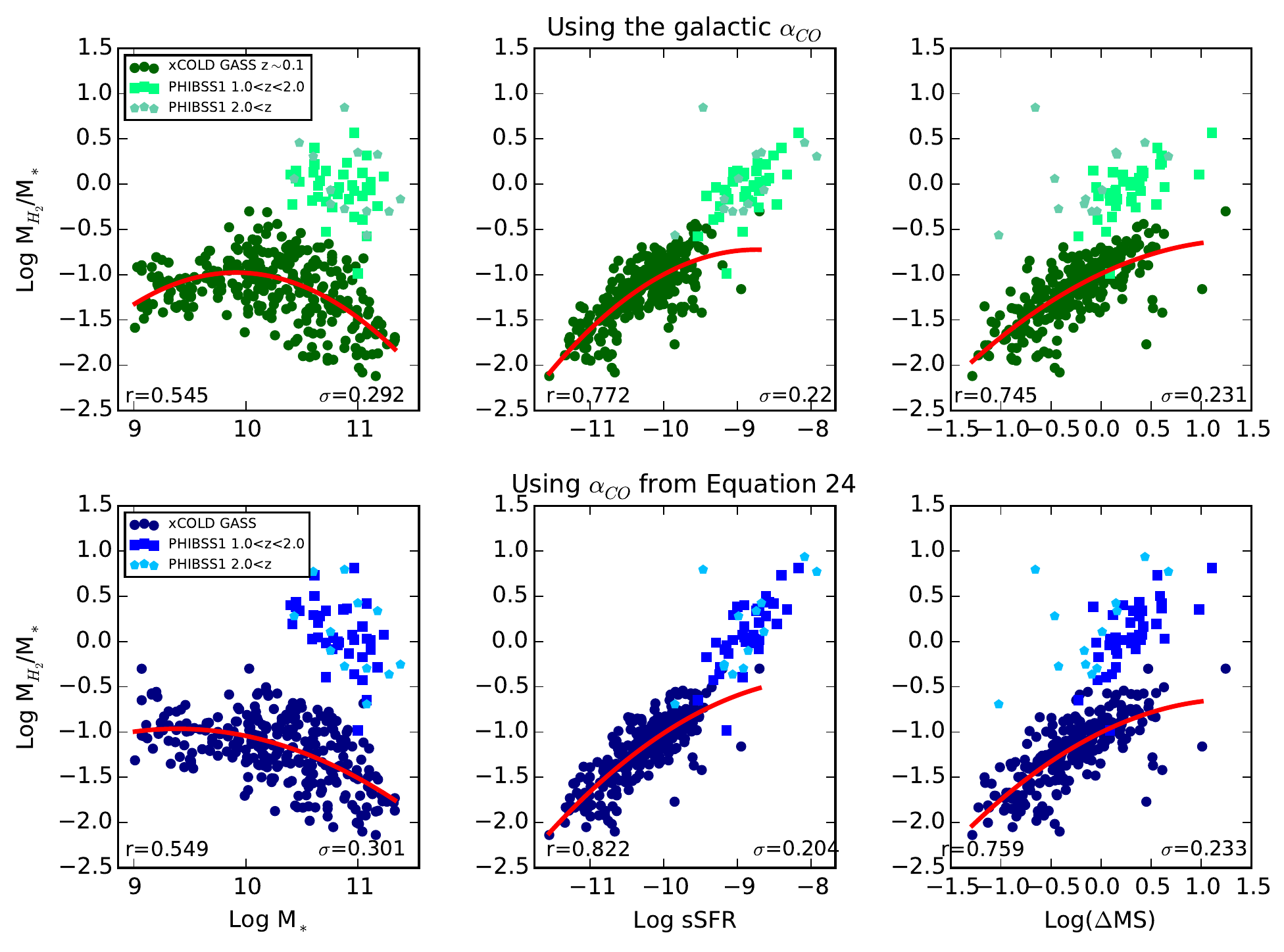}
\caption{{\textit{Top row:}} Scaling relations between molecular gas mass fraction and \mstar, sSFR and $\Delta$(MS) (from left to right) for the full xCOLD GASS and PHIBSS1 samples, when using a constant Galactic conversion factor. The red lines are second order fits to the xCOLD GASS data, with the correlation coefficient and the one sigma scatter also given. {\textit{Bottom row:}} Same scaling relations as above, but using the \aco\ conversion function presented in Eq. \ref{conversion_factor_equation}.}
\label{mh2_scaling_relation} 
\end{figure*}

\subsection{Molecular gas fractions up to z$\sim$2.0}
We present scaling relations for the CO-traced molecular gas mass fraction as a function of stellar mass, specific star formation rate and $\Delta$(MS) (from left to right) for the full xCOLD GASS and PHIBSS1 samples in Figure \ref{mh2_scaling_relation}. We aim to showcase our new prescription for $\alpha_{CO}$ and so, for now, only use the CO-detected galaxies in both samples. In a future publication a more thorough, quantitative, statistical treatment of the whole sample, including detections and non-detections, will be performed. 

The top row in Figure \ref{mh2_scaling_relation} presents the results obtained when using a constant galactic conversion factor; this had been previously explored for the COLD GASS sample in \citet{2011MNRAS.415...32S}, only for galaxies with stellar masses $>$ 10$^{10}$ M$_{\odot}$ and so we are extending the sample to include galaxies with stellar masses $>$ 10$^{9}$ M$_{\odot}$. This relation for the PHIBSS1 sample was first presented in \citet{2013ApJ...768...74T} and \citet{2015ApJ...800...20G}. The bottom row presents the results obtained when our prescription is used. We fit quadratic polynomials to the xCOLD GASS data to qualitatively show the differences in the trends; we also present the product moment correlation co-efficient and scatter for each fit, calculated using the using the Python SciPy routine {\sc linregress}.

\begin{figure*}
\includegraphics[scale=0.85]{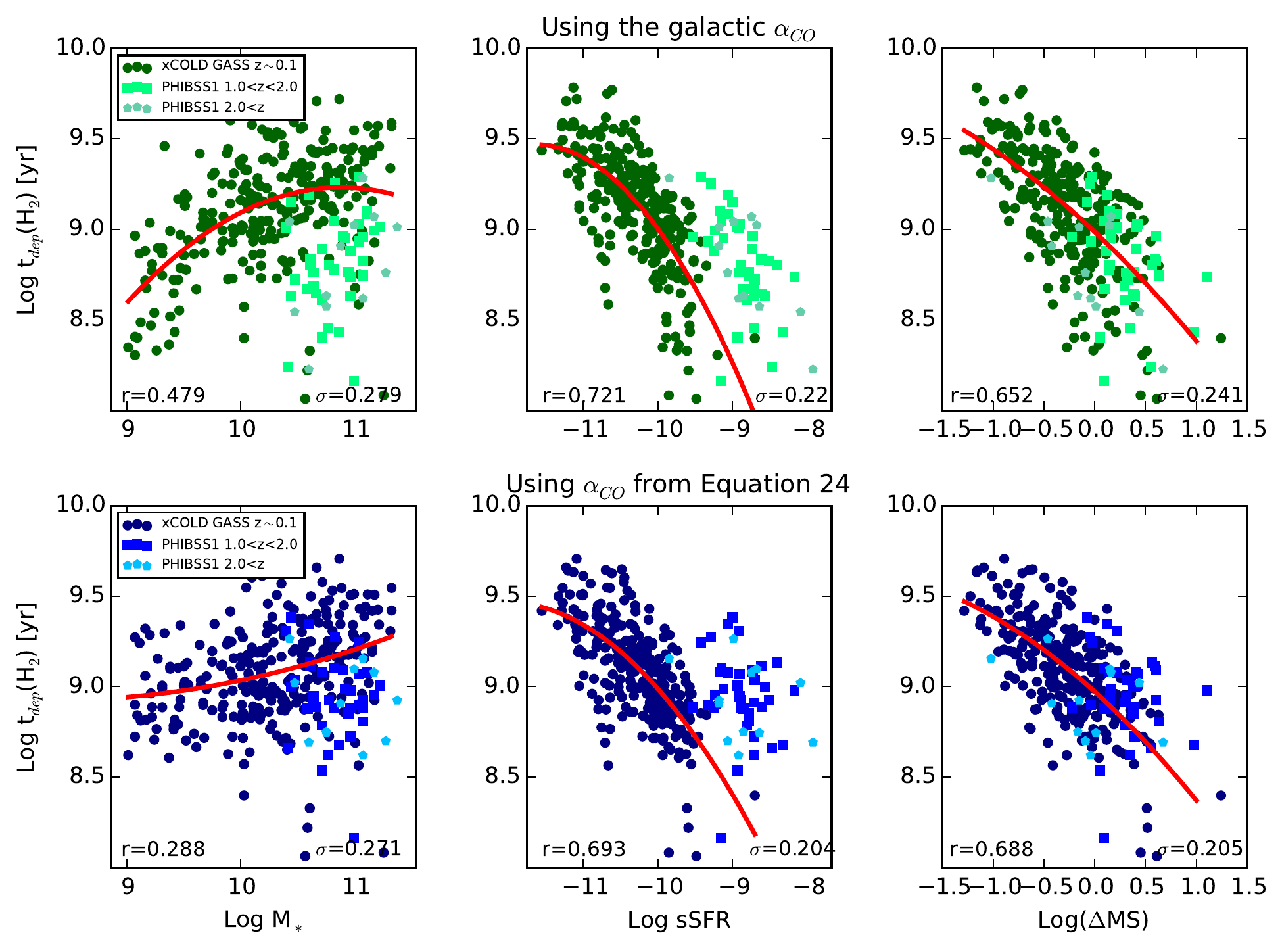}
\caption{Scaling relations for molecular gas depletion time using either a constant Galactic conversion factor (top row) or the new two-parameter conversion function (bottom row).  Symbols and lines are as described in Fig. \ref{mh2_scaling_relation}.}
\label{tdepletion_scaling_relation} 
\end{figure*}

As can been seen, the correlation with stellar mass remains unchanged for galaxies with M$_{*}$$>$10$^{10}$ M$_{\odot}$ at low redshift as the standard galactic conversion factor is a good approximation to the conversion function predicted from this work. However, due to the emergence of lower metallicities in galaxies with stellar masses less than 10$^{10}$ M$_{\odot}$ the trend changes. We start to see a flattening of the molecular gas mass fraction for stellar masses between 10$^{9}$ -10$^{10}$ M$_{\odot}$ for the low redshift galaxies, which is in excellent agreement with the trends found in \citet{2016arXiv160209077G} and the star formation models of \citet{2009ApJ...693..216K, 2008ApJ...689..865K} demonstrated in \citet{2015MNRAS.450..606L}. In the star formation models, this flattening occurs as depletion timescales are independent of stellar mass, something which we shall explore in Section \ref{tdep}. This flattening also validates part of the ideal gas regulator model (sSFR = constant; \citet{2013ApJ...772..119L}) for which no or very little dependence of gas fractions on stellar mass is expected, as we observe up to M$_{*}$$\sim$10$^{10}$M$_{\odot}$. This result is also of interest given the trend of increasing HI gas mass fraction with decreasing stellar mass, down to M$_{*}$$\sim$10$^{9}$M$_{\odot}$ \citep{2015MNRAS.452.2479B}, implying that low mass galaxies are less efficient at converting their HI gas reservoirs into molecular gas. For the high redshift objects the trend of increasing molecular gas fraction with decreasing stellar mass is still evident; this is true across all redshifts for M$_{*}$$>$10$^{10}$ \citep{2013ApJ...768...74T}. The decreasing gas fractions as stellar mass increases, at all redshifts, is driven by the flattening of the SFR-M$_{*}$ relation at high stellar masses.

The correlation with specific star formation rate strengthens when using the prescription presented here. A tight correlation between molecular gas and star formation is expected, confirming that the variable $\alpha_{CO}$ presented here outperforms a constant Galactic value. This is in agreement with the relations found in \citet{2016arXiv160705289S} showing that star formation activity in a galaxy is controlled by the total available gas. Interestingly, the high redshift galaxies simply extend the trends found from the low redshift sample.
 
We also plot the correlation with $\Delta$(MS) and find similar results to the trends with specific star formation rate for local universe galaxies. This is expected as sSFR$_{\mbox{MS}}$ is roughly constant in the local universe and therefore $\Delta$(MS)$\propto$sSFR at low redshift. The high redshift galaxies are offset from the trend seen in the low redshift sample as a direct consequence of the evolution of sSFR on the main sequence with redshift (e.g \citet{2011ApJ...730...61K}) and the increase in the gas supply rate.

\subsection{Molecular gas depletion times up to z$\sim$2.0}\label{tdep}
We present scaling relations for the CO-traced molecular gas depletion times, where t$_{dep}$ = M$_{H_{2}}$/SFR, as a function of stellar mass, specific star formation rate and $\Delta$(MS) for the full xCOLD GASS and PHIBSS1 samples in Figure \ref{tdepletion_scaling_relation}. While a positive correlation between t$_{dep}$ and M$_{\ast}$ is observed when applying a Galactic conversion factor, this trend becomes statistically insignificant with a correlation coefficient of 0.289 once our conversion function is used. The important consequence of this is that t$_{dep}$ (or equivalently, star formation efficiency) does not depend on stellar mass, as also seen in the PHIBSS1 sample. This is all in agreement with previous trends found in the low and high redshift universe \citep{leroy13,2013ApJ...768...74T,tacconi17,2014A&A...562A..30S,2014MNRAS.443.1329H,2014ApJ...793...19S,2015ApJ...800...20G}. This important result validates part of the equilibrium model \citep{2012MNRAS.421...98D, 2013ApJ...772..119L} which states that gas depletion times are independent or have very little dependence on stellar mass.

The correlation with sSFR, albeit weaker, is in agreement with the relations found in \citet{2011MNRAS.415...61S}, {\edit and confirms the conclusion from \citet{bothwell14} that it is robust against the choice of a specific metallicity-dependent \aco\ conversion function}. Our shallower trend is expected because our prescription is dependent on $\Delta$(MS), which is closely linked to sSFR. The trend observed here is in excellent agreement with that found in \citet{2015A&A...583A.114H}. Moreover the high redshift galaxies are offset as previously reported in \citet{2012ApJ...758...73S}. The trend between t$_{dep}$ and sSFR is redshift independent once accounting for the redshift evolution of the main sequence \citep[see bottom right panel of Fig. \ref{tdepletion_scaling_relation} and][]{2015ApJ...800...20G,tacconi17}. This suggests that the process of star formation on the main sequence is driven by similar physical mechanisms across cosmic time, independently of stellar mass.

\begin{figure*}
\includegraphics[scale=0.44]{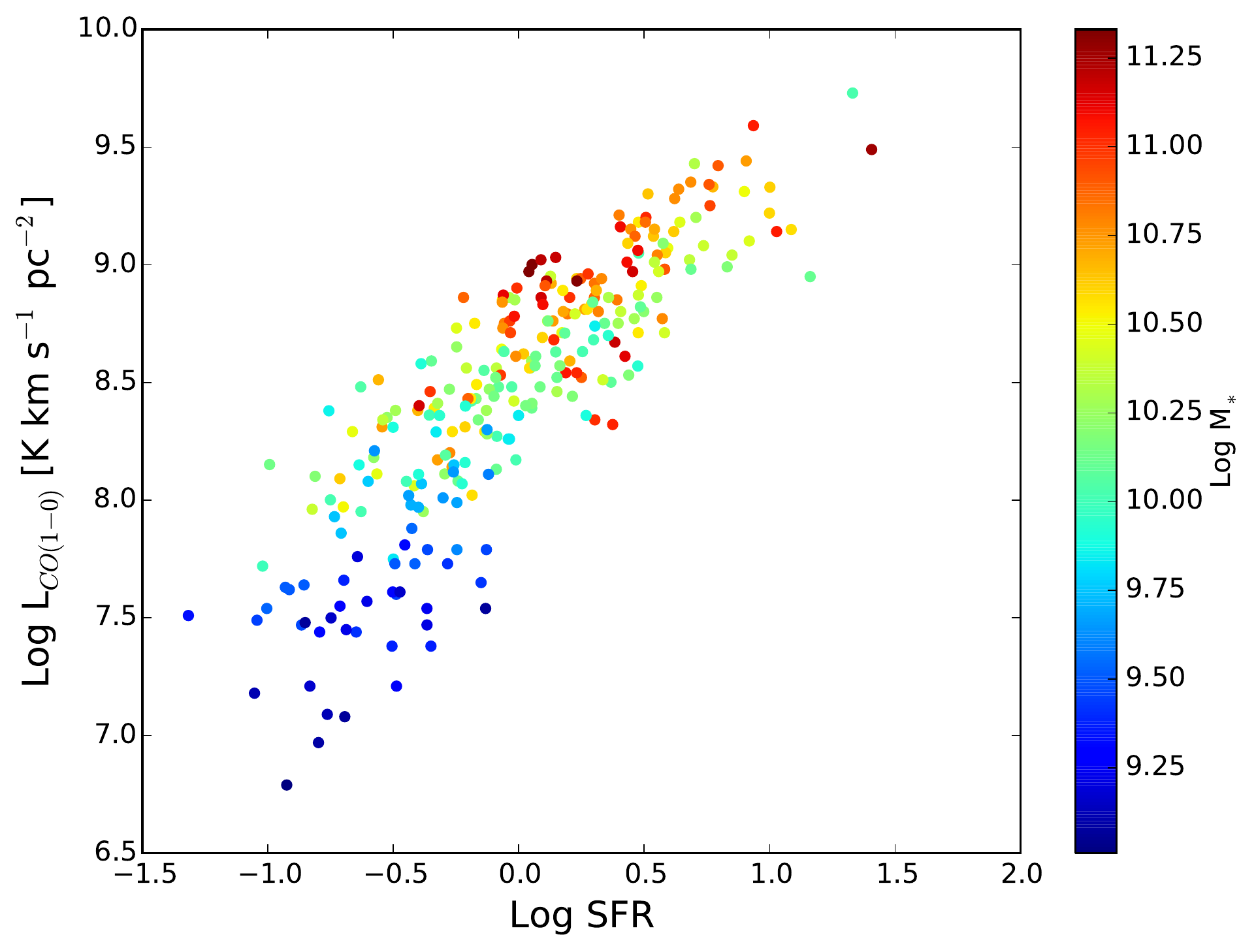}
\includegraphics[scale=0.44]{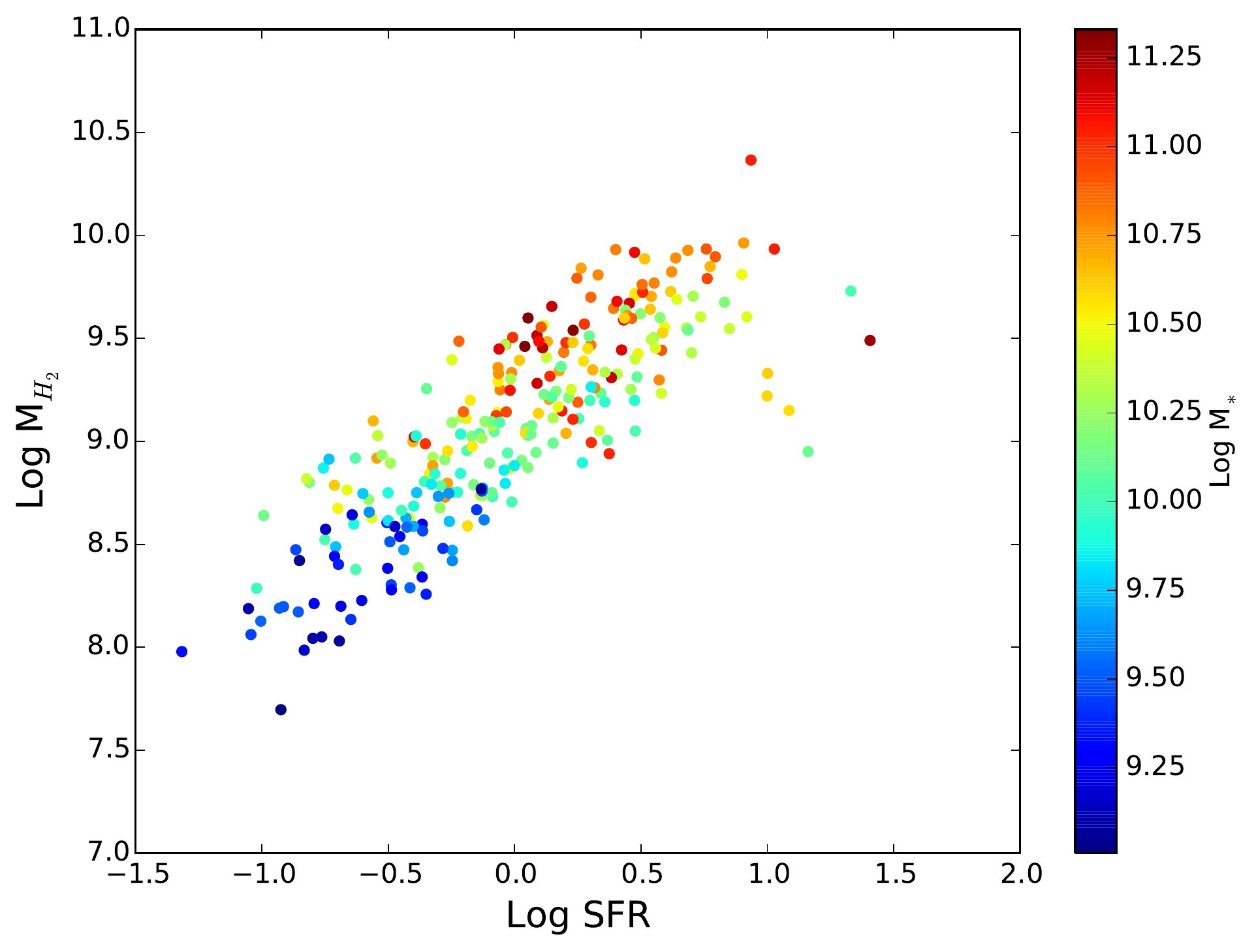}
\caption{{\bf Left:} CO luminosities of the full xCOLD GASS sample as a function of SFR, with the colour of each point denoting stellar mass. {\bf Right:} Relation between M$_{H_{2}}$ and star formation rate, using Eq. \ref{conversion_factor_equation} as the conversion function; the lowest mass galaxies which appeared under luminous in CO in the left-hand plot follow the same linear trend as the higher mass galaxies, indicating that they are on average equally efficient at forming stars .}
\label{mh2_versus_sfr} 
\end{figure*}

Finally, we plot in the left panel of Figure \ref{mh2_versus_sfr} CO(1-0) luminosity versus star formation rate for the whole xCOLD GASS sample. Low redshift galaxies with stellar mass $<$ 10$^{10}$ M$_{\odot}$ have much lower CO luminosities per unit star formation than their higher mass counterparts. Is this due to a higher star formation efficiency in lower mass galaxies or due to the photodissociation of CO leading to higher conversion factors? To this end we plot M$_{H_{2}}$ versus star formation rate, using our conversion function. The divergence away from the trend in the left plot is accounted for by the photodissociation of CO and low mass galaxies are, on average, as efficient as higher mass galaxies at forming stars.

It now becomes apparent why our trend for the depletion time versus sSFR, presented in Figure \ref{tdepletion_scaling_relation}, is in agreement with that presented in \citet{2015A&A...583A.114H}; their conversion function was derived under the assumption that the linear relation between L$_{CO}$ and SFR observed in high mass, solar metallicity galaxies extends to the regime of low mass, low metallicity galaxies, with any deviations being attributed to CO photodissociation effects. The results of this study find this assumption to have been correct.

\section{Summary \& Conclusions}\label{summary}

We present here results from xCOLD GASS, a legacy survey of CO(1-0) observations form the IRAM observatory, combined with {\it Herschel} observations of ionised carbon.  We provide the first scaling relations for the integrated $L_{\mbox{[C\scriptsize II]}}$/$L_{\mbox{CO(1-0)}}$ \normalsize ratio as a function of several galaxy parameters for a sample of low metallicity galaxies over 2 dex in $L_{\mbox{[C\scriptsize II]}}$/$L_{\mbox{CO(1-0)}}$. From this, we corrected for contaminant [C\scriptsize II\normalsize] emission from non-molecular phases of the ISM and provide scaling relations for the molecular region $L_{\mbox{[C\scriptsize II]}}$/$L_{\mbox{CO(1-0)}}$ \normalsize ratio. 

We show that the integrated and molecular-only $L_{\mbox{[C\scriptsize II]}}$/$L_{\mbox{CO(1-0)}}$ ratio depends most strongly on parameters which describe the strength of the UV radiation field and the ability of the CO molecule to shield itself, via dust, from the UV radiation impinging on the surface of the the molecular regions deep inside the PDR. A clear dependence on the colour (NUV-r) of the galaxies, specific star formation rate, UV field hardness, gas-phase metallicity and $\Delta$(MS) is observed. On the other hand, the L$_{\mbox{[C\scriptsize II\normalsize]}}$/L$_{\mbox{CO(1-0)}}$ ratio does not depend strongly on large scale parameters which describe the mass and structural properties of the galaxies. 

Through a Bayesian analysis we establish that only two parameters, metallicity and $\Delta$(MS), are needed to robustly parametrise a prescription for variations in L$_{\mbox{[C\scriptsize II\normalsize]}}$/L$_{\mbox{CO(1-0)}}$ across our combined xCOLD GASS and DGS sample. We use our parametrisation of L$_{\mbox{[C\scriptsize II\normalsize]}}$/L$_{\mbox{CO(1-0)}}$, alongside radiative transfer modelling, to present a novel conversion function where:
\begin{dmath}
\centering
\log \alpha_{\mbox{CO}} (\pm 0.165\ \mbox{dex}) = 15.623 - 1.732[12 + \log(\mbox{O}/\mbox{H})] + 0.051 \log \Delta(\mbox{MS}).
\end{dmath}
The prescription should only be applied to galaxies with $7.9<12 + \log(\mbox{O/H})<8.8$ and $-0.8<\log \Delta(\mbox{MS})<1.3$ and which are not in the high-pressure ``merger" regime where a lower $\alpha_{CO}$ value should be use. Our \aco\ conversion function is consistent with previous metallicity-only prescriptions, however we are now able to attribute deviations from this relation to a second order dependence on the offset of a galaxy from the star-forming main sequence. {\it The main interest of this new \aco\ conversion function is that it is calibrated independently of any assumption on the molecular gas depletion timescale, making it perfectly suited for the study of gas and star formation scaling relations}. 

We apply the new \aco\ to the full xCOLD GASS and PHIBSS1 samples and investigate such gas scaling relations. We find a complete flattening of the relation between molecular gas mass fraction and stellar mass as stellar mass decreases; this is expected based on the linearity of the M$_\ast$-SFR relation in this mass regime, and the close relation between molecular gas and star formation. As previously reported, there are systematic variations of the gas depletion timescale across the low and high redshift galaxy populations, although the dependence on sSFR is slightly weaker than previously reported. The depletion time, however, does not vary significantly with stellar mass. Instead, the low L$_{CO}$/SFR ratios in low mass galaxies are entirely due to photodissociation of CO, meaning that on average lower mass galaxies are equally efficient at forming stars than their high mass counterparts.

\section*{Acknowledgments}
GA would like to thank the UK Science and Technologies Facilities Council (STFC) for their support via a postgraduate Studentship.  AS acknowledges the support of the Royal Society through the award of a University Research Fellowship and of a Research Grant. BC is the recipient of an Australian Research Council Future Fellowship (FT120100660). BC and LC acknowledge support from the Australian Research CouncilÕs Discovery Projects funding scheme (DP150101734). This work is based on observations carried out with the IRAM 30m telescope. IRAM is supported by INSU/CNRS (France), MPG (Germany), and IGN (Spain). We thank the staff of the telescope for their help in conducting the COLD GASS observations. PACS, aboard Herschel, has been developed by MPE (Germany); UVIE (Austria); KU Leuven, CSL, IMEC (Belgium); CEA, LAM (France); MPIA (Germany); INAF- IFSI/OAA/OAP/OAT, LENS, SISSA (Italy); IAC (Spain). This development has been supported by BMVIT (Austria), ESA-PRODEX (Belgium), CEA/CNES (France), DLR (Germany), ASI/INAF (Italy), and CICYT/MCYT (Spain). 

\bibliographystyle{mnras}
\bibliography{ms.bib}

\begin{thebibliography}{}
\makeatletter
\relax
\def\mn@urlcharsother{\let\do\@makeother \do\$\do\&\do\#\do\^\do\_\do\%\do\~}
\def\mn@doi{\begingroup\mn@urlcharsother \@ifnextchar [ {\mn@doi@}
  {\mn@doi@[]}}
\def\mn@doi@[#1]#2{\def\@tempa{#1}\ifx\@tempa\@empty \href
  {http://dx.doi.org/#2} {doi:#2}\else \href {http://dx.doi.org/#2} {#1}\fi
  \endgroup}
\def\mn@eprint#1#2{\mn@eprint@#1:#2::\@nil}
\def\mn@eprint@arXiv#1{\href {http://arxiv.org/abs/#1} {{\tt arXiv:#1}}}
\def\mn@eprint@dblp#1{\href {http://dblp.uni-trier.de/rec/bibtex/#1.xml}
  {dblp:#1}}
\def\mn@eprint@#1:#2:#3:#4\@nil{\def\@tempa {#1}\def\@tempb {#2}\def\@tempc
  {#3}\ifx \@tempc \@empty \let \@tempc \@tempb \let \@tempb \@tempa \fi \ifx
  \@tempb \@empty \def\@tempb {arXiv}\fi \@ifundefined
  {mn@eprint@\@tempb}{\@tempb:\@tempc}{\expandafter \expandafter \csname
  mn@eprint@\@tempb\endcsname \expandafter{\@tempc}}}

\bibitem[\protect\citeauthoryear{{Abazajian} et~al.,}{{Abazajian}
  et~al.}{2009}]{2009ApJS..182..543A}
{Abazajian} K.~N.,  et~al., 2009, \mn@doi [\apjs]
  {10.1088/0067-0049/182/2/543}, \href
  {http://adsabs.harvard.edu/abs/2009ApJS..182..543A} {182, 543}

\bibitem[\protect\citeauthoryear{{Abdo}, {Ackermann}, {Ajello}, {Baldini},
  {Ballet}, {Barbiellini}  \& {Fermi/LAT Collaboration}}{{Abdo}
  et~al.}{2010}]{2010ApJ...710..133A}
{Abdo} A.~A.,  {Ackermann} M.,  {Ajello} M.,  {Baldini} L.,  {Ballet} J.,
  {Barbiellini} G.,   {Fermi/LAT Collaboration} 2010, \mn@doi [\apj]
  {10.1088/0004-637X/710/1/133}, \href
  {http://adsabs.harvard.edu/abs/2010ApJ...710..133A} {710, 133}

\bibitem[\protect\citeauthoryear{{Accurso}, {Saintonge}, {Bisbas}  \&
  {Viti}}{{Accurso} et~al.}{2017}]{accurso17}
{Accurso} G.,  {Saintonge} A.,  {Bisbas} T.~G.,   {Viti} S.,  2017, \mn@doi
  [\mnras] {10.1093/mnras/stw2580}, \href
  {http://adsabs.harvard.edu/abs/2017MNRAS.464.3315A} {464, 3315}

\bibitem[\protect\citeauthoryear{Akaike}{Akaike}{1981}]{RePEc:eee:econom:v:16:y:1981:i:1:p:3-14}
Akaike H.,  1981, Journal of Econometrics, 16, 3

\bibitem[\protect\citeauthoryear{{Arimoto}, {Sofue}  \& {Tsujimoto}}{{Arimoto}
  et~al.}{1996}]{1996PASJ...48..275A}
{Arimoto} N.,  {Sofue} Y.,   {Tsujimoto} T.,  1996, \mn@doi [\pasj]
  {10.1093/pasj/48.2.275}, \href
  {http://adsabs.harvard.edu/abs/1996PASJ...48..275A} {48, 275}

\bibitem[\protect\citeauthoryear{{Asplund}, {Grevesse}, {Sauval}  \&
  {Scott}}{{Asplund} et~al.}{2009}]{2009ARA&A..47..481A}
{Asplund} M.,  {Grevesse} N.,  {Sauval} A.~J.,   {Scott} P.,  2009, \mn@doi
  [\araa] {10.1146/annurev.astro.46.060407.145222}, \href
  {http://adsabs.harvard.edu/abs/2009ARA%26A..47..481A} {47, 481}

\bibitem[\protect\citeauthoryear{{Bisbas}, {Bell}, {Viti}, {Yates}  \&
  {Barlow}}{{Bisbas} et~al.}{2012}]{2012MNRAS.427.2100B}
{Bisbas} T.~G.,  {Bell} T.~A.,  {Viti} S.,  {Yates} J.,   {Barlow} M.~J.,
  2012, \mn@doi [\mnras] {10.1111/j.1365-2966.2012.22077.x}, \href
  {http://adsabs.harvard.edu/abs/2012MNRAS.427.2100B} {427, 2100}

\bibitem[\protect\citeauthoryear{{Bisbas}, {Papadopoulos}  \& {Viti}}{{Bisbas}
  et~al.}{2015}]{2015ApJ...803...37B}
{Bisbas} T.~G.,  {Papadopoulos} P.~P.,   {Viti} S.,  2015, \mn@doi [\apj]
  {10.1088/0004-637X/803/1/37}, \href
  {http://adsabs.harvard.edu/abs/2015ApJ...803...37B} {803, 37}

\bibitem[\protect\citeauthoryear{{Bolatto}, {Wolfire}  \& {Leroy}}{{Bolatto}
  et~al.}{2013}]{2013ARA&A..51..207B}
{Bolatto} A.~D.,  {Wolfire} M.,   {Leroy} A.~K.,  2013, \mn@doi [\araa]
  {10.1146/annurev-astro-082812-140944}, \href
  {http://adsabs.harvard.edu/abs/2013ARA%26A..51..207B} {51, 207}

\bibitem[\protect\citeauthoryear{{Boselli}, {Lequeux}  \& {Gavazzi}}{{Boselli}
  et~al.}{2002}]{2002A&A...384...33B}
{Boselli} A.,  {Lequeux} J.,   {Gavazzi} G.,  2002, \mn@doi [\aap]
  {10.1051/0004-6361:20011747}, \href
  {http://adsabs.harvard.edu/abs/2002A%26A...384...33B} {384, 33}

\bibitem[\protect\citeauthoryear{{Bothwell} et~al.,}{{Bothwell}
  et~al.}{2014}]{bothwell14}
{Bothwell} M.~S.,  et~al., 2014, \mn@doi [\mnras] {10.1093/mnras/stu1936},
  \href {http://adsabs.harvard.edu/abs/2014MNRAS.445.2599B} {445, 2599}

\bibitem[\protect\citeauthoryear{{Bouch{\'e}} et~al.,}{{Bouch{\'e}}
  et~al.}{2010}]{2010ApJ...718.1001B}
{Bouch{\'e}} N.,  et~al., 2010, \mn@doi [\apj] {10.1088/0004-637X/718/2/1001},
  \href {http://adsabs.harvard.edu/abs/2010ApJ...718.1001B} {718, 1001}

\bibitem[\protect\citeauthoryear{{Brown}, {Catinella}, {Cortese}, {Kilborn},
  {Haynes}  \& {Giovanelli}}{{Brown} et~al.}{2015}]{2015MNRAS.452.2479B}
{Brown} T.,  {Catinella} B.,  {Cortese} L.,  {Kilborn} V.,  {Haynes} M.~P.,
  {Giovanelli} R.,  2015, \mn@doi [\mnras] {10.1093/mnras/stv1311}, \href
  {http://adsabs.harvard.edu/abs/2015MNRAS.452.2479B} {452, 2479}

\bibitem[\protect\citeauthoryear{{Carter} et~al.,}{{Carter}
  et~al.}{2012}]{2012A&A...538A..89C}
{Carter} M.,  et~al., 2012, \mn@doi [\aap] {10.1051/0004-6361/201118452}, \href
  {http://adsabs.harvard.edu/abs/2012A%26A...538A..89C} {538, A89}

\bibitem[\protect\citeauthoryear{{Catinella}, {Schiminovich}, {Kauffmann},
  {Fabello}  \& {Wang}}{{Catinella} et~al.}{2010}]{2010MNRAS.403..683C}
{Catinella} B.,  {Schiminovich} D.,  {Kauffmann} G.,  {Fabello} S.,   {Wang}
  J.,  2010, \mn@doi [\mnras] {10.1111/j.1365-2966.2009.16180.x}, \href
  {http://adsabs.harvard.edu/abs/2010MNRAS.403..683C} {403, 683}

\bibitem[\protect\citeauthoryear{{Catinella} et~al.,}{{Catinella}
  et~al.}{2013}]{2013MNRAS.436...34C}
{Catinella} B.,  et~al., 2013, \mn@doi [\mnras] {10.1093/mnras/stt1417}, \href
  {http://adsabs.harvard.edu/abs/2013MNRAS.436...34C} {436, 34}

\bibitem[\protect\citeauthoryear{{Clark} \& {Glover}}{{Clark} \&
  {Glover}}{2015}]{2015MNRAS.452.2057C}
{Clark} P.~C.,  {Glover} S.~C.~O.,  2015, \mn@doi [\mnras]
  {10.1093/mnras/stv1369}, \href
  {http://adsabs.harvard.edu/abs/2015MNRAS.452.2057C} {452, 2057}

\bibitem[\protect\citeauthoryear{{Cormier} et~al.,}{{Cormier}
  et~al.}{2014}]{2014A&A...564A.121C}
{Cormier} D.,  et~al., 2014, \mn@doi [\aap] {10.1051/0004-6361/201322096},
  \href {http://adsabs.harvard.edu/abs/2014A%26A...564A.121C} {564, A121}

\bibitem[\protect\citeauthoryear{{Dav{\'e}}, {Finlator}  \&
  {Oppenheimer}}{{Dav{\'e}} et~al.}{2012}]{2012MNRAS.421...98D}
{Dav{\'e}} R.,  {Finlator} K.,   {Oppenheimer} B.~D.,  2012, \mn@doi [\mnras]
  {10.1111/j.1365-2966.2011.20148.x}, \href
  {http://adsabs.harvard.edu/abs/2012MNRAS.421...98D} {421, 98}

\bibitem[\protect\citeauthoryear{{Davis} et~al.,}{{Davis}
  et~al.}{2014}]{2014MNRAS.444.3427D}
{Davis} T.~A.,  et~al., 2014, \mn@doi [\mnras] {10.1093/mnras/stu570}, \href
  {http://adsabs.harvard.edu/abs/2014MNRAS.444.3427D} {444, 3427}

\bibitem[\protect\citeauthoryear{{Dickman}, {Snell}  \& {Schloerb}}{{Dickman}
  et~al.}{1986}]{1986ApJ...309..326D}
{Dickman} R.~L.,  {Snell} R.~L.,   {Schloerb} F.~P.,  1986, \mn@doi [\apj]
  {10.1086/164604}, \href {http://adsabs.harvard.edu/abs/1986ApJ...309..326D}
  {309, 326}

\bibitem[\protect\citeauthoryear{{Dom{\'{\i}}nguez}, {Siana}, {Brooks},
  {Christensen}, {Bruzual}, {Stark}  \& {Alavi}}{{Dom{\'{\i}}nguez}
  et~al.}{2015}]{2015MNRAS.451..839D}
{Dom{\'{\i}}nguez} A.,  {Siana} B.,  {Brooks} A.~M.,  {Christensen} C.~R.,
  {Bruzual} G.,  {Stark} D.~P.,   {Alavi} A.,  2015, \mn@doi [\mnras]
  {10.1093/mnras/stv1001}, \href
  {http://adsabs.harvard.edu/abs/2015MNRAS.451..839D} {451, 839}

\bibitem[\protect\citeauthoryear{{Ercolano}, {Barlow}, {Storey}  \&
  {Liu}}{{Ercolano} et~al.}{2003}]{2003MNRAS.340.1136E}
{Ercolano} B.,  {Barlow} M.~J.,  {Storey} P.~J.,   {Liu} X.-W.,  2003, \mn@doi
  [\mnras] {10.1046/j.1365-8711.2003.06371.x}, \href
  {http://adsabs.harvard.edu/abs/2003MNRAS.340.1136E} {340, 1136}

\bibitem[\protect\citeauthoryear{{Ercolano}, {Barlow}  \& {Storey}}{{Ercolano}
  et~al.}{2005}]{2005MNRAS.362.1038E}
{Ercolano} B.,  {Barlow} M.~J.,   {Storey} P.~J.,  2005, \mn@doi [\mnras]
  {10.1111/j.1365-2966.2005.09381.x}, \href
  {http://adsabs.harvard.edu/abs/2005MNRAS.362.1038E} {362, 1038}

\bibitem[\protect\citeauthoryear{{Foster}, {Mandel}, {Pineda}, {Covey}, {Arce}
  \& {Goodman}}{{Foster} et~al.}{2013}]{2013MNRAS.428.1606F}
{Foster} J.~B.,  {Mandel} K.~S.,  {Pineda} J.~E.,  {Covey} K.~R.,  {Arce}
  H.~G.,   {Goodman} A.~A.,  2013, \mn@doi [\mnras] {10.1093/mnras/sts144},
  \href {http://adsabs.harvard.edu/abs/2013MNRAS.428.1606F} {428, 1606}

\bibitem[\protect\citeauthoryear{{Galametz} et~al.,}{{Galametz}
  et~al.}{2010}]{2010A&A...518L..55G}
{Galametz} M.,  et~al., 2010, \mn@doi [\aap] {10.1051/0004-6361/201014604},
  \href {http://adsabs.harvard.edu/abs/2010A%26A...518L..55G} {518, L55}

\bibitem[\protect\citeauthoryear{{Genzel}, {Tacconi}, {Combes}, {Bolatto},
  {Neri}  \& {Sternberg}}{{Genzel} et~al.}{2012}]{2012ApJ...746...69G}
{Genzel} R.,  {Tacconi} L.~J.,  {Combes} F.,  {Bolatto} A.,  {Neri} R.,
  {Sternberg} A.,  2012, \mn@doi [\apj] {10.1088/0004-637X/746/1/69}, \href
  {http://adsabs.harvard.edu/abs/2012ApJ...746...69G} {746, 69}

\bibitem[\protect\citeauthoryear{{Genzel} et~al.,}{{Genzel}
  et~al.}{2015}]{2015ApJ...800...20G}
{Genzel} R.,  et~al., 2015, \mn@doi [\apj] {10.1088/0004-637X/800/1/20}, \href
  {http://adsabs.harvard.edu/abs/2015ApJ...800...20G} {800, 20}

\bibitem[\protect\citeauthoryear{{Giovanelli} et~al.,}{{Giovanelli}
  et~al.}{2005}]{2005AJ....130.2598G}
{Giovanelli} R.,  et~al., 2005, \mn@doi [\aj] {10.1086/497431}, \href
  {http://adsabs.harvard.edu/abs/2005AJ....130.2598G} {130, 2598}

\bibitem[\protect\citeauthoryear{Goodman \& Weare}{Goodman \&
  Weare}{2010}]{emcee}
Goodman J.,  Weare J.,  2010, \mn@doi [Comm. App. Math. and Comp. Sci.]
  {10.2140/camcos.2010.5.65}, 5

\bibitem[\protect\citeauthoryear{{Grenier}, {Casandjian}  \&
  {Terrier}}{{Grenier} et~al.}{2005}]{2005Sci...307.1292G}
{Grenier} I.~A.,  {Casandjian} J.-M.,   {Terrier} R.,  2005, \mn@doi [Science]
  {10.1126/science.1106924}, \href
  {http://adsabs.harvard.edu/abs/2005Sci...307.1292G} {307, 1292}

\bibitem[\protect\citeauthoryear{{Grossi} et~al.,}{{Grossi}
  et~al.}{2016}]{2016arXiv160209077G}
{Grossi} M.,  et~al., 2016, preprint, \href
  {http://adsabs.harvard.edu/abs/2016arXiv160209077G} {} (\mn@eprint {arXiv}
  {1602.09077})

\bibitem[\protect\citeauthoryear{{Guo}, {Zheng}, {Wang}  \& {Fu}}{{Guo}
  et~al.}{2015}]{2015ApJ...808L..49G}
{Guo} K.,  {Zheng} X.~Z.,  {Wang} T.,   {Fu} H.,  2015, \mn@doi [\apjl]
  {10.1088/2041-8205/808/2/L49}, \href
  {http://adsabs.harvard.edu/abs/2015ApJ...808L..49G} {808, L49}

\bibitem[\protect\citeauthoryear{{Ho}, {Kudritzki}, {Kewley}, {Zahid},
  {Dopita}, {Bresolin}  \& {Rupke}}{{Ho} et~al.}{2015}]{2015MNRAS.448.2030H}
{Ho} I.-T.,  {Kudritzki} R.-P.,  {Kewley} L.~J.,  {Zahid} H.~J.,  {Dopita}
  M.~A.,  {Bresolin} F.,   {Rupke} D.~S.~N.,  2015, \mn@doi [\mnras]
  {10.1093/mnras/stv067}, \href
  {http://adsabs.harvard.edu/abs/2015MNRAS.448.2030H} {448, 2030}

\bibitem[\protect\citeauthoryear{{Huang} \& {Kauffmann}}{{Huang} \&
  {Kauffmann}}{2014}]{2014MNRAS.443.1329H}
{Huang} M.-L.,  {Kauffmann} G.,  2014, \mn@doi [\mnras]
  {10.1093/mnras/stu1232}, \href
  {http://adsabs.harvard.edu/abs/2014MNRAS.443.1329H} {443, 1329}

\bibitem[\protect\citeauthoryear{{Hughes} et~al.,}{{Hughes}
  et~al.}{2015}]{2015A&A...575A..17H}
{Hughes} T.~M.,  et~al., 2015, \mn@doi [\aap] {10.1051/0004-6361/201424732},
  \href {http://adsabs.harvard.edu/abs/2015A%26A...575A..17H} {575, A17}

\bibitem[\protect\citeauthoryear{{Hunt} et~al.,}{{Hunt}
  et~al.}{2015}]{2015A&A...583A.114H}
{Hunt} L.~K.,  et~al., 2015, \mn@doi [\aap] {10.1051/0004-6361/201526553},
  \href {http://adsabs.harvard.edu/abs/2015A%26A...583A.114H} {583, A114}

\bibitem[\protect\citeauthoryear{{Israel}}{{Israel}}{1997}]{1997A&A...328..471I}
{Israel} F.~P.,  1997, \aap, \href
  {http://adsabs.harvard.edu/abs/1997A%26A...328..471I} {328, 471}

\bibitem[\protect\citeauthoryear{{Israel} \& {Baas}}{{Israel} \&
  {Baas}}{2003}]{2003A&A...404..495I}
{Israel} F.~P.,  {Baas} F.,  2003, \mn@doi [\aap] {10.1051/0004-6361:20030479},
  \href {http://adsabs.harvard.edu/abs/2003A%26A...404..495I} {404, 495}

\bibitem[\protect\citeauthoryear{{Karim} et~al.,}{{Karim}
  et~al.}{2011}]{2011ApJ...730...61K}
{Karim} A.,  et~al., 2011, \mn@doi [\apj] {10.1088/0004-637X/730/2/61}, \href
  {http://adsabs.harvard.edu/abs/2011ApJ...730...61K} {730, 61}

\bibitem[\protect\citeauthoryear{{Kauffmann} et~al.,}{{Kauffmann}
  et~al.}{2003}]{2003MNRAS.341...33K}
{Kauffmann} G.,  et~al., 2003, \mn@doi [\mnras]
  {10.1046/j.1365-8711.2003.06291.x}, \href
  {http://adsabs.harvard.edu/abs/2003MNRAS.341...33K} {341, 33}

\bibitem[\protect\citeauthoryear{{Kaufman}, {Wolfire}, {Hollenbach}  \&
  {Luhman}}{{Kaufman} et~al.}{1999}]{1999ApJ...527..795K}
{Kaufman} M.~J.,  {Wolfire} M.~G.,  {Hollenbach} D.~J.,   {Luhman} M.~L.,
  1999, \mn@doi [\apj] {10.1086/308102}, \href
  {http://adsabs.harvard.edu/abs/1999ApJ...527..795K} {527, 795}

\bibitem[\protect\citeauthoryear{{Kelly}}{{Kelly}}{2007}]{2007ApJ...665.1489K}
{Kelly} B.~C.,  2007, \mn@doi [\apj] {10.1086/519947}, \href
  {http://adsabs.harvard.edu/abs/2007ApJ...665.1489K} {665, 1489}

\bibitem[\protect\citeauthoryear{{Kennicutt} \& {Evans}}{{Kennicutt} \&
  {Evans}}{2012}]{2012ARA&A..50..531K}
{Kennicutt} R.~C.,  {Evans} N.~J.,  2012, \mn@doi [\araa]
  {10.1146/annurev-astro-081811-125610}, \href
  {http://adsabs.harvard.edu/abs/2012ARA%26A..50..531K} {50, 531}

\bibitem[\protect\citeauthoryear{{Kewley} \& {Ellison}}{{Kewley} \&
  {Ellison}}{2008}]{2008ApJ...681.1183K}
{Kewley} L.~J.,  {Ellison} S.~L.,  2008, \mn@doi [\apj] {10.1086/587500}, \href
  {http://adsabs.harvard.edu/abs/2008ApJ...681.1183K} {681, 1183}

\bibitem[\protect\citeauthoryear{{Kramer} et~al.,}{{Kramer}
  et~al.}{2013}]{2013A&A...553A.114K}
{Kramer} C.,  et~al., 2013, \mn@doi [\aap] {10.1051/0004-6361/201220683}, \href
  {http://adsabs.harvard.edu/abs/2013A%26A...553A.114K} {553, A114}

\bibitem[\protect\citeauthoryear{{Krumholz}, {McKee}  \&
  {Tumlinson}}{{Krumholz} et~al.}{2008}]{2008ApJ...689..865K}
{Krumholz} M.~R.,  {McKee} C.~F.,   {Tumlinson} J.,  2008, \mn@doi [\apj]
  {10.1086/592490}, \href {http://adsabs.harvard.edu/abs/2008ApJ...689..865K}
  {689, 865}

\bibitem[\protect\citeauthoryear{{Krumholz}, {McKee}  \&
  {Tumlinson}}{{Krumholz} et~al.}{2009}]{2009ApJ...693..216K}
{Krumholz} M.~R.,  {McKee} C.~F.,   {Tumlinson} J.,  2009, \mn@doi [\apj]
  {10.1088/0004-637X/693/1/216}, \href
  {http://adsabs.harvard.edu/abs/2009ApJ...693..216K} {693, 216}

\bibitem[\protect\citeauthoryear{{Leitherer} et~al.,}{{Leitherer}
  et~al.}{1999}]{1999ApJS..123....3L}
{Leitherer} C.,  et~al., 1999, \mn@doi [\apjs] {10.1086/313233}, \href
  {http://adsabs.harvard.edu/abs/1999ApJS..123....3L} {123, 3}

\bibitem[\protect\citeauthoryear{{Leitherer}, {Ortiz Ot{\'a}lvaro}, {Bresolin},
  {Kudritzki}, {Lo Faro}, {Pauldrach}, {Pettini}  \& {Rix}}{{Leitherer}
  et~al.}{2010}]{2010ApJS..189..309L}
{Leitherer} C.,  {Ortiz Ot{\'a}lvaro} P.~A.,  {Bresolin} F.,  {Kudritzki}
  R.-P.,  {Lo Faro} B.,  {Pauldrach} A.~W.~A.,  {Pettini} M.,   {Rix} S.~A.,
  2010, \mn@doi [\apjs] {10.1088/0067-0049/189/2/309}, \href
  {http://adsabs.harvard.edu/abs/2010ApJS..189..309L} {189, 309}

\bibitem[\protect\citeauthoryear{{Leroy}, {Walter}, {Brinks}, {Bigiel}, {de
  Blok}, {Madore}  \& {Thornley}}{{Leroy} et~al.}{2008}]{2008AJ....136.2782L}
{Leroy} A.~K.,  {Walter} F.,  {Brinks} E.,  {Bigiel} F.,  {de Blok} W.~J.~G.,
  {Madore} B.,   {Thornley} M.~D.,  2008, \mn@doi [\aj]
  {10.1088/0004-6256/136/6/2782}, \href
  {http://adsabs.harvard.edu/abs/2008AJ....136.2782L} {136, 2782}

\bibitem[\protect\citeauthoryear{{Leroy} et~al.,}{{Leroy}
  et~al.}{2009}]{2009AJ....137.4670L}
{Leroy} A.~K.,  et~al., 2009, \mn@doi [\aj] {10.1088/0004-6256/137/6/4670},
  \href {http://adsabs.harvard.edu/abs/2009AJ....137.4670L} {137, 4670}

\bibitem[\protect\citeauthoryear{{Leroy} et~al.,}{{Leroy}
  et~al.}{2011}]{2011ApJ...737...12L}
{Leroy} A.~K.,  et~al., 2011, \mn@doi [\apj] {10.1088/0004-637X/737/1/12},
  \href {http://adsabs.harvard.edu/abs/2011ApJ...737...12L} {737, 12}

\bibitem[\protect\citeauthoryear{{Leroy} et~al.,}{{Leroy}
  et~al.}{2013}]{leroy13}
{Leroy} A.~K.,  et~al., 2013, \mn@doi [\aj] {10.1088/0004-6256/146/2/19}, \href
  {http://adsabs.harvard.edu/abs/2013AJ....146...19L} {146, 19}

\bibitem[\protect\citeauthoryear{{Lilly}, {Carollo}, {Pipino}, {Renzini}  \&
  {Peng}}{{Lilly} et~al.}{2013}]{2013ApJ...772..119L}
{Lilly} S.~J.,  {Carollo} C.~M.,  {Pipino} A.,  {Renzini} A.,   {Peng} Y.,
  2013, \mn@doi [\apj] {10.1088/0004-637X/772/2/119}, \href
  {http://adsabs.harvard.edu/abs/2013ApJ...772..119L} {772, 119}

\bibitem[\protect\citeauthoryear{{Lu}, {Mo}  \& {Lu}}{{Lu}
  et~al.}{2015}]{2015MNRAS.450..606L}
{Lu} Z.,  {Mo} H.~J.,   {Lu} Y.,  2015, \mn@doi [\mnras]
  {10.1093/mnras/stv671}, \href
  {http://adsabs.harvard.edu/abs/2015MNRAS.450..606L} {450, 606}

\bibitem[\protect\citeauthoryear{{Madden}, {Poglitsch}, {Geis}, {Stacey}  \&
  {Townes}}{{Madden} et~al.}{1997}]{1997ApJ...483..200M}
{Madden} S.~C.,  {Poglitsch} A.,  {Geis} N.,  {Stacey} G.~J.,   {Townes} C.~H.,
   1997, \apj, \href {http://adsabs.harvard.edu/abs/1997ApJ...483..200M} {483,
  200}

\bibitem[\protect\citeauthoryear{{Madden}, {R{\'e}my-Ruyer}, {Galametz},
  {Cormier}, {Lebouteiller}  \& {Galliano}}{{Madden}
  et~al.}{2013}]{2013PASP..125..600M}
{Madden} S.~C.,  {R{\'e}my-Ruyer} A.,  {Galametz} M.,  {Cormier} D.,
  {Lebouteiller} V.,   {Galliano} F.,  2013, \mn@doi [\pasp] {10.1086/671138},
  \href {http://adsabs.harvard.edu/abs/2013PASP..125..600M} {125, 600}

\bibitem[\protect\citeauthoryear{{Magnelli} et~al.,}{{Magnelli}
  et~al.}{2012}]{2012A&A...548A..22M}
{Magnelli} B.,  et~al., 2012, \mn@doi [\aap] {10.1051/0004-6361/201220074},
  \href {http://adsabs.harvard.edu/abs/2012A%26A...548A..22M} {548, A22}

\bibitem[\protect\citeauthoryear{{Magrini}, {Coccato}, {Stanghellini},
  {Casasola}  \& {Galli}}{{Magrini} et~al.}{2016}]{2016A&A...588A..91M}
{Magrini} L.,  {Coccato} L.,  {Stanghellini} L.,  {Casasola} V.,   {Galli} D.,
  2016, \mn@doi [\aap] {10.1051/0004-6361/201527799}, \href
  {http://adsabs.harvard.edu/abs/2016A%26A...588A..91M} {588, A91}

\bibitem[\protect\citeauthoryear{{Makrymallis} \& {Viti}}{{Makrymallis} \&
  {Viti}}{2014}]{2014ApJ...794...45M}
{Makrymallis} A.,  {Viti} S.,  2014, \mn@doi [\apj]
  {10.1088/0004-637X/794/1/45}, \href
  {http://adsabs.harvard.edu/abs/2014ApJ...794...45M} {794, 45}

\bibitem[\protect\citeauthoryear{{Mandel}, {Narayan}  \& {Kirshner}}{{Mandel}
  et~al.}{2011}]{2011ApJ...731..120M}
{Mandel} K.~S.,  {Narayan} G.,   {Kirshner} R.~P.,  2011, \mn@doi [\apj]
  {10.1088/0004-637X/731/2/120}, \href
  {http://adsabs.harvard.edu/abs/2011ApJ...731..120M} {731, 120}

\bibitem[\protect\citeauthoryear{{Mannucci}, {Cresci}, {Maiolino}, {Marconi}
  \& {Gnerucci}}{{Mannucci} et~al.}{2010}]{2010MNRAS.408.2115M}
{Mannucci} F.,  {Cresci} G.,  {Maiolino} R.,  {Marconi} A.,   {Gnerucci} A.,
  2010, \mn@doi [\mnras] {10.1111/j.1365-2966.2010.17291.x}, \href
  {http://adsabs.harvard.edu/abs/2010MNRAS.408.2115M} {408, 2115}

\bibitem[\protect\citeauthoryear{{Martin}, {Fanson}, {Schiminovich},
  {Morrissey}, {Friedman}  \& {Barlow}}{{Martin}
  et~al.}{2005}]{2005ApJ...619L...1M}
{Martin} D.~C.,  {Fanson} J.,  {Schiminovich} D.,  {Morrissey} P.,  {Friedman}
  P.~G.,   {Barlow} T.~A.,  2005, \mn@doi [\apjl] {10.1086/426387}, \href
  {http://adsabs.harvard.edu/abs/2005ApJ...619L...1M} {619, L1}

\bibitem[\protect\citeauthoryear{{Mookerjea} et~al.,}{{Mookerjea}
  et~al.}{2011}]{2011A&A...532A.152M}
{Mookerjea} B.,  et~al., 2011, \mn@doi [\aap] {10.1051/0004-6361/201116447},
  \href {http://adsabs.harvard.edu/abs/2011A%26A...532A.152M} {532, A152}

\bibitem[\protect\citeauthoryear{{Oberst} et~al.,}{{Oberst}
  et~al.}{2006}]{2006ApJ...652L.125O}
{Oberst} T.~E.,  et~al., 2006, \mn@doi [\apjl] {10.1086/510289}, \href
  {http://adsabs.harvard.edu/abs/2006ApJ...652L.125O} {652, L125}

\bibitem[\protect\citeauthoryear{{Obreschkow} \& {Rawlings}}{{Obreschkow} \&
  {Rawlings}}{2009}]{2009MNRAS.394.1857O}
{Obreschkow} D.,  {Rawlings} S.,  2009, \mn@doi [\mnras]
  {10.1111/j.1365-2966.2009.14497.x}, \href
  {http://adsabs.harvard.edu/abs/2009MNRAS.394.1857O} {394, 1857}

\bibitem[\protect\citeauthoryear{{Ott}}{{Ott}}{2010}]{2010ASPC..434..139O}
{Ott} S.,  2010, in {Mizumoto} Y.,  {Morita} K.-I.,   {Ohishi} M.,  eds,
  Astronomical Society of the Pacific Conference Series Vol. 434, Astronomical
  Data Analysis Software and Systems XIX. p.~139 (\mn@eprint {arXiv}
  {1011.1209})

\bibitem[\protect\citeauthoryear{{Pettini} \& {Pagel}}{{Pettini} \&
  {Pagel}}{2004}]{2004MNRAS.348L..59P}
{Pettini} M.,  {Pagel} B.~E.~J.,  2004, \mn@doi [\mnras]
  {10.1111/j.1365-2966.2004.07591.x}, \href
  {http://adsabs.harvard.edu/abs/2004MNRAS.348L..59P} {348, L59}

\bibitem[\protect\citeauthoryear{{Pilbratt}, {Riedinger}, {Passvogel}, {Crone},
  {Doyle}  \& {Gageur}}{{Pilbratt} et~al.}{2010}]{2010A&A...518L...1P}
{Pilbratt} G.~L.,  {Riedinger} J.~R.,  {Passvogel} T.,  {Crone} G.,  {Doyle}
  D.,   {Gageur} U.,  2010, \mn@doi [\aap] {10.1051/0004-6361/201014759}, \href
  {http://adsabs.harvard.edu/abs/2010A%26A...518L...1P} {518, L1}

\bibitem[\protect\citeauthoryear{{Pineda}, {Langer}, {Velusamy}  \&
  {Goldsmith}}{{Pineda} et~al.}{2013}]{2013A&A...554A.103P}
{Pineda} J.~L.,  {Langer} W.~D.,  {Velusamy} T.,   {Goldsmith} P.~F.,  2013,
  \mn@doi [\aap] {10.1051/0004-6361/201321188}, \href
  {http://adsabs.harvard.edu/abs/2013A%26A...554A.103P} {554, A103}

\bibitem[\protect\citeauthoryear{{Poglitsch}, {Krabbe}, {Madden}, {Nikola},
  {Geis}, {Johansson}, {Stacey}  \& {Sternberg}}{{Poglitsch}
  et~al.}{1995}]{1995ApJ...454..293P}
{Poglitsch} A.,  {Krabbe} A.,  {Madden} S.~C.,  {Nikola} T.,  {Geis} N.,
  {Johansson} L.~E.~B.,  {Stacey} G.~J.,   {Sternberg} A.,  1995, \mn@doi
  [\apj] {10.1086/176482}, \href
  {http://adsabs.harvard.edu/abs/1995ApJ...454..293P} {454, 293}

\bibitem[\protect\citeauthoryear{{Poglitsch}, {Waelkens}, {Geis},
  {Feuchtgruber}, {Vandenbussche}, {Rodriguez}  \& {Krause}}{{Poglitsch}
  et~al.}{2010}]{2010A&A...518L...2P}
{Poglitsch} A.,  {Waelkens} C.,  {Geis} N.,  {Feuchtgruber} H.,
  {Vandenbussche} B.,  {Rodriguez} L.,   {Krause} O.,  2010, \mn@doi [\aap]
  {10.1051/0004-6361/201014535}, \href
  {http://adsabs.harvard.edu/abs/2010A%26A...518L...2P} {518, L2}

\bibitem[\protect\citeauthoryear{{R{\'e}my-Ruyer}, {Madden}, {Galliano},
  {Galametz}, {Takeuchi}  \& {Asano}}{{R{\'e}my-Ruyer}
  et~al.}{2014}]{2014A&A...563A..31R}
{R{\'e}my-Ruyer} A.,  {Madden} S.~C.,  {Galliano} F.,  {Galametz} M.,
  {Takeuchi} T.~T.,   {Asano} R.~S.,  2014, \mn@doi [\aap]
  {10.1051/0004-6361/201322803}, \href
  {http://adsabs.harvard.edu/abs/2014A%26A...563A..31R} {563, A31}

\bibitem[\protect\citeauthoryear{{R{\"o}llig}, {Ossenkopf}, {Jeyakumar},
  {Stutzki}  \& {Sternberg}}{{R{\"o}llig} et~al.}{2006}]{2006A&A...451..917R}
{R{\"o}llig} M.,  {Ossenkopf} V.,  {Jeyakumar} S.,  {Stutzki} J.,   {Sternberg}
  A.,  2006, \mn@doi [\aap] {10.1051/0004-6361:20053845}, \href
  {http://adsabs.harvard.edu/abs/2006A%26A...451..917R} {451, 917}

\bibitem[\protect\citeauthoryear{{Saintonge}, {Kauffmann}, {Kramer}, {Tacconi},
  {Buchbender}, {Catinella}  \& {Fabello}}{{Saintonge}
  et~al.}{2011a}]{2011MNRAS.415...32S}
{Saintonge} A.,  {Kauffmann} G.,  {Kramer} C.,  {Tacconi} L.~J.,  {Buchbender}
  C.,  {Catinella} B.,   {Fabello} S.,  2011a, \mn@doi [\mnras]
  {10.1111/j.1365-2966.2011.18677.x}, \href
  {http://adsabs.harvard.edu/abs/2011MNRAS.415...32S} {415, 32}

\bibitem[\protect\citeauthoryear{{Saintonge}, {Kauffmann}, {Wang}, {Kramer},
  {Tacconi}, {Buchbender}  \& {Catinella}}{{Saintonge}
  et~al.}{2011b}]{2011MNRAS.415...61S}
{Saintonge} A.,  {Kauffmann} G.,  {Wang} J.,  {Kramer} C.,  {Tacconi} L.~J.,
  {Buchbender} C.,   {Catinella} B.,  2011b, \mn@doi [\mnras]
  {10.1111/j.1365-2966.2011.18823.x}, \href
  {http://adsabs.harvard.edu/abs/2011MNRAS.415...61S} {415, 61}

\bibitem[\protect\citeauthoryear{{Saintonge} et~al.,}{{Saintonge}
  et~al.}{2012}]{2012ApJ...758...73S}
{Saintonge} A.,  et~al., 2012, \mn@doi [\apj] {10.1088/0004-637X/758/2/73},
  \href {http://adsabs.harvard.edu/abs/2012ApJ...758...73S} {758, 73}

\bibitem[\protect\citeauthoryear{{Saintonge} et~al.,}{{Saintonge}
  et~al.}{2013}]{2013ApJ...778....2S}
{Saintonge} A.,  et~al., 2013, \mn@doi [\apj] {10.1088/0004-637X/778/1/2},
  \href {http://adsabs.harvard.edu/abs/2013ApJ...778....2S} {778, 2}

\bibitem[\protect\citeauthoryear{{Saintonge} et~al.,}{{Saintonge}
  et~al.}{2016}]{2016arXiv160705289S}
{Saintonge} A.,  et~al., 2016, preprint, \href
  {http://adsabs.harvard.edu/abs/2016arXiv160705289S} {} (\mn@eprint {arXiv}
  {1607.05289})

\bibitem[\protect\citeauthoryear{{Sandstrom}, {Leroy}, {Walter}, {Bolatto},
  {Croxall}  \& et al.}{{Sandstrom} et~al.}{2012}]{sandstrom13}
{Sandstrom} K.~M.,  {Leroy} A.~K.,  {Walter} F.,  {Bolatto} A.~D.,  {Croxall}
  K.~V.,   et al. 2012, arXiv:1212.120, \href
  {http://adsabs.harvard.edu/abs/2012arXiv1212.1208S} {}

\bibitem[\protect\citeauthoryear{{Sandstrom} et~al.,}{{Sandstrom}
  et~al.}{2013}]{2013ApJ...777....5S}
{Sandstrom} K.~M.,  et~al., 2013, \mn@doi [\apj] {10.1088/0004-637X/777/1/5},
  \href {http://adsabs.harvard.edu/abs/2013ApJ...777....5S} {777, 5}

\bibitem[\protect\citeauthoryear{{Santini} et~al.,}{{Santini}
  et~al.}{2014}]{2014A&A...562A..30S}
{Santini} P.,  et~al., 2014, \mn@doi [\aap] {10.1051/0004-6361/201322835},
  \href {http://adsabs.harvard.edu/abs/2014A%26A...562A..30S} {562, A30}

\bibitem[\protect\citeauthoryear{{Sargent} et~al.,}{{Sargent}
  et~al.}{2014}]{2014ApJ...793...19S}
{Sargent} M.~T.,  et~al., 2014, \mn@doi [\apj] {10.1088/0004-637X/793/1/19},
  \href {http://adsabs.harvard.edu/abs/2014ApJ...793...19S} {793, 19}

\bibitem[\protect\citeauthoryear{{Schruba} et~al.,}{{Schruba}
  et~al.}{2012}]{2012AJ....143..138S}
{Schruba} A.,  et~al., 2012, \mn@doi [\aj] {10.1088/0004-6256/143/6/138}, \href
  {http://adsabs.harvard.edu/abs/2012AJ....143..138S} {143, 138}

\bibitem[\protect\citeauthoryear{Schwarz}{Schwarz}{1978}]{schwarz1978}
Schwarz G.,  1978, \mn@doi [Ann. Statist.] {10.1214/aos/1176344136}, 6, 461

\bibitem[\protect\citeauthoryear{{Scoville}}{{Scoville}}{2013}]{2013seg..book..491S}
{Scoville} N.~Z.,  2013, {Evolution of star formation and gas}.
p.~491

\bibitem[\protect\citeauthoryear{{Shetty}, {Kelly}  \& {Bigiel}}{{Shetty}
  et~al.}{2013}]{2013MNRAS.430..288S}
{Shetty} R.,  {Kelly} B.~C.,   {Bigiel} F.,  2013, \mn@doi [\mnras]
  {10.1093/mnras/sts617}, \href
  {http://adsabs.harvard.edu/abs/2013MNRAS.430..288S} {430, 288}

\bibitem[\protect\citeauthoryear{{Solomon}, {Rivolo}, {Barrett}  \&
  {Yahil}}{{Solomon} et~al.}{1987}]{1987ApJ...319..730S}
{Solomon} P.~M.,  {Rivolo} A.~R.,  {Barrett} J.,   {Yahil} A.,  1987, \mn@doi
  [\apj] {10.1086/165493}, \href
  {http://adsabs.harvard.edu/abs/1987ApJ...319..730S} {319, 730}

\bibitem[\protect\citeauthoryear{{Solomon}, {Downes}, {Radford}  \&
  {Barrett}}{{Solomon} et~al.}{1997}]{1997ApJ...478..144S}
{Solomon} P.~M.,  {Downes} D.,  {Radford} S.~J.~E.,   {Barrett} J.~W.,  1997,
  \apj, \href {http://adsabs.harvard.edu/abs/1997ApJ...478..144S} {478, 144}

\bibitem[\protect\citeauthoryear{{Stacey}, {Geis}, {Genzel}, {Lugten},
  {Poglitsch}, {Sternberg}  \& {Townes}}{{Stacey}
  et~al.}{1991}]{1991ApJ...373..423S}
{Stacey} G.~J.,  {Geis} N.,  {Genzel} R.,  {Lugten} J.~B.,  {Poglitsch} A.,
  {Sternberg} A.,   {Townes} C.~H.,  1991, \mn@doi [\apj] {10.1086/170062},
  \href {http://adsabs.harvard.edu/abs/1991ApJ...373..423S} {373, 423}

\bibitem[\protect\citeauthoryear{{Stoughton} et~al.,}{{Stoughton}
  et~al.}{2002}]{2002AJ....123..485S}
{Stoughton} C.,  et~al., 2002, \mn@doi [\aj] {10.1086/324741}, \href
  {http://adsabs.harvard.edu/abs/2002AJ....123..485S} {123, 485}

\bibitem[\protect\citeauthoryear{{Strong} \& {Mattox}}{{Strong} \&
  {Mattox}}{1996}]{1996A&A...308L..21S}
{Strong} A.~W.,  {Mattox} J.~R.,  1996, \aap, \href
  {http://adsabs.harvard.edu/abs/1996A%26A...308L..21S} {308, L21}

\bibitem[\protect\citeauthoryear{{Tacconi} et~al.,}{{Tacconi}
  et~al.}{2013}]{2013ApJ...768...74T}
{Tacconi} L.~J.,  et~al., 2013, \mn@doi [\apj] {10.1088/0004-637X/768/1/74},
  \href {http://adsabs.harvard.edu/abs/2013ApJ...768...74T} {768, 74}

\bibitem[\protect\citeauthoryear{{Tacconi} et~al.,}{{Tacconi}
  et~al.}{2017}]{tacconi17}
{Tacconi} L.~J.,  et~al., 2017, preprint, \href
  {http://adsabs.harvard.edu/abs/2017arXiv170201140T} {} (\mn@eprint {arXiv}
  {1702.01140})

\bibitem[\protect\citeauthoryear{{Tielens} \& {Hollenbach}}{{Tielens} \&
  {Hollenbach}}{1985}]{1985ApJ...291..722T}
{Tielens} A.~G.~G.~M.,  {Hollenbach} D.,  1985, \mn@doi [\apj]
  {10.1086/163111}, \href {http://adsabs.harvard.edu/abs/1985ApJ...291..722T}
  {291, 722}

\bibitem[\protect\citeauthoryear{{Tissera}, {Pedrosa}, {Sillero}  \&
  {Vilchez}}{{Tissera} et~al.}{2016}]{2016MNRAS.456.2982T}
{Tissera} P.~B.,  {Pedrosa} S.~E.,  {Sillero} E.,   {Vilchez} J.~M.,  2016,
  \mn@doi [\mnras] {10.1093/mnras/stv2736}, \href
  {http://adsabs.harvard.edu/abs/2016MNRAS.456.2982T} {456, 2982}

\bibitem[\protect\citeauthoryear{{Tremonti} et~al.,}{{Tremonti}
  et~al.}{2004}]{2004ApJ...613..898T}
{Tremonti} C.~A.,  et~al., 2004, \mn@doi [\apj] {10.1086/423264}, \href
  {http://adsabs.harvard.edu/abs/2004ApJ...613..898T} {613, 898}

\bibitem[\protect\citeauthoryear{{Wang}, {Overzier}, {Kauffmann}, {von der
  Linden}  \& {Kong}}{{Wang} et~al.}{2010}]{2010MNRAS.401..433W}
{Wang} J.,  {Overzier} R.,  {Kauffmann} G.,  {von der Linden} A.,   {Kong} X.,
  2010, \mn@doi [\mnras] {10.1111/j.1365-2966.2009.15653.x}, \href
  {http://adsabs.harvard.edu/abs/2010MNRAS.401..433W} {401, 433}

\bibitem[\protect\citeauthoryear{{Whitaker}, {van Dokkum}, {Brammer}  \&
  {Franx}}{{Whitaker} et~al.}{2012}]{2012ApJ...754L..29W}
{Whitaker} K.~E.,  {van Dokkum} P.~G.,  {Brammer} G.,   {Franx} M.,  2012,
  \mn@doi [\apjl] {10.1088/2041-8205/754/2/L29}, \href
  {http://adsabs.harvard.edu/abs/2012ApJ...754L..29W} {754, L29}

\bibitem[\protect\citeauthoryear{{White} \& {Frenk}}{{White} \&
  {Frenk}}{1991}]{1991ApJ...379...52W}
{White} S.~D.~M.,  {Frenk} C.~S.,  1991, \mn@doi [\apj] {10.1086/170483}, \href
  {http://adsabs.harvard.edu/abs/1991ApJ...379...52W} {379, 52}

\bibitem[\protect\citeauthoryear{{Wilson}}{{Wilson}}{1995}]{1995ApJ...448L..97W}
{Wilson} C.~D.,  1995, \mn@doi [\apjl] {10.1086/309615}, \href
  {http://adsabs.harvard.edu/abs/1995ApJ...448L..97W} {448, L97}

\bibitem[\protect\citeauthoryear{{Wolfire}, {Hollenbach}  \& {McKee}}{{Wolfire}
  et~al.}{2010}]{2010ApJ...716.1191W}
{Wolfire} M.~G.,  {Hollenbach} D.,   {McKee} C.~F.,  2010, \mn@doi [\apj]
  {10.1088/0004-637X/716/2/1191}, \href
  {http://adsabs.harvard.edu/abs/2010ApJ...716.1191W} {716, 1191}

\bibitem[\protect\citeauthoryear{{Wright}, {Eisenhardt}, {Mainzer}, {Ressler},
  {Cutri}  \& {Jarrett}}{{Wright} et~al.}{2010}]{2010AJ....140.1868W}
{Wright} E.~L.,  {Eisenhardt} P.~R.~M.,  {Mainzer} A.~K.,  {Ressler} M.~E.,
  {Cutri} R.~M.,   {Jarrett} T.,  2010, \mn@doi [\aj]
  {10.1088/0004-6256/140/6/1868}, \href
  {http://adsabs.harvard.edu/abs/2010AJ....140.1868W} {140, 1868}

\bibitem[\protect\citeauthoryear{{Wuyts} et~al.,}{{Wuyts}
  et~al.}{2016}]{2016arXiv160301139W}
{Wuyts} E.,  et~al., 2016, preprint, \href
  {http://adsabs.harvard.edu/abs/2016arXiv160301139W} {} (\mn@eprint {arXiv}
  {1603.01139})

\bibitem[\protect\citeauthoryear{{Young} \& {Scoville}}{{Young} \&
  {Scoville}}{1991}]{1991ARA&A..29..581Y}
{Young} J.~S.,  {Scoville} N.~Z.,  1991, \mn@doi [\araa]
  {10.1146/annurev.aa.29.090191.003053}, \href
  {http://adsabs.harvard.edu/abs/1991ARA%26A..29..581Y} {29, 581}

\bibitem[\protect\citeauthoryear{{van Dishoeck} \& {Black}}{{van Dishoeck} \&
  {Black}}{1986}]{1986ApJS...62..109V}
{van Dishoeck} E.~F.,  {Black} J.~H.,  1986, \mn@doi [\apjs] {10.1086/191135},
  \href {http://adsabs.harvard.edu/abs/1986ApJS...62..109V} {62, 109}

\bibitem[\protect\citeauthoryear{{van Dishoeck} \& {Black}}{{van Dishoeck} \&
  {Black}}{1988}]{1988ApJ...334..771V}
{van Dishoeck} E.~F.,  {Black} J.~H.,  1988, \mn@doi [\apj] {10.1086/166877},
  \href {http://adsabs.harvard.edu/abs/1988ApJ...334..771V} {334, 771}

\makeatother
\end{thebibliography}

\bsp
\appendix
\section{Herschel PACS and IRAM beam sizes and reduced spectra} 
\label{AppA}
Below we present the available Herschel and IRAM data we have assembled for each of the 24 xCOLD GASS galaxies, similar to that in Figure \ref{example_galaxy_herschel_iram_data}.
\begin{figure}
  \centering
\includegraphics[scale=0.48]{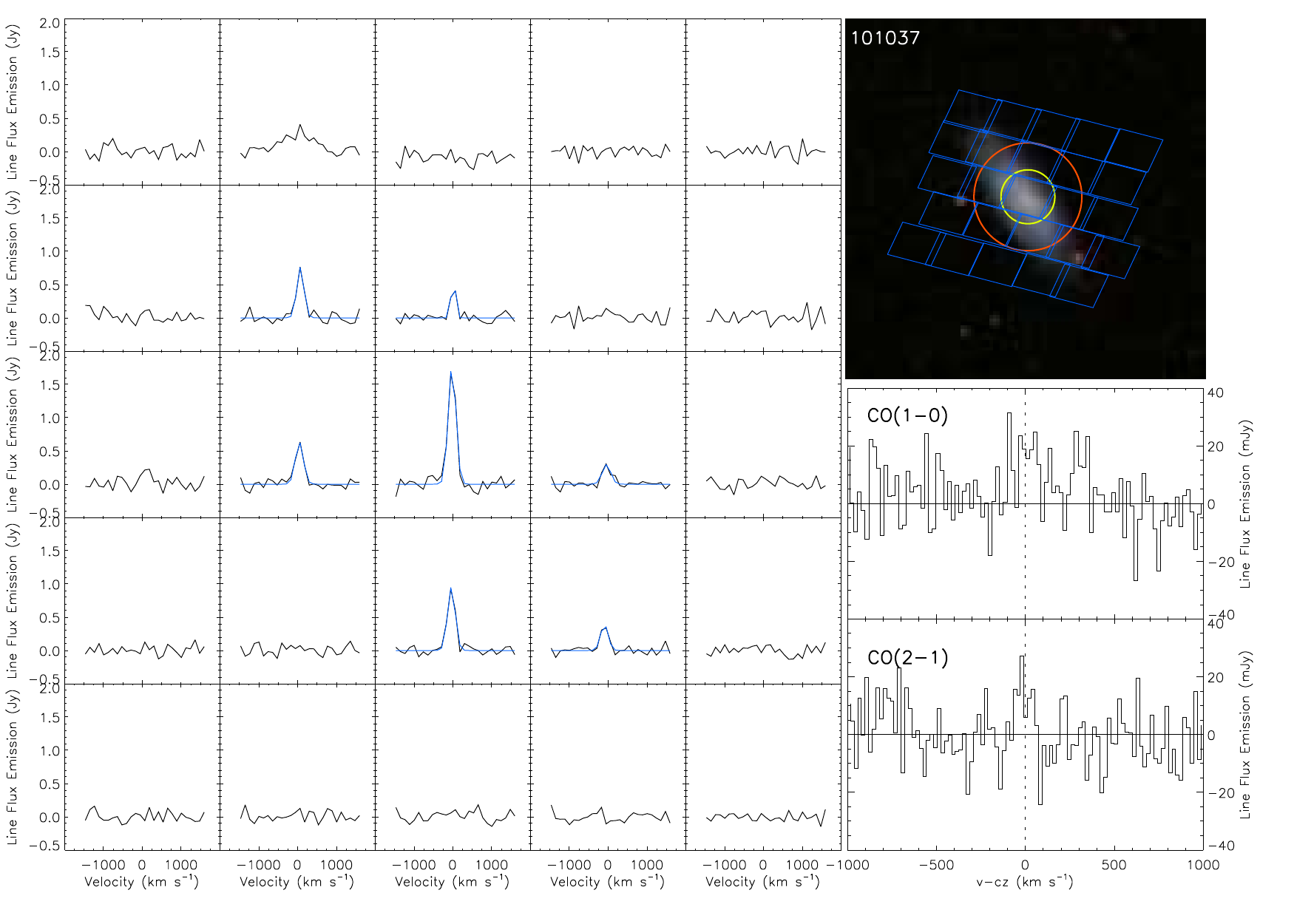}
\includegraphics[scale=0.48]{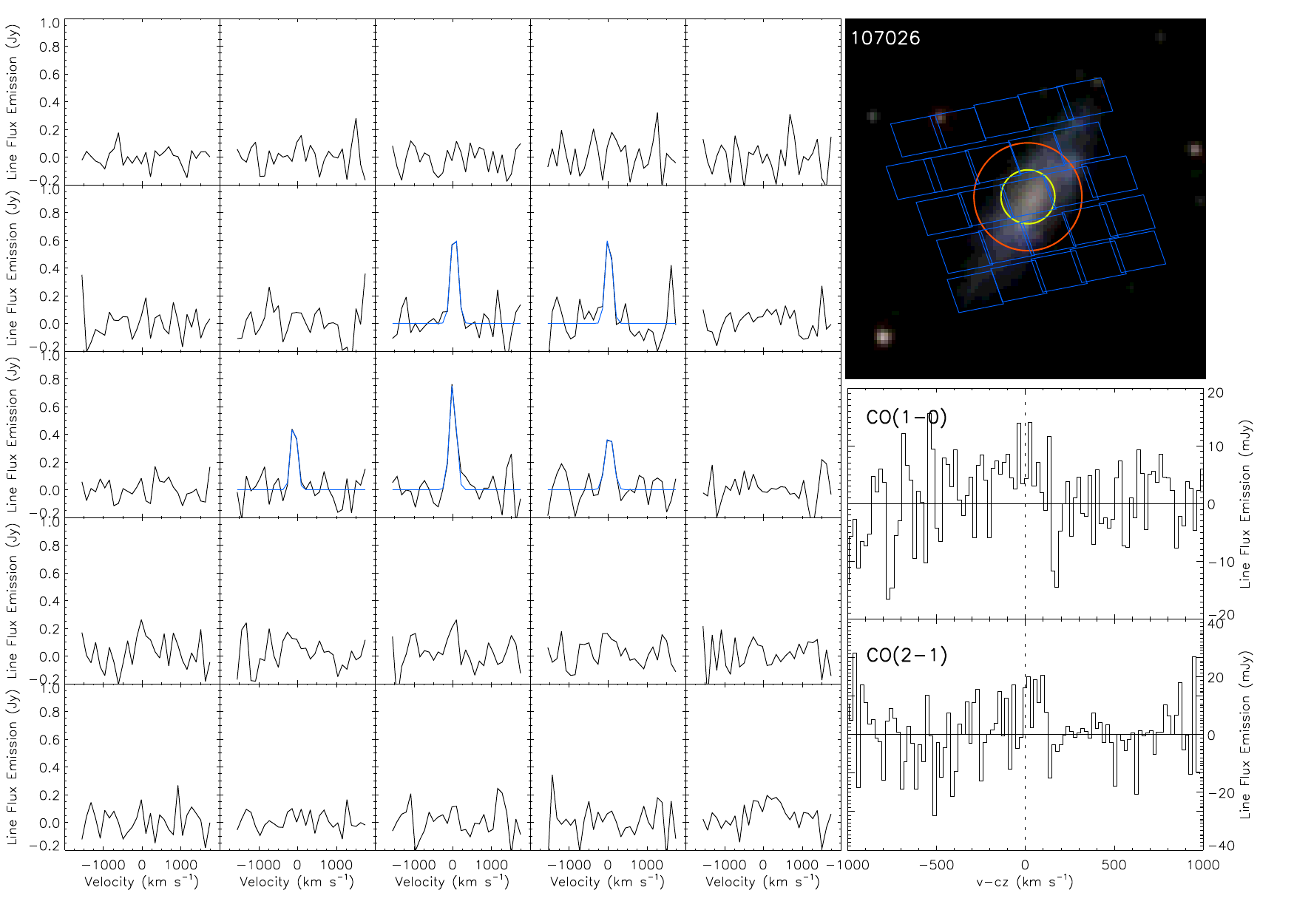}
\includegraphics[scale=0.48]{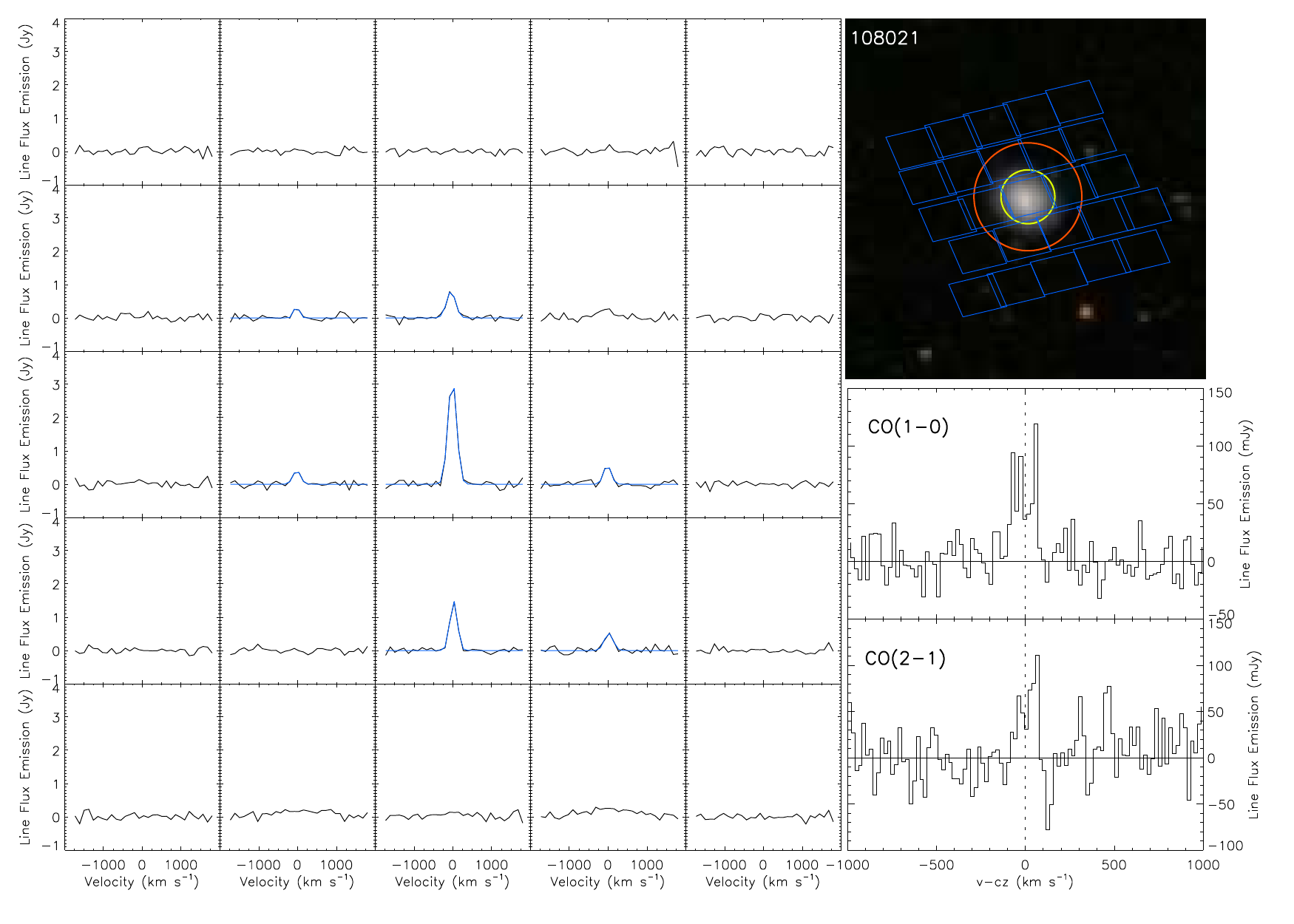}
\includegraphics[scale=0.48]{gass108045_my_reduction.pdf}
\label{} 
\end{figure}


\begin{figure}
  \centering
\includegraphics[scale=0.48]{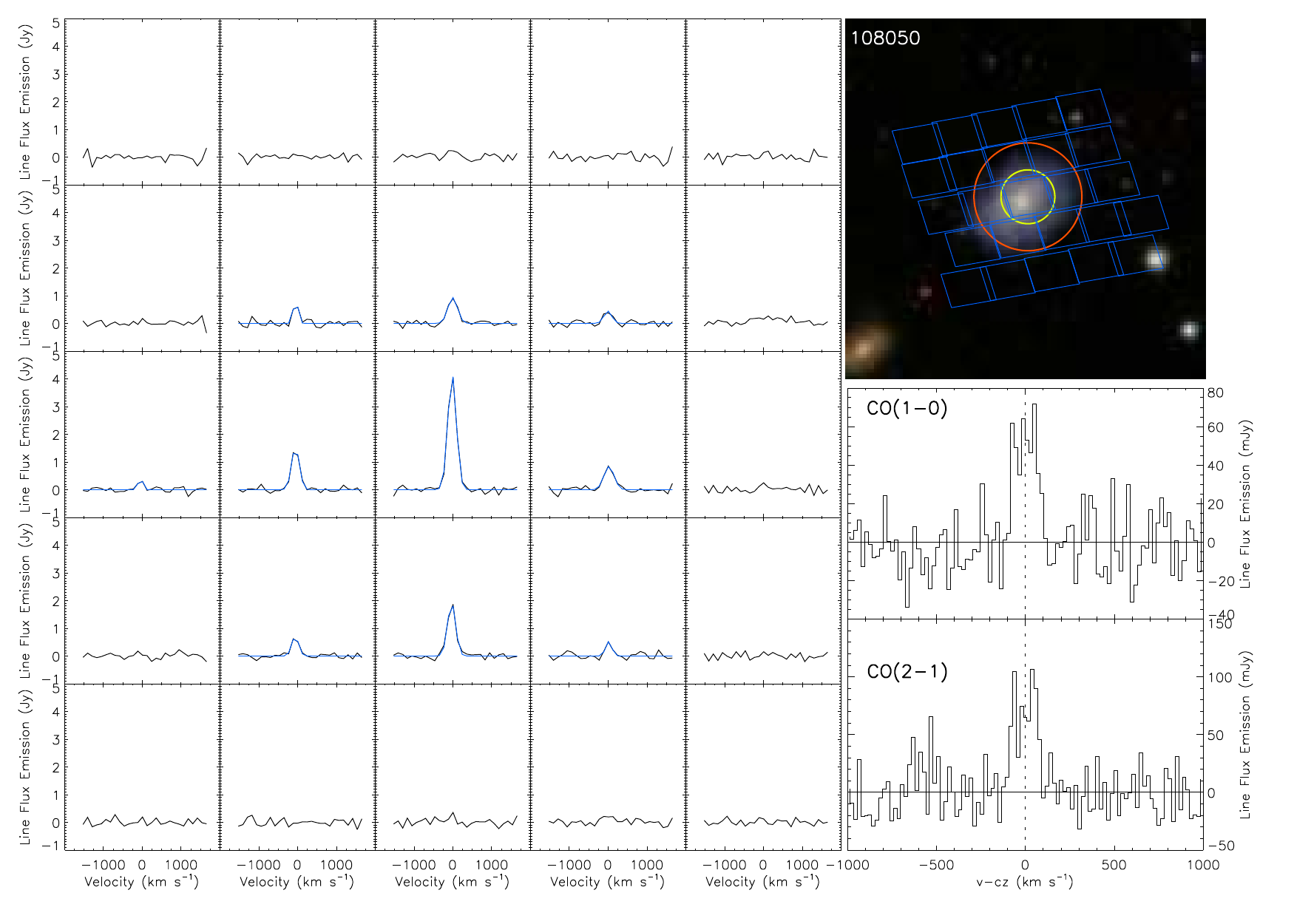}
\includegraphics[scale=0.48]{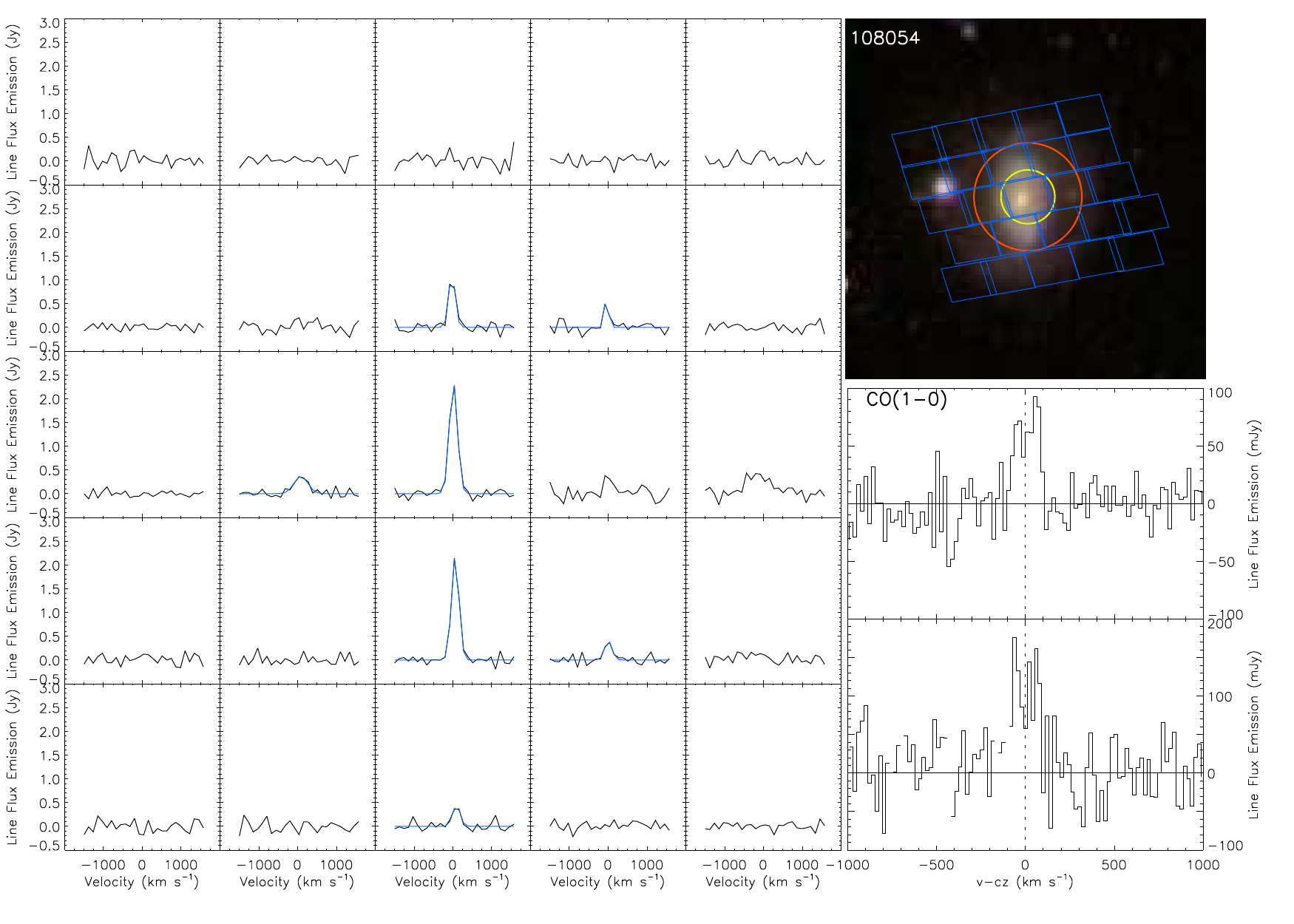}
\includegraphics[scale=0.48]{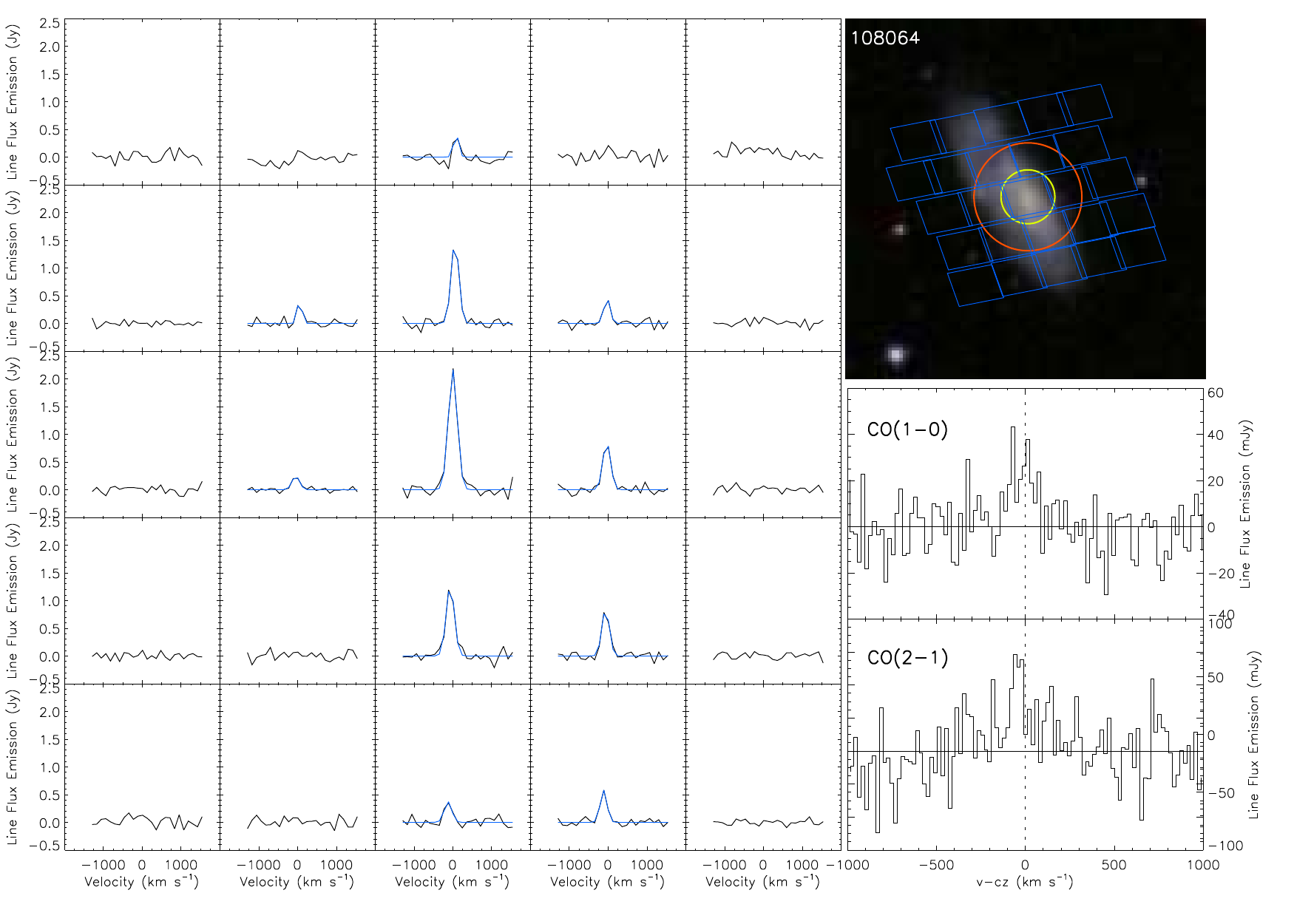}
\includegraphics[scale=0.48]{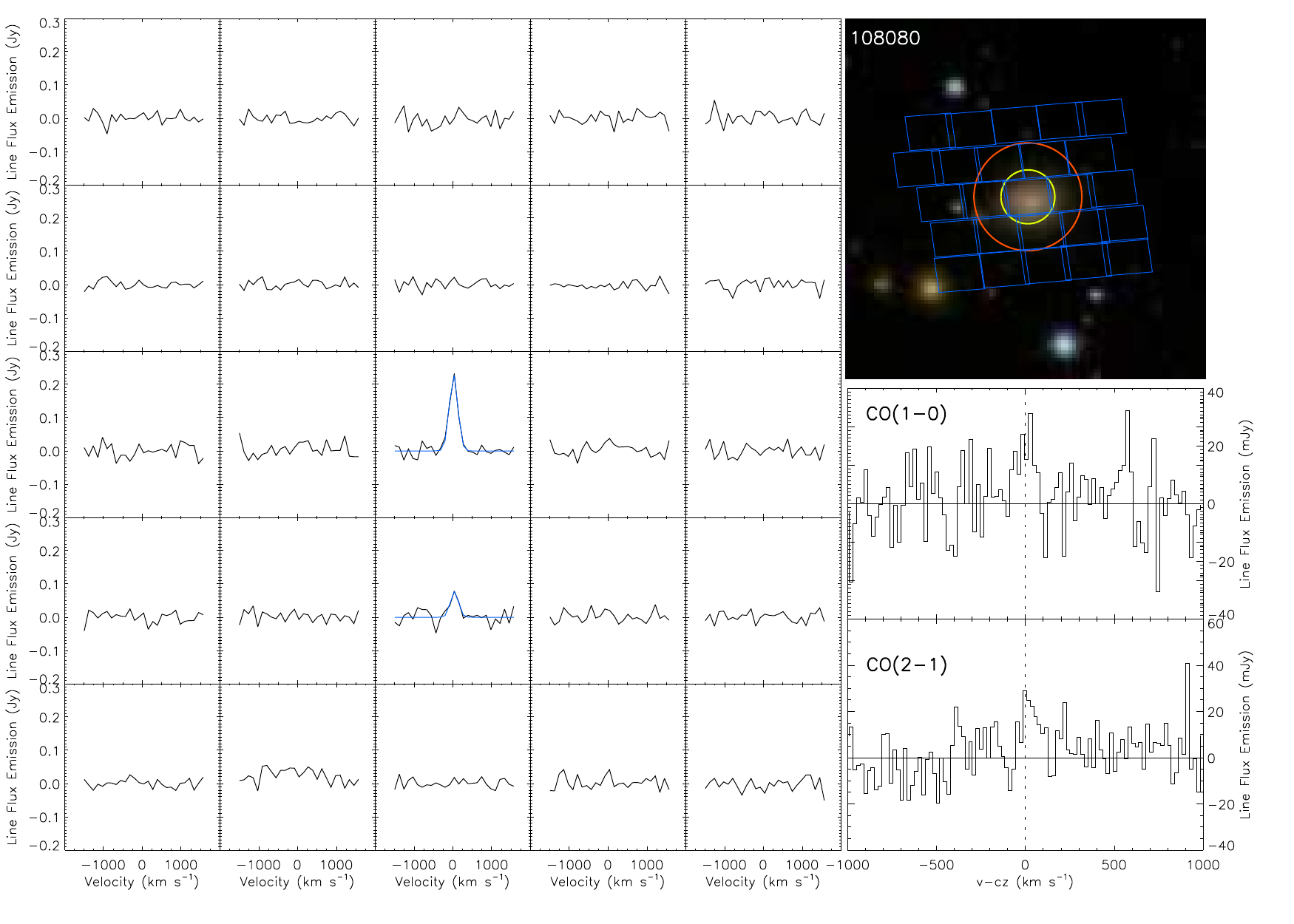}
\label{} 
\end{figure}

\begin{figure}
  \centering
\includegraphics[scale=0.48]{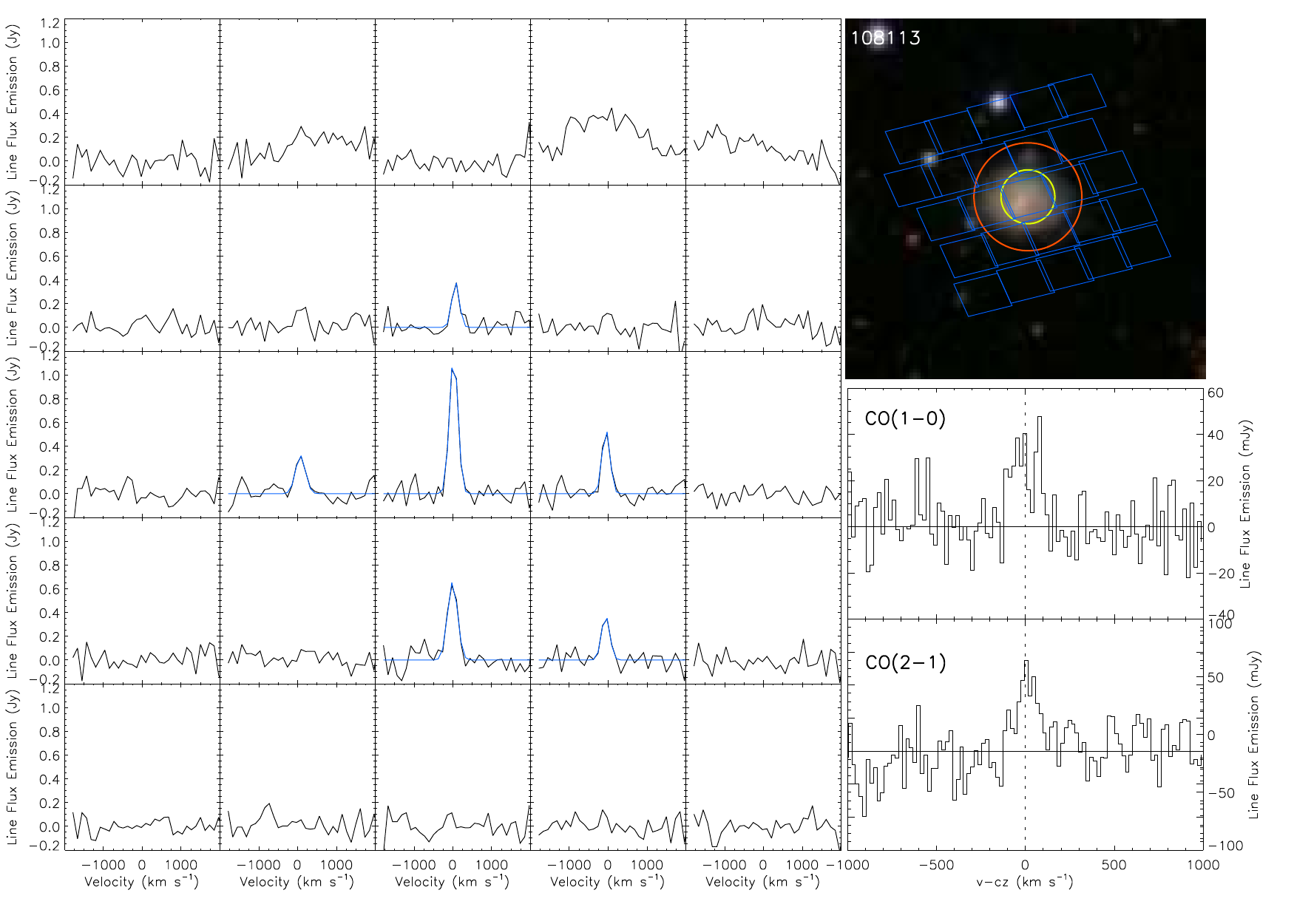}
\includegraphics[scale=0.48]{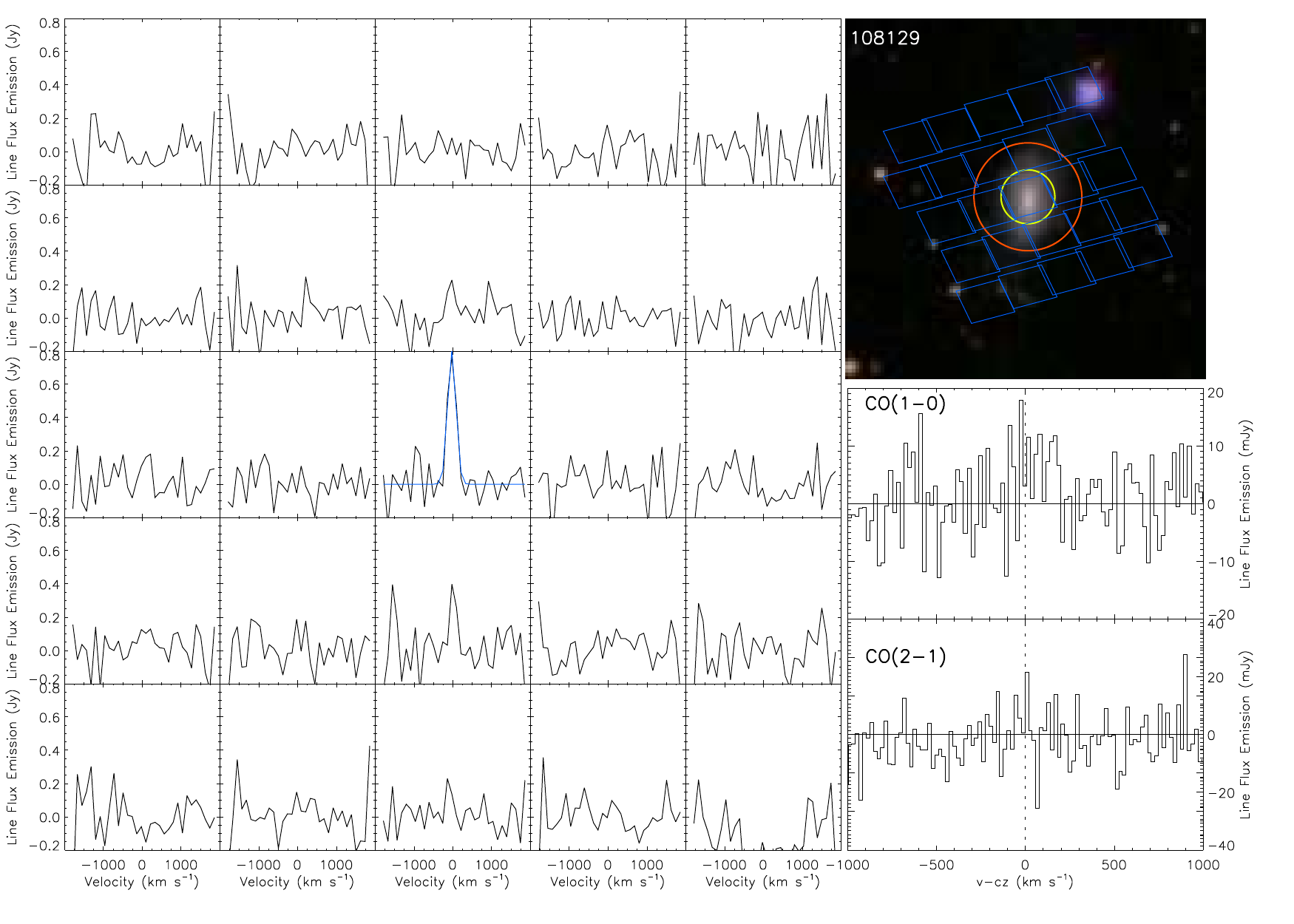}
\includegraphics[scale=0.48]{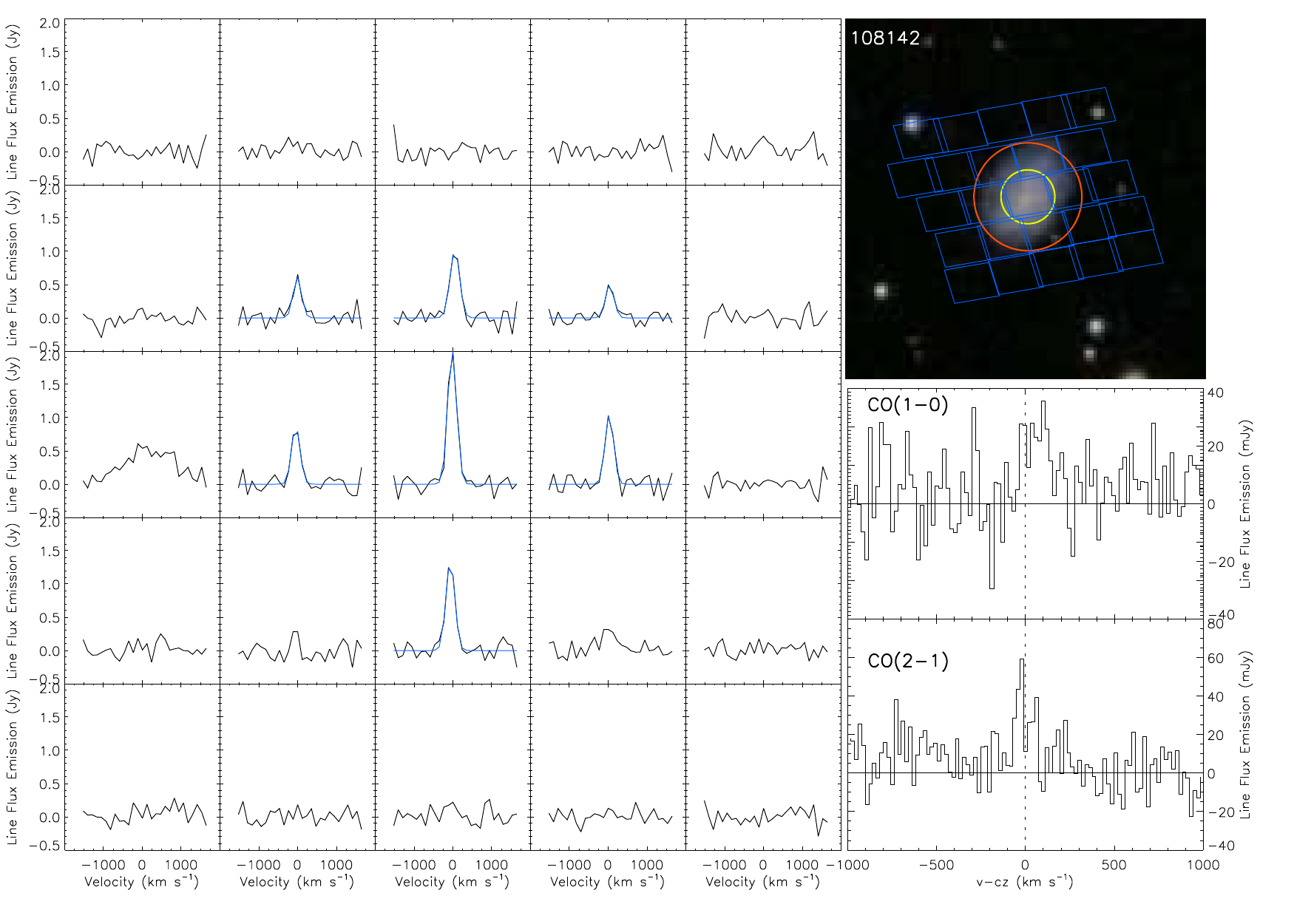}
\includegraphics[scale=0.48]{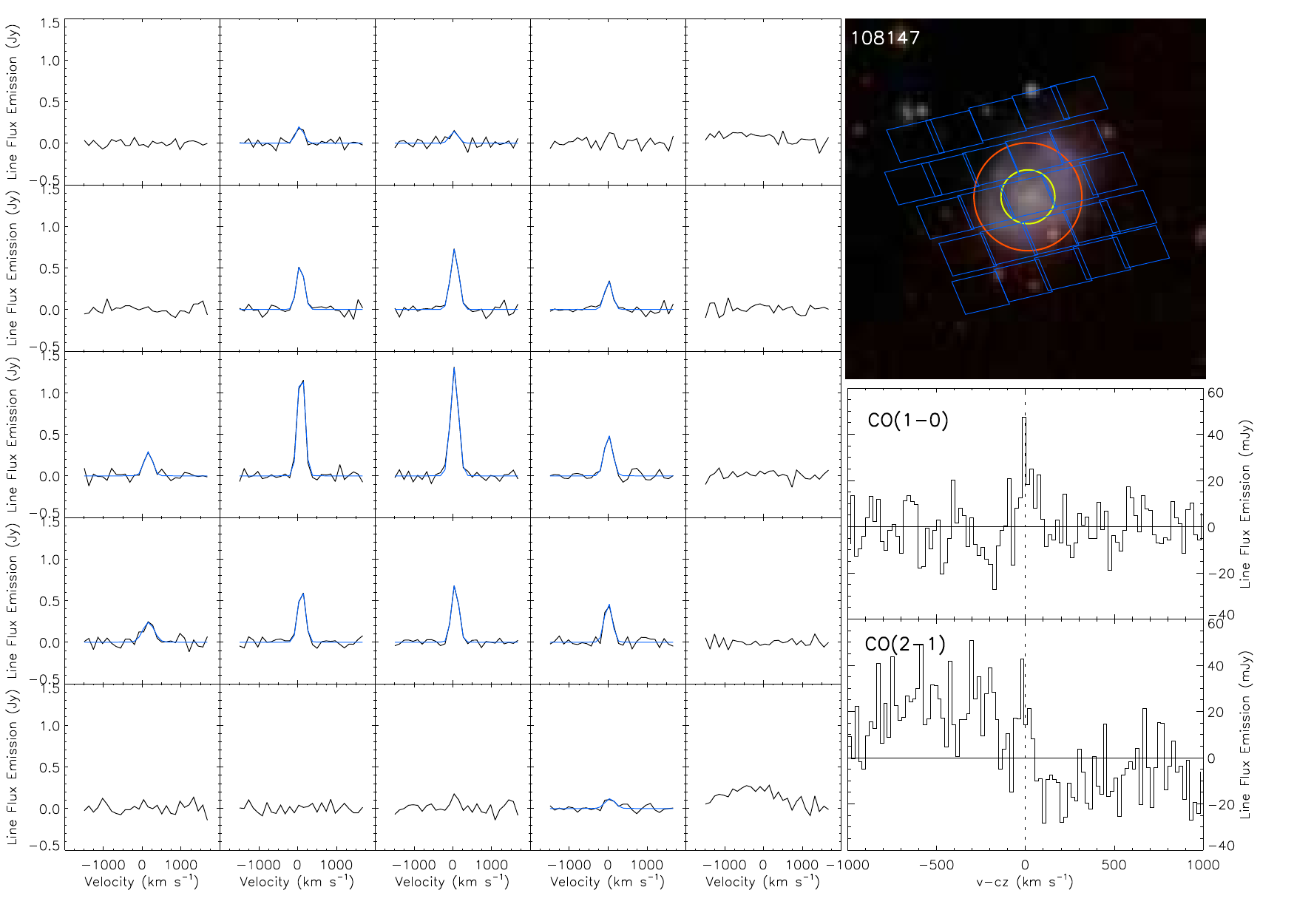}
\label{} 
\end{figure}

\begin{figure}
  \centering
\includegraphics[scale=0.48]{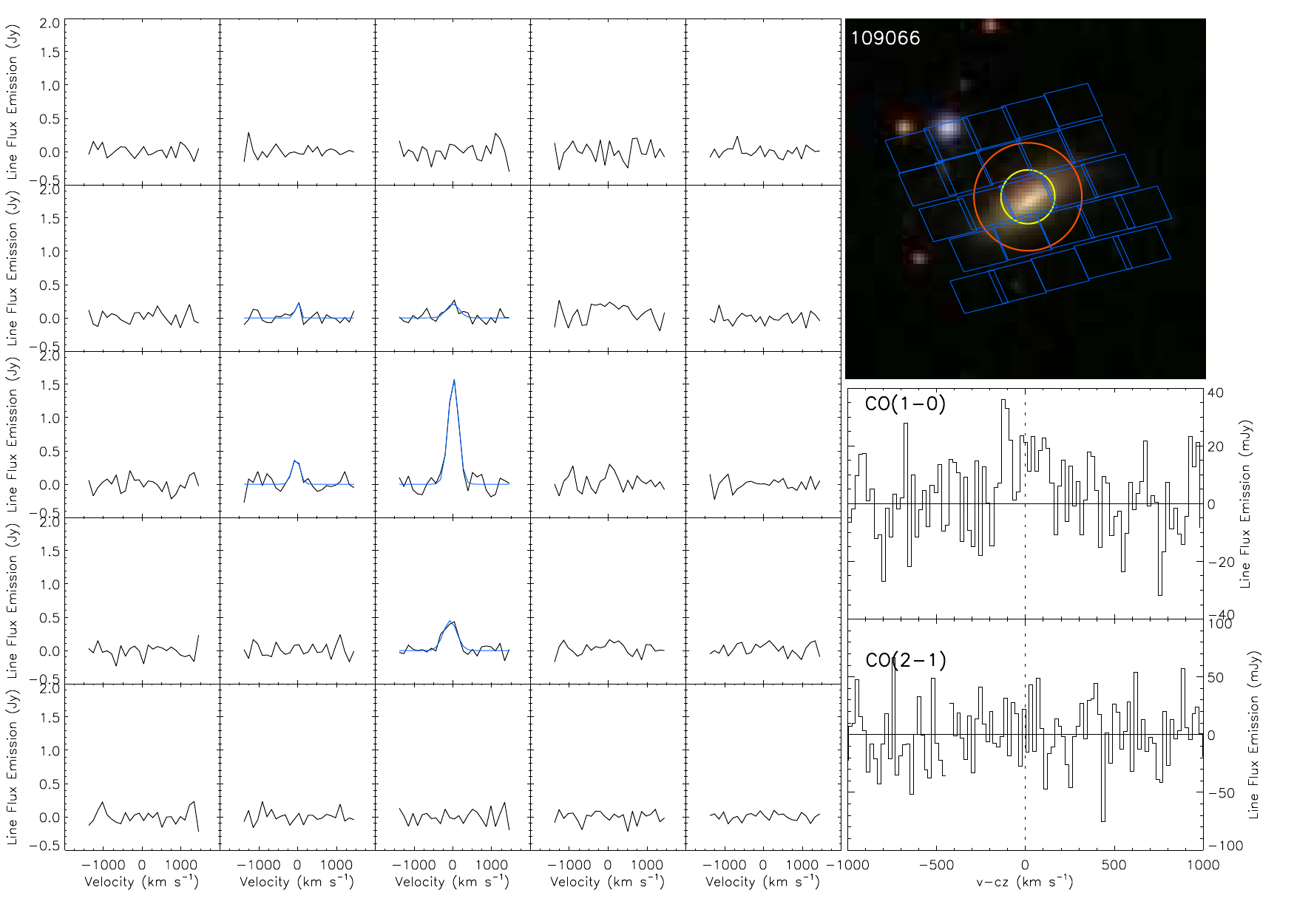}
\includegraphics[scale=0.48]{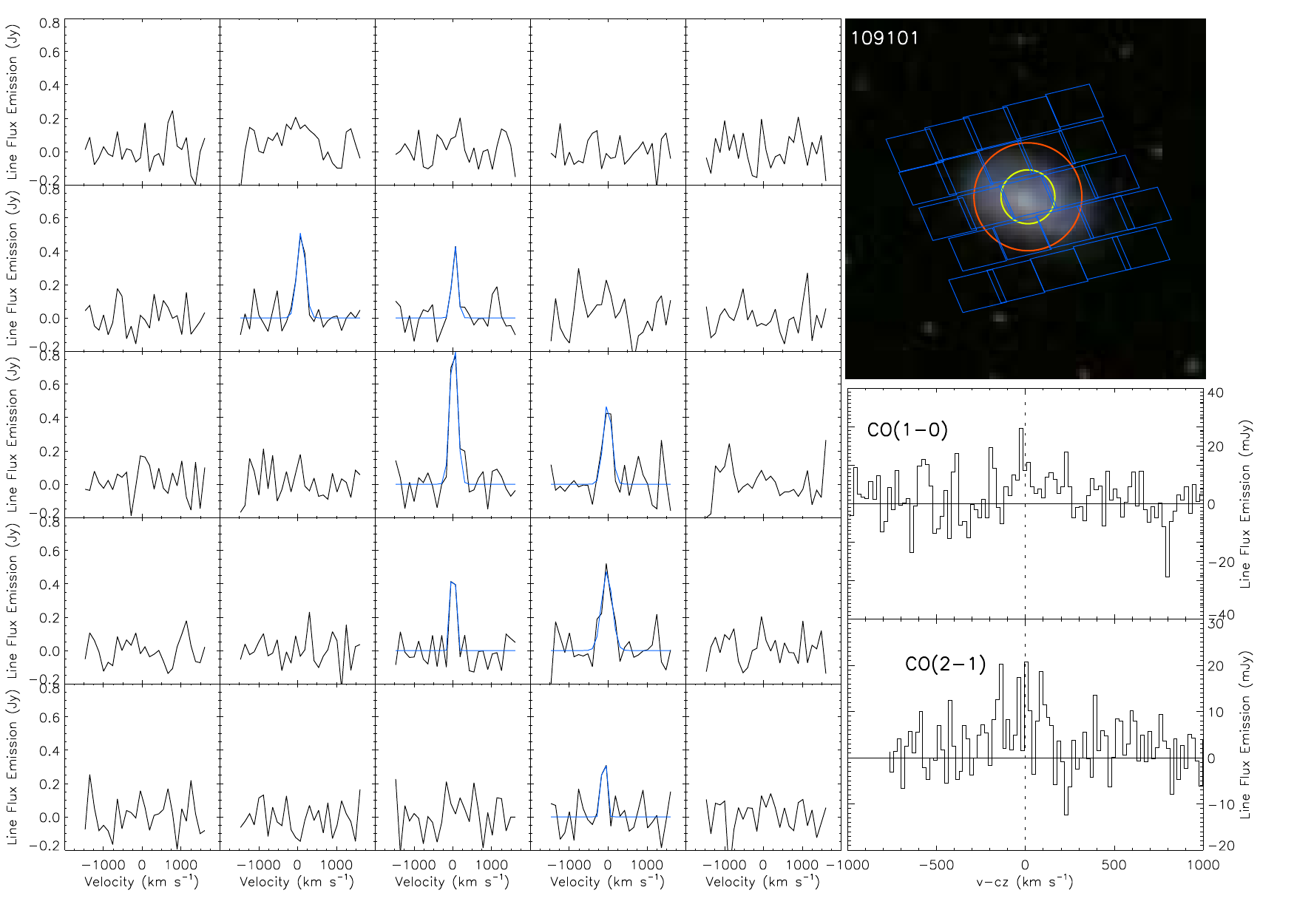}
\includegraphics[scale=0.48]{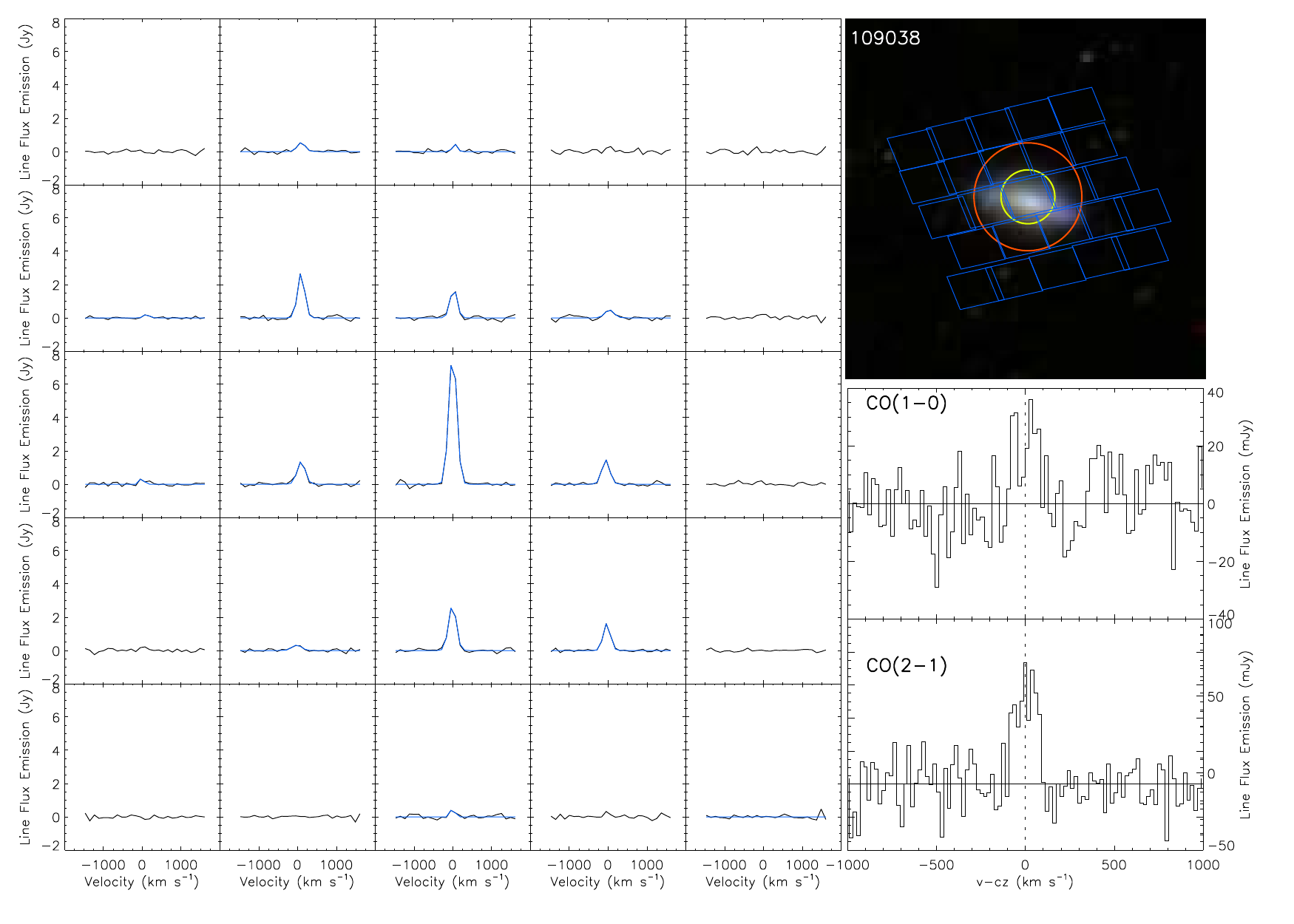}
\includegraphics[scale=0.48]{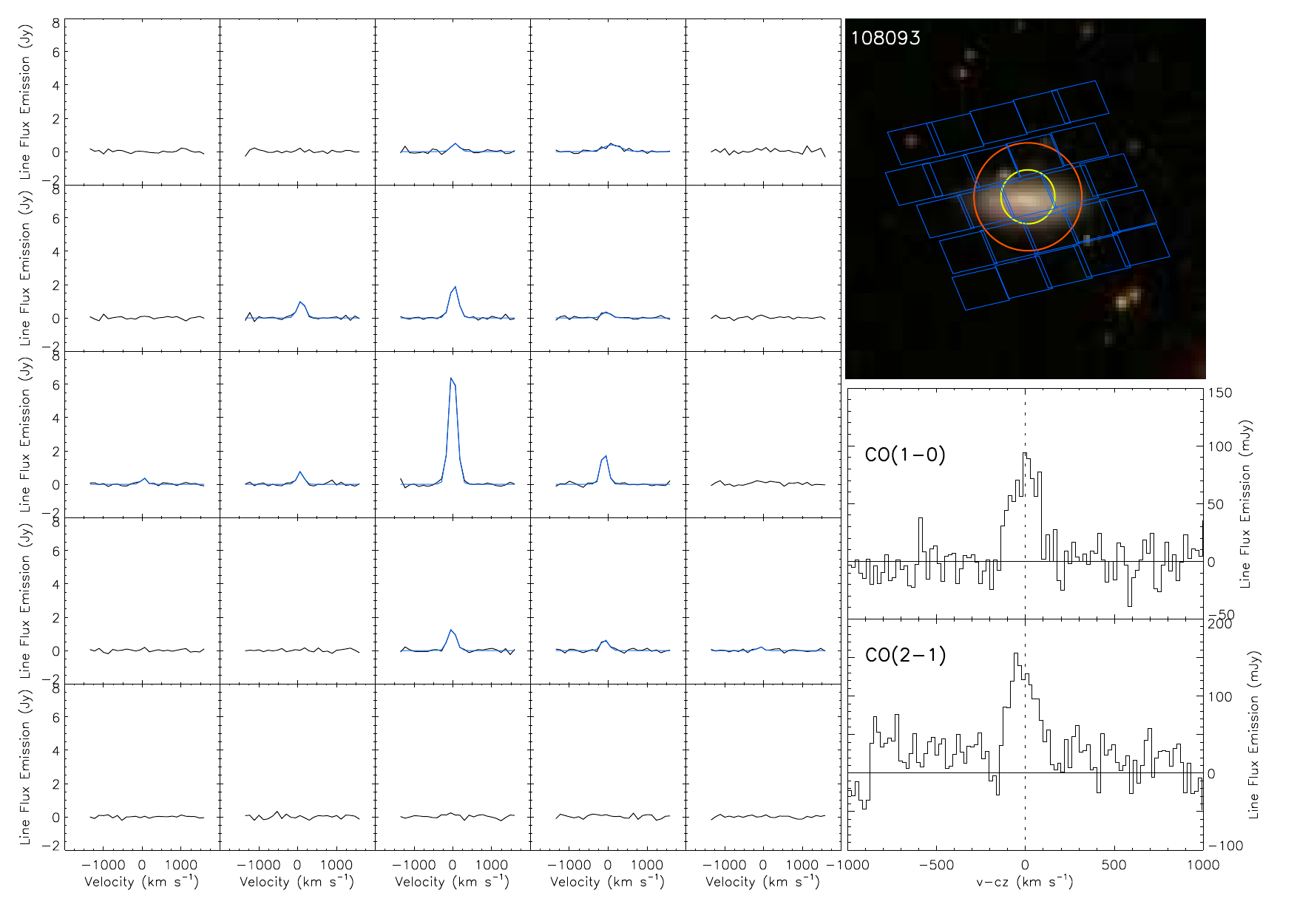}
\label{} 
\end{figure}

\begin{figure}
  \centering
\includegraphics[scale=0.48]{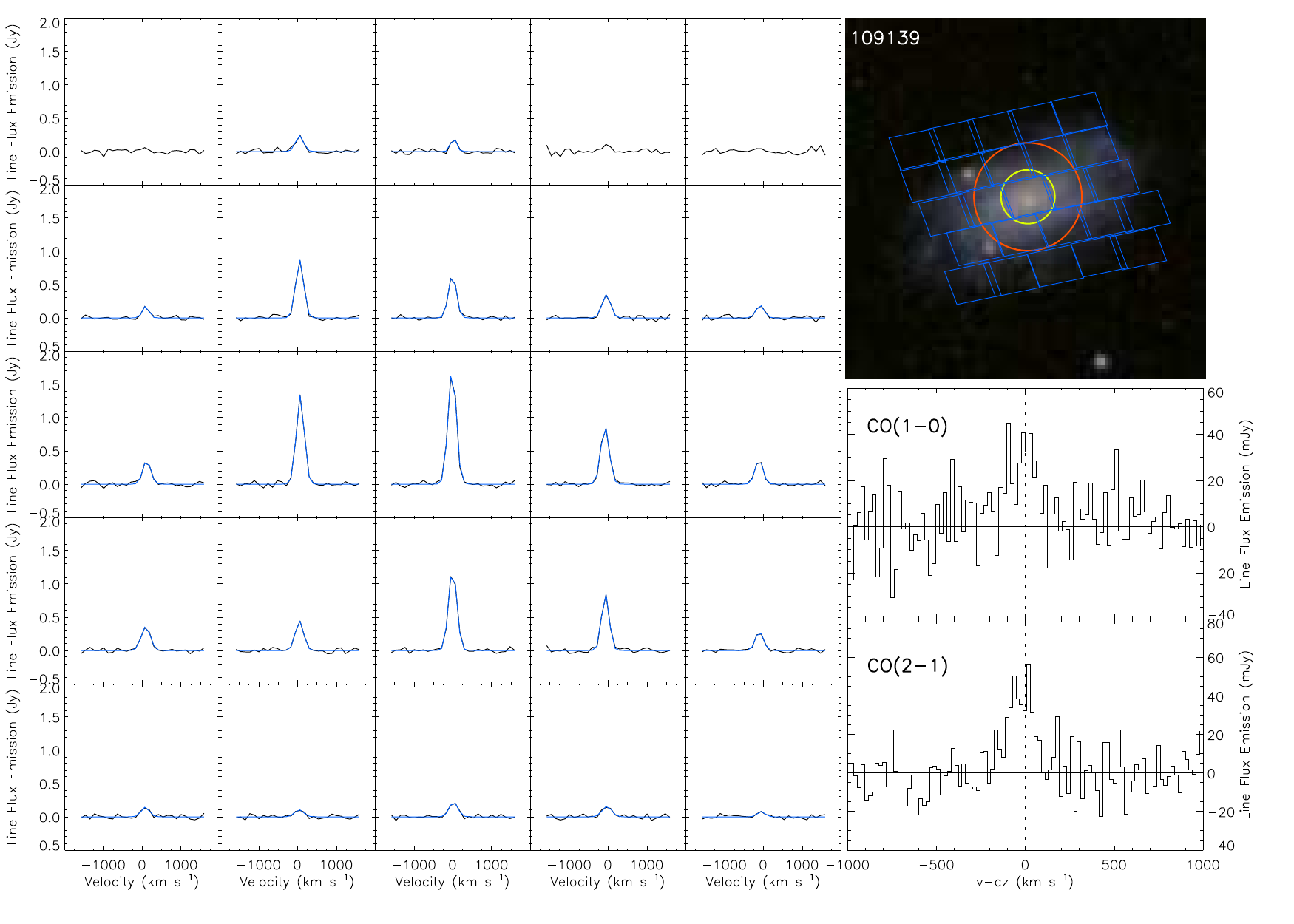}
\includegraphics[scale=0.48]{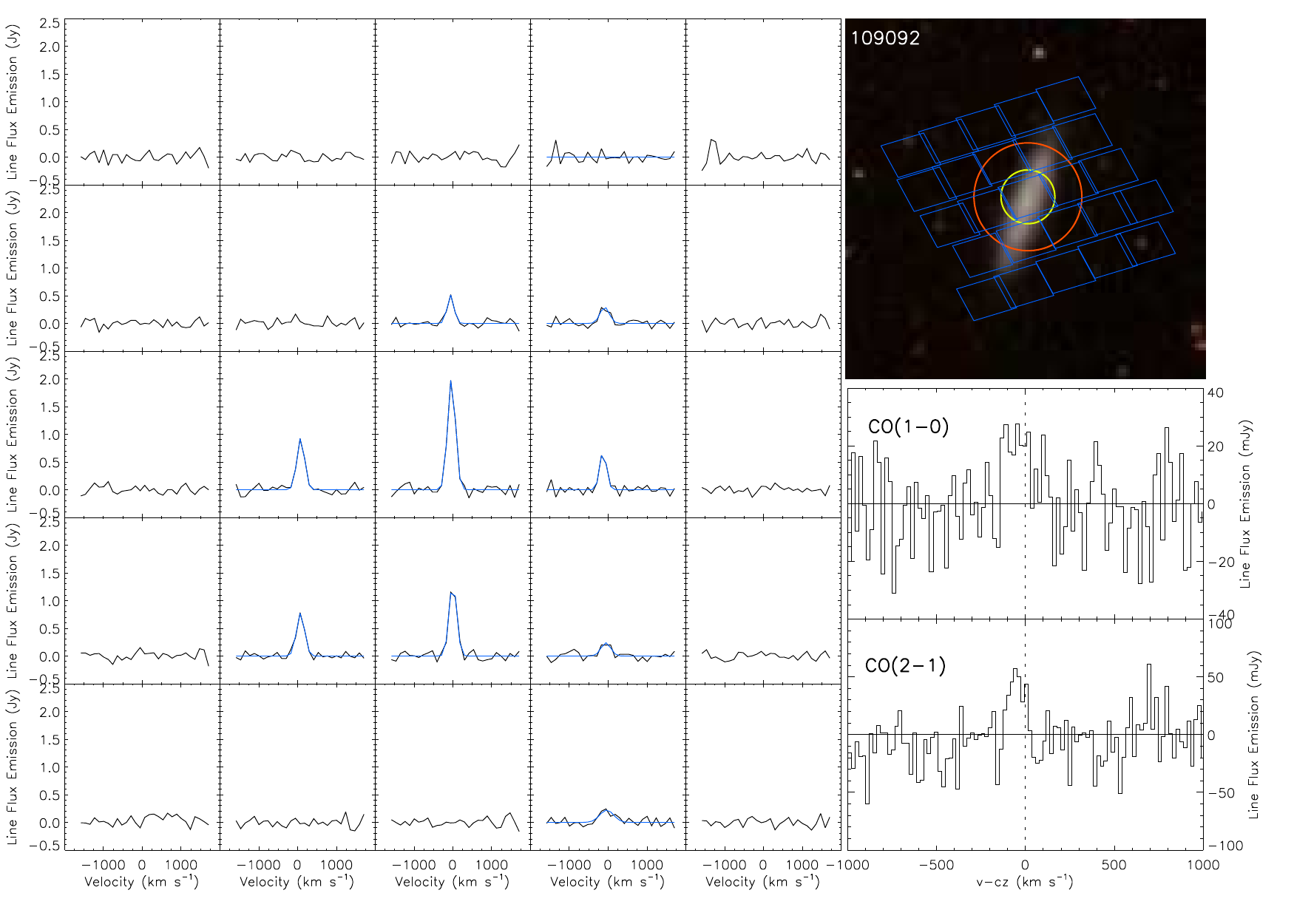}
\includegraphics[scale=0.48]{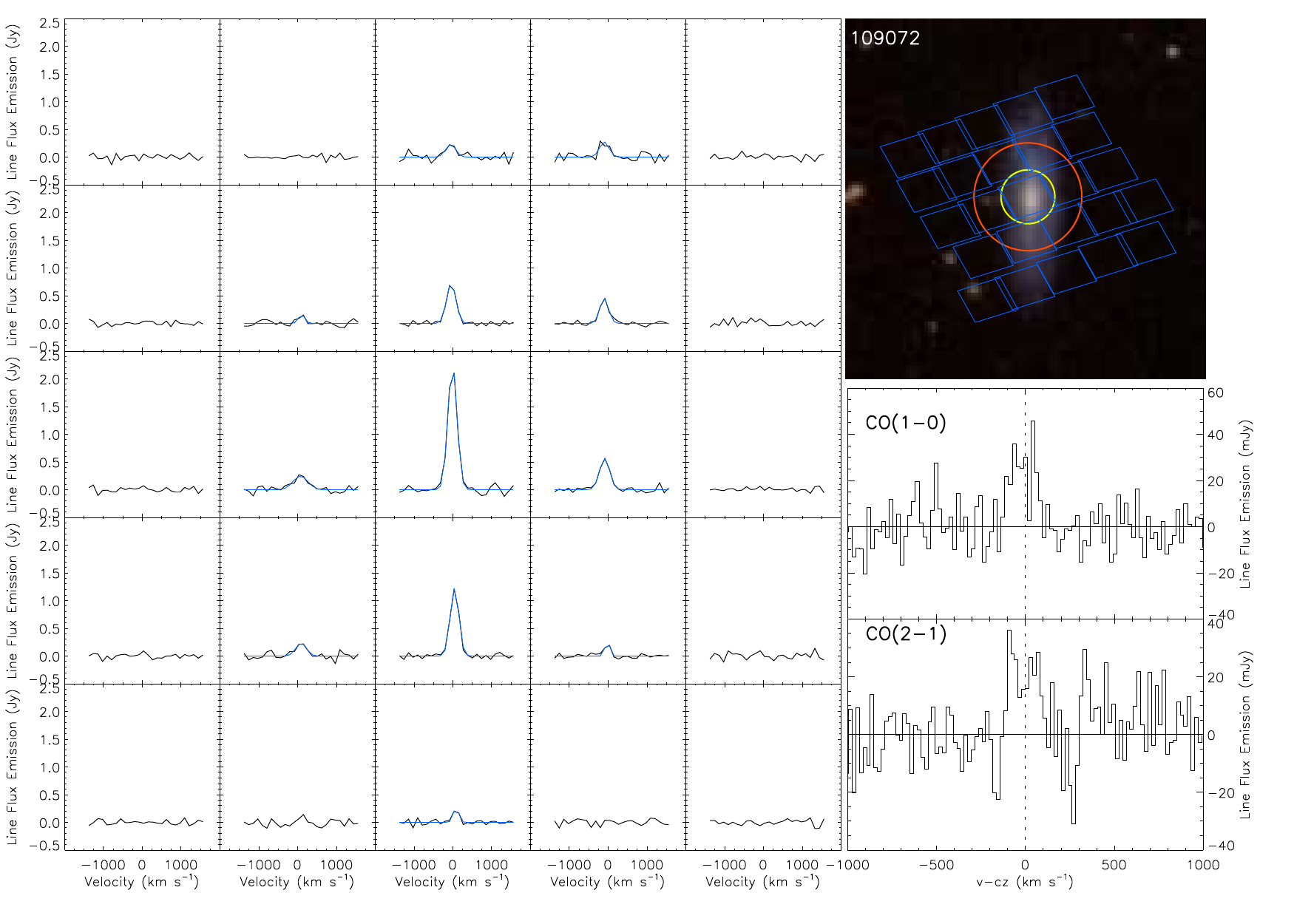}
\includegraphics[scale=0.48]{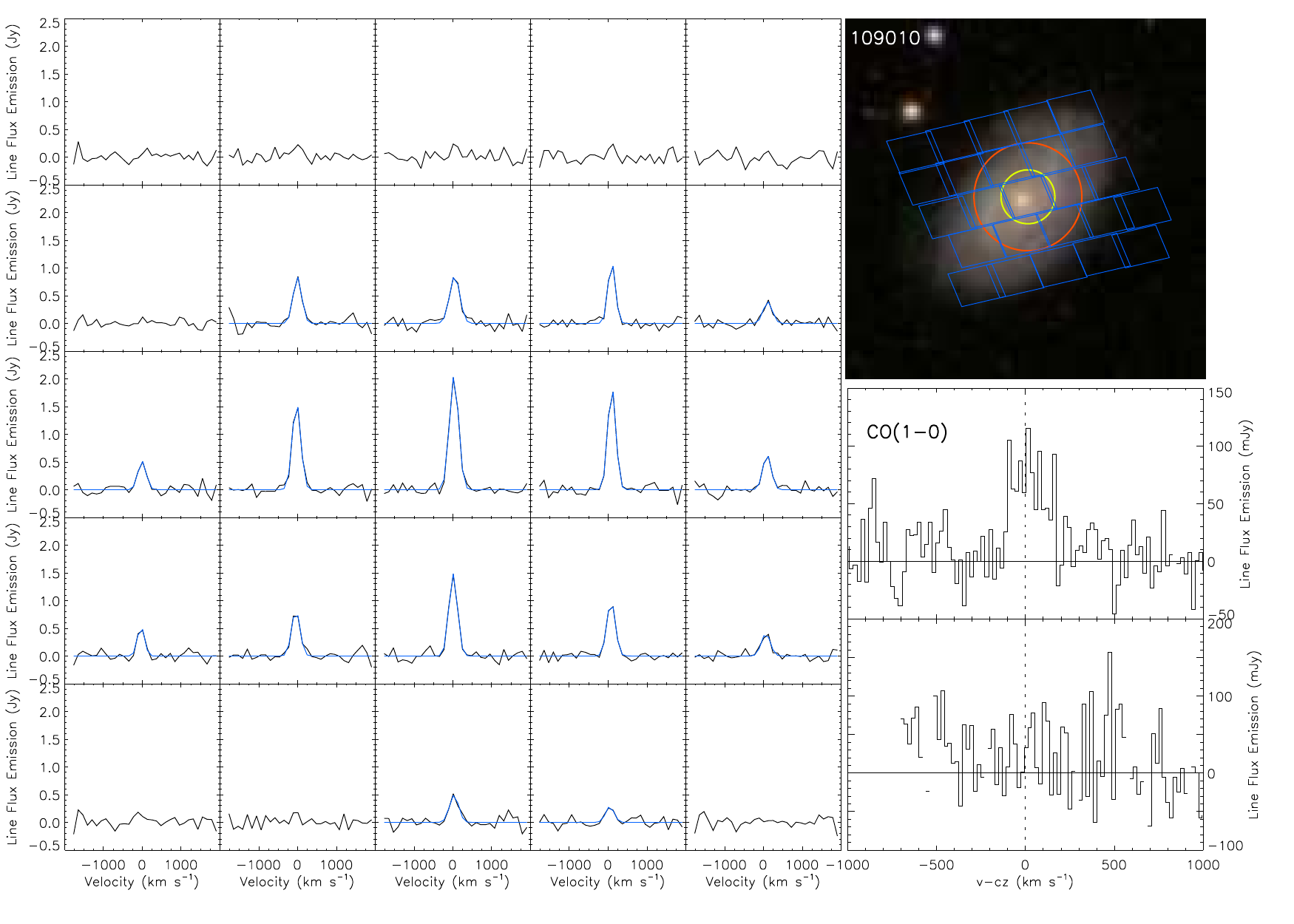}
\label{} 
\end{figure}

\begin{figure}
  \centering
\includegraphics[scale=0.48]{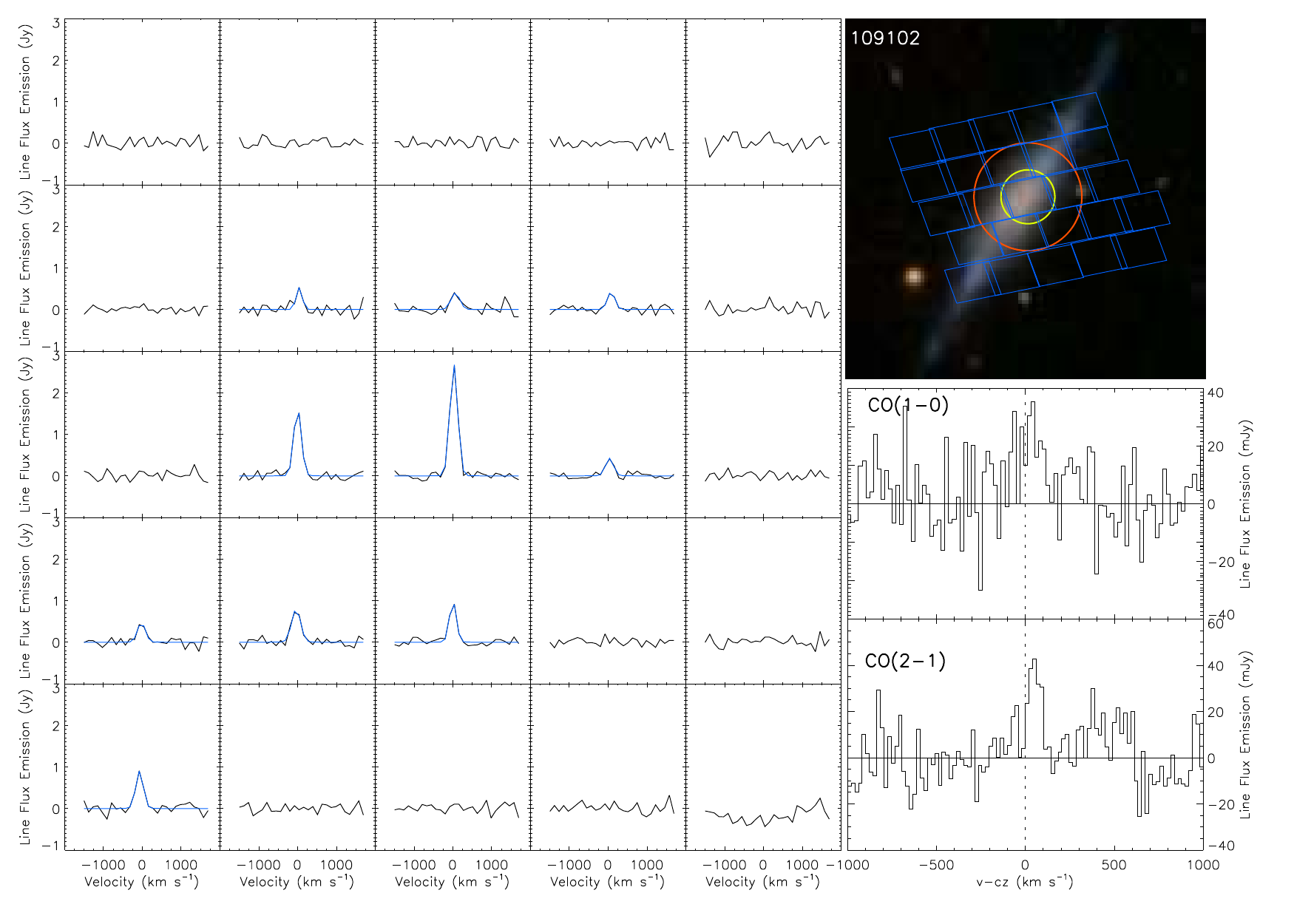}
\includegraphics[scale=0.48]{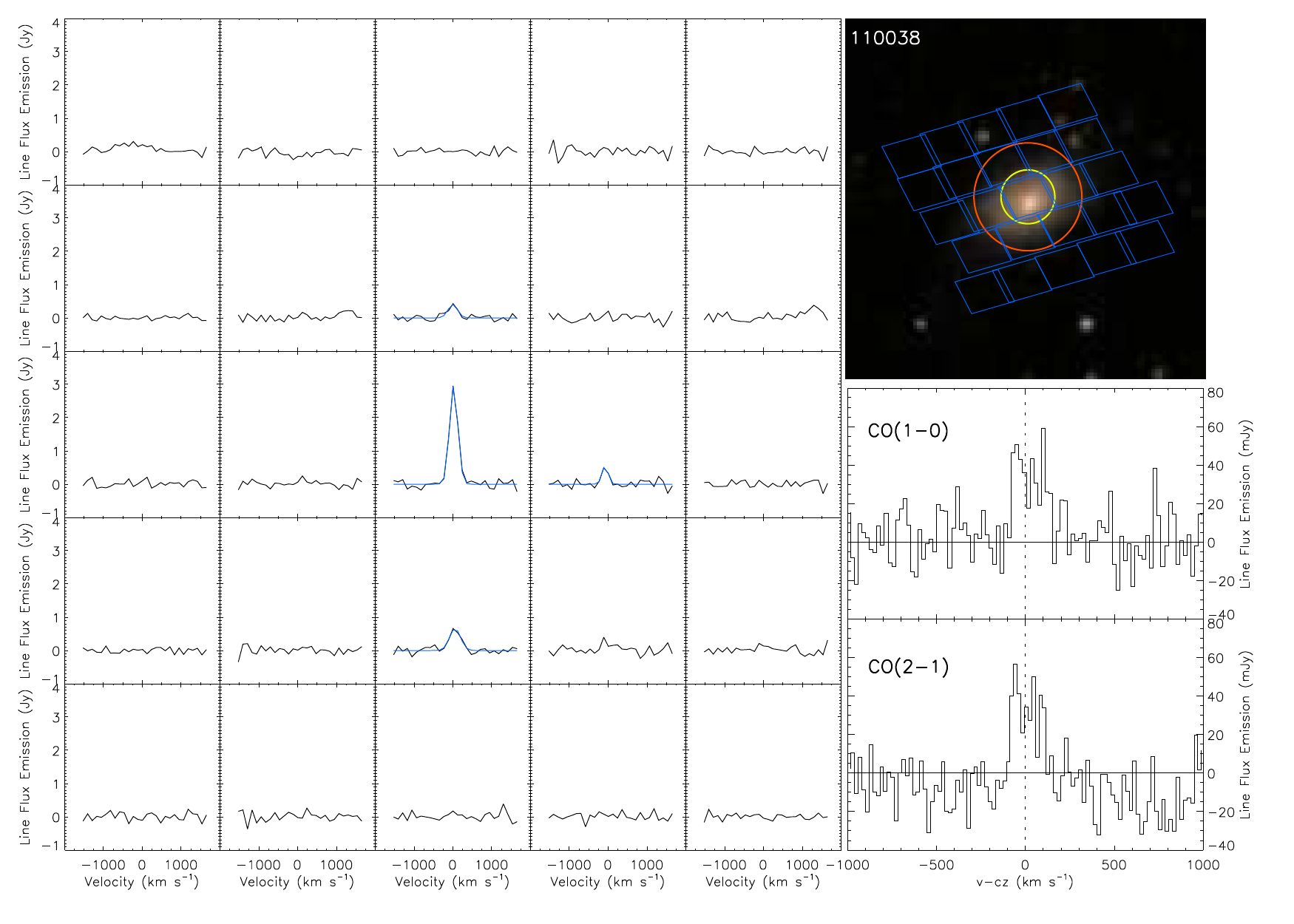}
\includegraphics[scale=0.48]{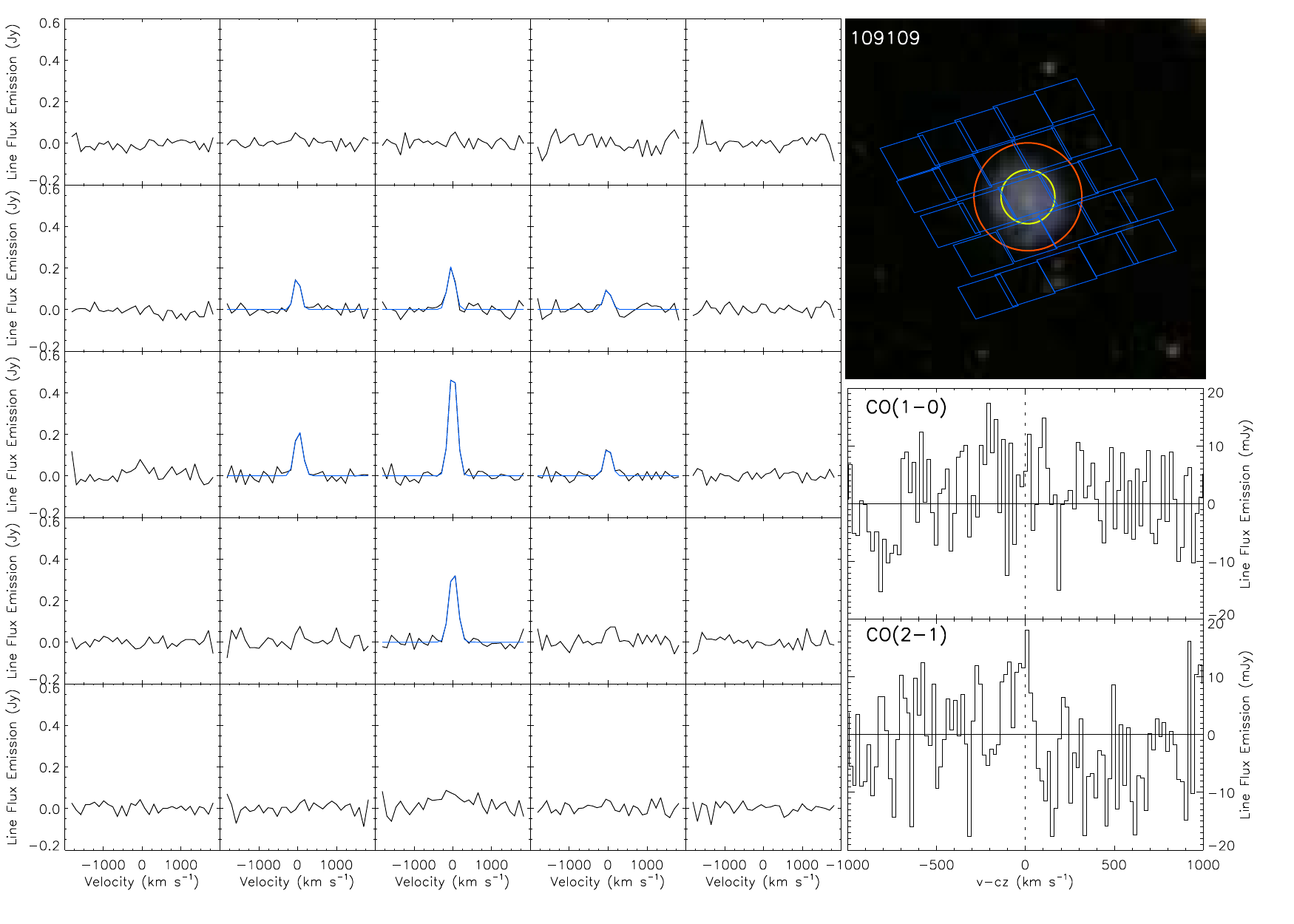}
\includegraphics[scale=0.48]{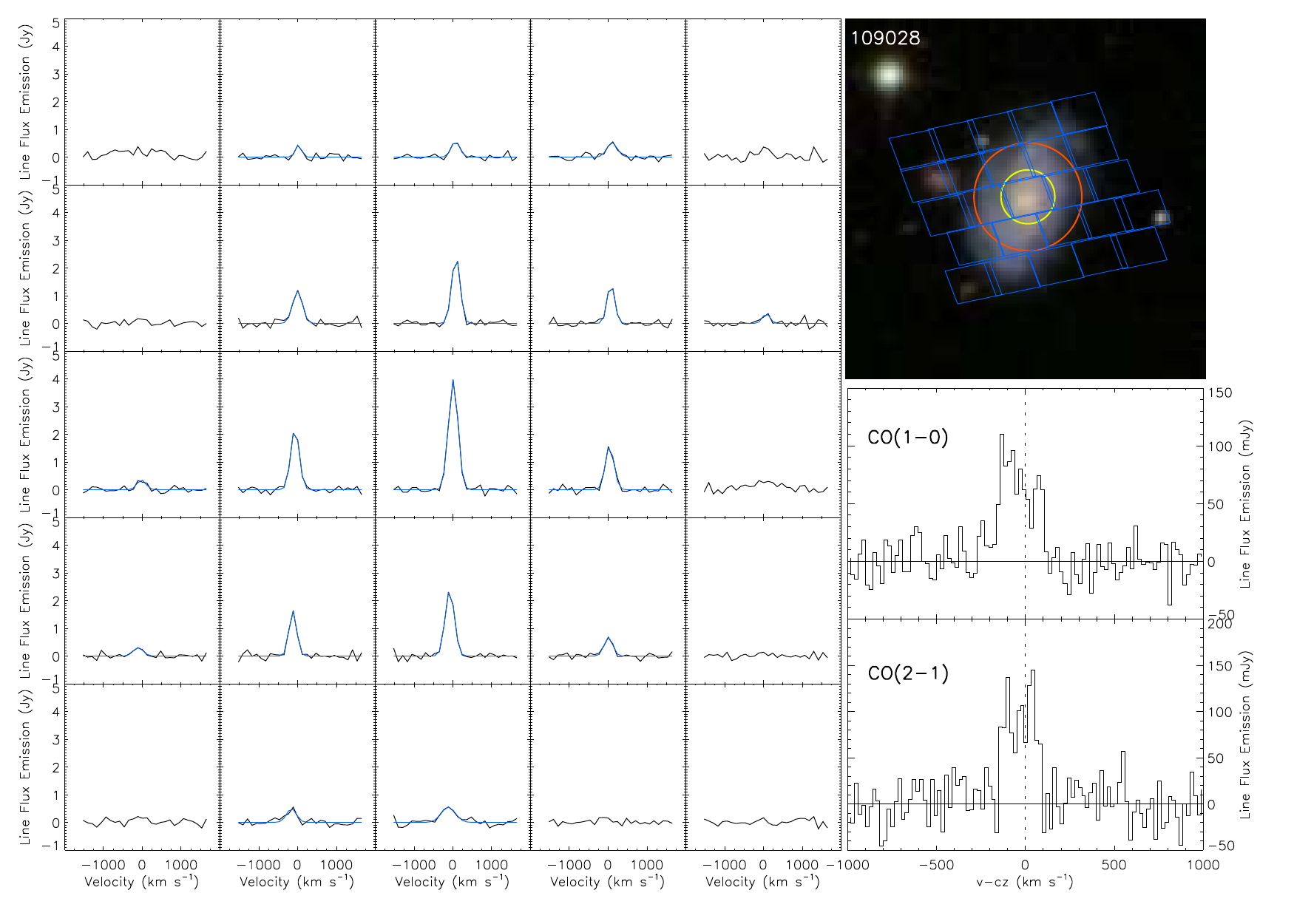}
\label{} 
\end{figure}

\end{document}